\documentclass[prxquantum,aps,floatfix,twocolumn,nopacs,superscriptaddress]{revtex4}

\usepackage[utf8]{inputenc}
\usepackage[american,]{babel}
\usepackage[T1]{fontenc}
\usepackage[pdftex]{graphicx}  
\usepackage{graphicx, xcolor}
\usepackage{dcolumn}
\usepackage{bm}
\usepackage{amsmath,amsthm,amssymb}
\usepackage{amsfonts}
\usepackage{bbold}
\usepackage{hyperref}
\usepackage{xcolor}
\hypersetup{colorlinks,bookmarksopen,bookmarksnumbered,
citecolor=blue,
linkcolor=blue,
pdfstartview=false,
urlcolor=blue}
\usepackage{graphicx}
\usepackage{braket}
\usepackage{wrapfig}
\usepackage{soul}
\usepackage{mathtools}
\usepackage{float}
\usepackage{natbib}

\begin{document}

\title{Entanglement Dynamics in Monitored Systems and the Role of Quantum Jumps}

\author{Youenn Le Gal}
\affiliation{JEIP, UAR 3573 CNRS, Coll\`ege de France,   PSL  Research  University, 11,  place  Marcelin  Berthelot,75231 Paris Cedex 05, France}
\author{Xhek Turkeshi}
\affiliation{JEIP, UAR 3573 CNRS, Coll\`ege de France,   PSL  Research  University, 11,  place  Marcelin  Berthelot,75231 Paris Cedex 05, France}
\affiliation{Institut f\"ur Theoretische Physik, Universit\"at zu K\"oln, Z\"ulpicher Strasse 77, 50937 K\"oln, Germany}
\author{Marco Schir\`o}
\affiliation{JEIP, UAR 3573 CNRS, Coll\`ege de France,   PSL  Research  University, 11,  place  Marcelin  Berthelot,75231 Paris Cedex 05, France}

\begin{abstract}
Monitored quantum many-body systems display a rich pattern of entanglement dynamics, which is unique to this non-unitary setting. 
This work studies the effect of quantum jumps on the entanglement dynamics beyond the no-click limit corresponding to a deterministic non-Hermitian evolution. To this aim we introduce a new tool which looks at the statistics of entanglement entropy gain and loss after and in-between quantum jumps. 
This insight allows us to build a simple stochastic model of a random walk with partial resetting, which reproduces the entanglement dynamics, and to dissect the mutual role of jumps and non-Hermitian evolution on the entanglement scaling.  We apply these ideas to the study of measurement-induced transitions in monitored fermions. We demonstrate that significant deviations from the no-click limit arise whenever quantum jumps strongly renormalize the non-Hermitian dynamics, as in the case of models with $U(1)$ symmetry at weak monitoring. On the other hand, we show that the weak monitoring phase of the Ising chain leads to a robust sub-volume logarithmic phase due to weakly renormalized non-Hermitian dynamics. 
\end{abstract}

\date{\today}
\maketitle

\section{Introduction}

The spreading of quantum entanglement under unitary dynamics displays remarkable robustness and universality~\cite{Calabrese_2005,liu2014entanglement,nahum2017quantum}. For example, in clean systems with short-ranged interactions, the entanglement entropy is generally expected to grow linearly in time and to saturate to a volume law~\cite{DeChiara_2006,kim2013ballistic}; the violation of this behavior is often taken as a smoking gun of non-ergodic dynamics~\cite{bardarson2012unbounded}. On the other hand, non-unitary processes such as quantum measurements can strongly affect how entanglement spreads throughout the system. Out of this competition, a novel type of measurement-induced phase transition (MIPT) in the entanglement content of the system has been discovered~\cite{fisher2023randomquantumcircuits,potter2022quantumsciencesandtechnology,lunt2022quantumsciencesandtechnology}.

Entanglement transitions due to measurements have been studied broadly in two somewhat different settings. On the one hand, in the stochastic dynamics encoded in a quantum many-body trajectory describing the system's evolution conditioned to a set of measurement outcomes. In this setting, the criticality is hidden in the rare fluctuations of the measurement process, probed by a non-linear functional of the state such as the entanglement entropy or the purity in a dynamical purification protocol~\cite{gullans2020dynamicalpurificationphase}, while conventional observables averaged over the noise are usually transparent to it. 
This makes the theoretical description and experimental detection of MIPT particularly challenging, even though recent progress has been made~\cite{noel2021measurementinducedquantum,koh2022experimentalrealizationof,hoke2023quantuminformationphases}. On the theoretical front volume-to-area law entanglement transitions have been reported in monitored random circuits~\cite{li2018quantumzenoeffect,li2019measurementdrivenentanglement,skinner2019measurementinducedphase,szyniszewski2019entanglementtransitionfrom,jian2020measurementinducedcriticality,choi2020quantumerrorcorrection,zabalo2022operatorscalingdimensions,sierant2022universalbehaviorbeyond,sierant2022measurementinducedphase,klocke2023majorana} and non-integrable Hamiltonian~\cite{fuji2020measurementinducedquantum,lunt2020measurementinducedentanglement,dogger2022generalizedquantummeasurements,xing2023interactions,altland2022dynamics} with projective or weak measurements.
Monitored non-interacting systems, on the other hand, are not expected to sustain a volume law phase~\cite{cao2019entanglementina,fidkowski2021howdynamicalquantum}. Still, a critical sub-volume phase, whose origin and stability are currently under debate~\cite{coppola2022growthofentanglement,loio2023purificationtimescalesin,poboiko2023theoryoffree,jian2023measurementinducedentanglement,fava2023nonlinearsigmamodels,carisch2023quantifying,jin2023measurementinduced}, has been numerically found in several works~\cite{alberton2021entanglementtransitionin,vanregemortel2021entanglement,turkeshi2021measurementinducedentanglement,botzung2021engineereddissipationinduced,bao2021symmetryenrichedphases,turkeshi2022entanglementtransitionsfrom,piccitto2022entanglementtransitionsin,kells2021topologicaltransitionswith,paviglianiti2023multipartite,muller2022measurementinduced,buchhold2022revealing}.

A different limit of the measurement problem is obtained by post-selecting atypical trajectories corresponding to specific measurement outcomes. To fix the ideas, consider, for example, the quantum jump (QJ) dynamics corresponding to a photo-counting monitoring protocol~\cite{dalibard1992wavefunction,plenio1998quantum,wiseman2009quantummeasurementand}: here, abrupt random quantum jumps (clicks) intersperse the deterministic evolution driven by a non-Hermitian Hamiltonian, which accounts for the measurement back-action. Post-selecting on the trajectory where no click has happened corresponds to purely non-Hermitian dynamics.
In this limit, several works have reported
measurement-induced entanglement transitions and highlighted their relation with the spectral properties of the non-Hermitian Hamiltonian~\cite{biella2021manybodyquantumzeno,gopalakrishnan2021entanglementandpurification,jian2021yangleeedge,turkeshi2023entanglementandcorrelation,legal2023volumetoarea,kawabata2023entanglementphasetransition,orito2023entanglement,zerba2023measurement,granet2023volume,su2023dynamics,banerjee2023entanglement,lee2023entanglement}.

The relation between these two limits of the measurement problem, particularly concerning the entanglement dynamics, is not well understood. Is the no-click dynamics stable enough to the inclusion of QJs, which should be seen as irrelevant perturbations? Or, on the contrary, do QJs completely change the entanglement structure of the monitored system? These questions ultimately go beyond measurement-induced transitions and touch upon the relevance of non-Hermitian Hamiltonians in the description of open quantum systems~\cite{ashida2020review,lau2018fundamental,minganti2019quantum,naghiloo2019quantum,minganti2020hybrid,han2023exceptional,ehrhardt2023exploring}.

To understand the role of QJs and non-Hermitian evolution on the entanglement dynamics in monitored systems we 
introduce a new tool: the statistics of the entanglement entropy changes after and in-between QJs.
We show that, on average, QJs induce an entanglement entropy loss while the non-Hermitian evolution causes a gain, although the statistics display rather broad tails and fluctuations. Using the full statistics of entanglement gain and loss we propose a classical stochastic random walk model with partial resetting for the entanglement dynamics, that we show to reproduce the full QJ dynamics.  More importantly, the entanglement gain and loss picture clarifies the mutual role of quantum jumps and non-Hermitian evolution in the scaling of entanglement entropy in the steady-state, with direct implications for the associated MIPT.

We showcase our new method by studying three models of monitored free-fermions: the Ising chain, the monitored SSH-chain and a model of free-fermions with $U(1)$ symmetry. These models are interesting for two reasons: first, they differ in their global symmetry, a fact which is believed to play a critical role in the stability of the MIPT;  second, their entanglement transition in the no-click limit has been studied in detail~\cite{turkeshi2023entanglementandcorrelation,legal2023volumetoarea}. By solving the full stochastic QJ dynamics for these models and computing the entanglement entropy, we show that QJs have remarkably different impacts on their phase diagrams, as compared to the no-click limit.
We understand this difference through a detailed 
analysis of the entanglement gain and loss, which reveals the crucial role of quantum jumps in \emph{renormalizing} the effective non-Hermitian dynamics with respect to the bare no-click limit. We show that this renormalization, or its lack thereof, can naturally explain the different behaviors of quantum jumps in the three models and provide a criterion for the relevance of the no-click limit. Finally, we demonstrate how the the entanglement gain/loss picture and the associated classical model remain valid even in presence of interactions breaking gaussianity.

The paper is structured as follows. In Sec.~\ref{sec:background} we review the background material: we introduce the
QJ measurement protocol and the models we will considered throughout this work as well as the quantities we will use to characterize them.  
In Sec.~\ref{sec:entanglement_loss_gain} we present the main quantity of interest for this work,  namely the statistics of entanglement gain and loss, that we use to construct a classical stochastic model for the entanglement dynamics. In Sec.~\ref{sec:applications} we present several applications of this tool to the study of MIPT in models of monitored free fermions. In particular
we use the insights from this model to discuss the role of jumps and non-Hermitian dynamics on the scaling of the entanglement and the MIPT. Finally Sec.~\ref{sec:conclusion} contains our conclusions and future perspectives. In the Appendixes we provide further methodological details and results relevant for our work.

\begin{figure}[!ht]     \includegraphics[width=0.95\columnwidth]{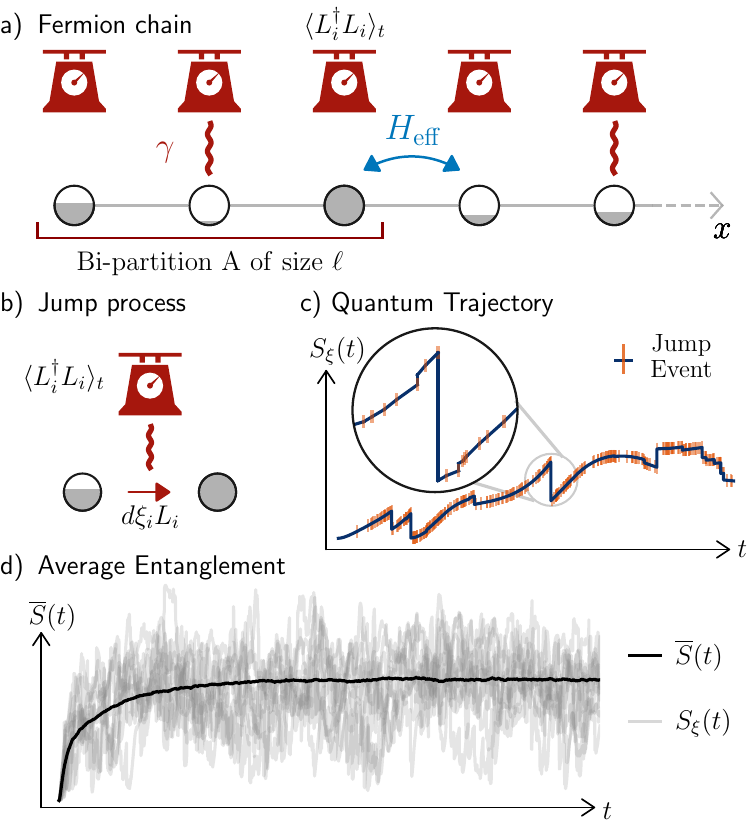}
    \caption{\label{fig:sketch}  Panel (a-b): cartoon of the setup. A monitored fermionic chain evolving under quantum jump protocol, characterized by a deterministic evolution driven by a non-Hermitian Hamiltonian $H_{\rm eff}$ (a) and stochastic Quantum Jumps  (b).  Panel (c): We are interested in the stochastic dynamics of the wave function and in particular in its entanglement structure, as measured from the bipartite entanglement entropy $S_{\xi}(t)$ (see panel a for the partition). Panel (d): typical quantum trajectory evolutions (10 realisations) for the entanglement entropy and its average over the measurement noise $\xi$.}
\end{figure}
\section{Background}
\label{sec:background}
In this Section we summarize relevant background material, including the Quantum Jump protocol to continuously monitor a quantum system and its sampling
via Montecarlo Wave Function method and the Waiting Time-Distribution. We then define the models we focus on throughout the paper and the main quantities we use to characterize their dynamics.

\subsection{Monitoring by Quantum Jumps}

In this work we are interested in the dynamics of continuously monitored quantum systems~\cite{wiseman2009quantummeasurementand}. The set up we have in mind is sketched in Fig.~(\ref{fig:sketch}), namely a fermionic chain where each lattice site is coupled to a measurement apparatus that weakly and continuously monitor some local observable of the chain. We are interested in the dynamics of the system wavefunction conditioned to the measurement outcomes, which realizes a so called quantum trajectory. Different measurement protocols give rise to different types of stochastic evolutions. Here we focus on the Quantum Jump (QJ) protocol~\cite{ueda1990,dalibard1992wavefunction,gardiner1992wave}, corresponding to the experimentally relevant photo-counting process, and which is described by the following stochastic Schrodinger equation
\begin{align}\label{eqn:SSE}
     d| \Psi(t) \rangle  &= 
-idt\left\{ H   - \frac{i}{2}\sum_{i} (L_{i}^{\dagger}L_{i} - \langle L_{i}^{\dagger}L_{i}\rangle_t ) \right\} | \Psi(t) \rangle \nonumber\\
&\ + \sum_{i} d\xi_{i}  \left\{  \frac{L_{i}}{\sqrt{\langle L_{i}^{\dagger}L_{i}\rangle}}-1 \right\} | \Psi(t) \rangle.
\end{align}
where $H$ is the Hamiltonian of the fermionic chain (we will give explicit expression in Sec.~\ref{sec:models}), $L_i,L^{\dagger}_i$ are the jump operators defined on each lattice site which describe the measurement process and $d\xi_i \in \{0,1\}$ is an increment for the in-homogeneous Poisson process with average ${P(d\xi_{i} = 1) = dt\langle \psi(t)| L_{i}^{\dagger} L_{i} |\psi(t) \rangle}.$

The dynamics in Eq.~(\ref{eqn:SSE}) is composed of a deterministic part (for $d\xi_i=0$) which corresponds to an effective non-Hermitian Hamiltonian $H_{\rm eff}$ given by
\begin{equation}\label{eqn:Heff}
  H_\mathrm{eff} = H   - \frac{i}{2}\sum_{i} L_{i}^{\dagger}L_{i} 
\end{equation}
and a stochastic one, the last term in Eq.~(\ref{eqn:SSE}), due to the action of QJs. The effective Hamiltonian $H_{\rm eff}$ is non-Hermitian because of the measurement back-action, encoded in the last term of Eq.~(\ref{eqn:Heff}). Its role is to control the dynamics in the so called \emph{no-click} limit, when no jumps happen during the quantum trajectory, as well as the evolution in-between two subsequent quantum jumps. We note that the evolution of the system is state-dependent (thus nonlinear); see the counter-term appearing in Eq.~(\ref{eqn:SSE}), to ensure the normalization.

The QJ evolution~\eqref{eqn:SSE} is solved using Monte Carlo methods~\cite{daley2014quantum}, either via a first-order integration scheme which introduces an explicit discretization $dt$ to sample the QJs or using higher-order schemes. 
The former approach has the drawback of not having a natural way to control the accuracy of the simulation, which is instead empirically benchmarked by considering different choices of $d t$ and checking that there is not qualitative difference in disorder averages. We overcome this limitation considering a higher-order scheme that is based on sampling the time at which subsequent jumps happen~\cite{breuer2002thetheoryof,daley2014quantum,radaelli2023gillespie}, i.e., using the cumulative waiting-time distribution defined as 
\begin{align}
F[\Psi,\tau]=1-\langle\Psi\vert 
e^{iH^{\dagger}_{\rm eff}\tau}e^{-iH_{\rm eff}\tau}\vert\Psi\rangle.\label{eq:wtdevo}
\end{align}
Specifically, we proceed by iterating the following loop : (i) Assume a normalised state $\vert \Psi(t_i)\rangle$ is reached at time $t_i$ (possibly the initial time of the dynamics, in which case $\vert \Psi(t_i)\rangle$ is the initial state). (ii) Extract a random number $r$ uniformly distributed in $[0,1]$ and find the random waiting time $\tau$ for the next jump by solving the equation $r=1-F[\Psi(t),\tau]$.

(iii) Within the time interval $[t,t+\tau]$ propagate the deterministic non-Hermitian evolution 
\begin{equation}
    |\Psi(t+\tau)\rangle = \frac{e^{-i H_\mathrm{eff} \tau}|\Psi(t)\rangle}{||e^{-iH_{\rm eff}\tau}\vert\Psi(t)\rangle||}.
\end{equation}
(iv) At time $t+\tau$ a quantum jump occurs. As before, the output channel is chosen splitting the [0,1] interval into segments of size $\langle \Psi(t+\tau)\vert L^{\dagger}_i L_i\vert \Psi(t+\tau) \rangle$ and checking in which one a uniformly drawn random number falls. The jump is immediate and the post-measurement state after a jump at site $j$ reads
\begin{equation}
    |\Psi(t+\tau^+)\rangle = \frac{L_{j}|\Psi(t+\tau)\rangle}{\sqrt{\langle \Psi(t+\tau)|L_{j}^\dagger L_{j}|\Psi(t+\tau)\rangle}}.
\end{equation}
In a nutshell, a quantum trajectory is specified as a sequence of non-Hermitian quantum quenches interspersed with discontinuous jumps (see Fig.~\ref{fig:sketch} (c)), that set the initial conditions for the forecoming integration steps.

An advantage of this approach is to access directly the waiting-time distribution (WTD) or delay function~\cite{Cohen-Tannoudji_1986,daley2014quantum,landi2023current}, which encodes the probability distribution of the times between QJs. Its behavior will be discussed for the the specific model of interest later on in the manuscript.

\subsection{Average vs Conditional Dynamics and Entanglement Entropy}

The stochastic Schrödinger equation~\eqref{eqn:SSE} describes the evolution of the conditional state, also called a quantum trajectory. By averaging the density matrix $\rho_{\xi}(t)=\vert\psi_{\xi}(t)\rangle
\langle\psi_{\xi}(t)\vert $ over the stochastic noise $\xi$ in Eq.~\eqref{eqn:SSE} one recovers a Lindblad master equation for the average state (This time dependent random variable $\mathbf{\xi}$ is used to label the different trajectories). These two descriptions are equivalent for what concerns simple observables,  which are linear functionals of the conditional state, and indeed quantum trajectories are also called \emph{unravellings} of the Lindblad master equation~\cite{carmichael1999statisticalmethodsin}. The stochastic dynamics however contains a richer information: if one is interested in quantities which are non-linear in the state and sensitive to higher moments of the density matrix, then averaging over the measurement noise gives rise to physics which is not captured by the Lindblad/average state. A relevant example is provided by the (von Neumann) entanglement entropy, defined as ~\cite{calabrese2004entanglemententropyand,amico2008entanglementinmanybody}
\begin{align}
	S_\mathbf{\xi}(t )  = -\mathrm{tr}_A\left[ \rho_\mathbf{\xi}^A(t)\ln \rho_\mathbf{\xi}^A(t)\right]\;,
	\label{eq:5v0}
\end{align}
where we have introduced a partition $A\cup B$ in the system (cf. Fig.~\ref{fig:sketch}) and the reduced density matrix $\rho_\mathbf{\xi}^A(t) = \mathrm{tr}_B |\Psi_\mathbf{\xi}(t)\rangle\langle\Psi_\mathbf{\xi}(t)|$.   Therefore the entanglement entropy will be a fluctuating quantity evolving stochastically, see Fig.~\ref{fig:sketch}. In this work we will be mainly concerned with the average entanglement entropy, given by 
\begin{equation}\label{eqn:aver_s}
    \overline{S} (t) = \int \mathcal{D}\mathbf{\xi} P(\mathbf{\xi}) S_\mathbf{\xi}(t).
\end{equation}
where the average is taken over the measurement noise $\xi$. In particular we will discuss how the steady-state entanglement entropy scale with the size of the sub-system $\ell$. In thermal equilibrium or for unitary dynamics this is known to sharply characterize the nature of a given phase~\cite{amico2008entanglementinmanybody}, depending on whether this scaling is proportional to the volume of the sub-system ($S\sim \ell$ in one dimension) or to its area ($S\sim \mathrm{const.}$ in one dimension), possibly with logarithmic corrections as for quantum critical states.

\subsection{Models of Monitored Fermions}\label{sec:models}

In this work we consider three models of monitored free fermions with different internal symmetries and measurement operators, leading to different non-Hermitian Hamiltonian. Furthermore for one of these cases we also discuss the role of interactions breaking gaussianity. The first model is the Quantum Ising chain in a transverse field
\begin{equation}
  H_\mathrm{Ising} = - \sum_{i=1}^L \left[ J\sigma_i^x\sigma_{i+1}^x  + h\sigma_i^z  \right]
\end{equation}
which is mapped via a Jordan-Wigner transformation to
\begin{equation}
  H_\mathrm{Ising} = - \sum_{i=1}^L \left[  J(c_i^\dagger c_{i+1} + c_i^\dagger c_{i+1}^\dagger + h.c.) + h(1 - 2 n_i)  \right].
\end{equation}
We choose the jump operators
\begin{equation}\label{eqn:jumps_Ising}
  L_{i} = \sqrt{2\gamma} n_{i},
\end{equation}
corresponding to monitoring of the local density $n_i=c^{\dagger}_ic_i$,
and get the effective non-Hermitian Hamiltonian  
\begin{equation}\label{eqn:Ising_nh}
  H_\mathrm{eff} = H_\mathrm{Ising}   - i \gamma   \sum_{i=1}^{L}  n_{i},
\end{equation}
describing a non-Hermitian Ising model in a complex-valued transverse field~\cite{hickey2013time,biella2021manybodyquantumzeno,turkeshi2023entanglementandcorrelation}.
We contrast the results for the Ising case with models with $U(1)$ symmetry. Specifically we consider a fermionic Su-Schrieffer-Heeger (SSH) chain with two different sublattices $A$ and $B$ and Hamiltonian
\begin{equation}
  H_\mathrm{SSH} =  -\sum_{j=1}^{L}  \left[(J-h)c_{A,j}^{\dagger}c_{B,j-1} + (J + h) c_{A,j}^{\dagger}c_{B,j} +\mathrm{h.c.}\right].
\end{equation}
We choose the jump operators as
\begin{equation}
  L_{A,i} = \sqrt{2\gamma} n_{A,i} \qquad L_{B,i} = \sqrt{2\gamma}(1-n_{B,i}),
\end{equation}
namely we independently and continuously monitor the local density of particles on sublattice $A$, $n_{A,i}=c^\dagger_{A,i} c_{A,i}$, and the local density of holes, $1-n_{B,i}= c_{B,i} c^\dagger_{B,i}$, on sublattice $B$.  Using these quantum jump operators we obtain an effective non-Hermitian Hamiltonian of the form 
 \begin{equation}\label{eqn:ssh_nh}
     H_\mathrm{eff}   =   H_\mathrm{SSH} - i \gamma \sum_{i=1}^L( c^\dagger_{A,i} c_{A,i} +c_{B,i} c^\dagger_{B,i} ).
 \end{equation} 
This non-Hermitian SSH model has been studied in Ref.~\onlinecite{legal2023volumetoarea}.
Finally we consider a model of one-dimensional lattice fermions  with $U(1)$ symmetry and  next-neighbor interactions, described by a Hamiltonian
\begin{equation}
 H =  -\sum_{j=1}^{L}  c_{j}^{\dagger}c_{j+1} +\mathrm{h.c.}+V\sum_{i=1}^L (1-2n_i)(1-2 n_{i+1}) 
\end{equation}
and monitoring of the local density by the jump operators
\begin{equation}
  L_{i} = \sqrt{2\gamma} n_{i}\,.
\end{equation}
The associated non-Hermitian  Hamiltonian reads
 \begin{equation}\label{eqn:XXZ_nh}
     H_\mathrm{eff}   =   H - i \gamma \sum_{i=1}^L c^\dagger_{i} c_{i} 
 \end{equation} 
We note that for $V=0$ the model reduces to monitored free-fermions which have been extensively studied, while a finite $V$ allows us to  discuss the role of many-body interactions breaking gaussianity. In both cases we note that the non-Hermitian Hamiltonian associated to this problem, Eq.~(\ref{eqn:XXZ_nh}), is special in the sense that its imaginary part commutes with its real-part. As such the evolution in between quantum jumps is unitary. On the other hand, the first two example have a non-trivial non-Hermitian evolution and the entanglement transition associated to the no-click limit has been discussed in Ref.~\cite{turkeshi2023entanglementandcorrelation,legal2023volumetoarea}. The key features of the no-click problem are summarised for completeness in Appendix~\ref{app:noclick}

\begin{figure*}[!ht] 
	 
\includegraphics[width=0.99\textwidth]{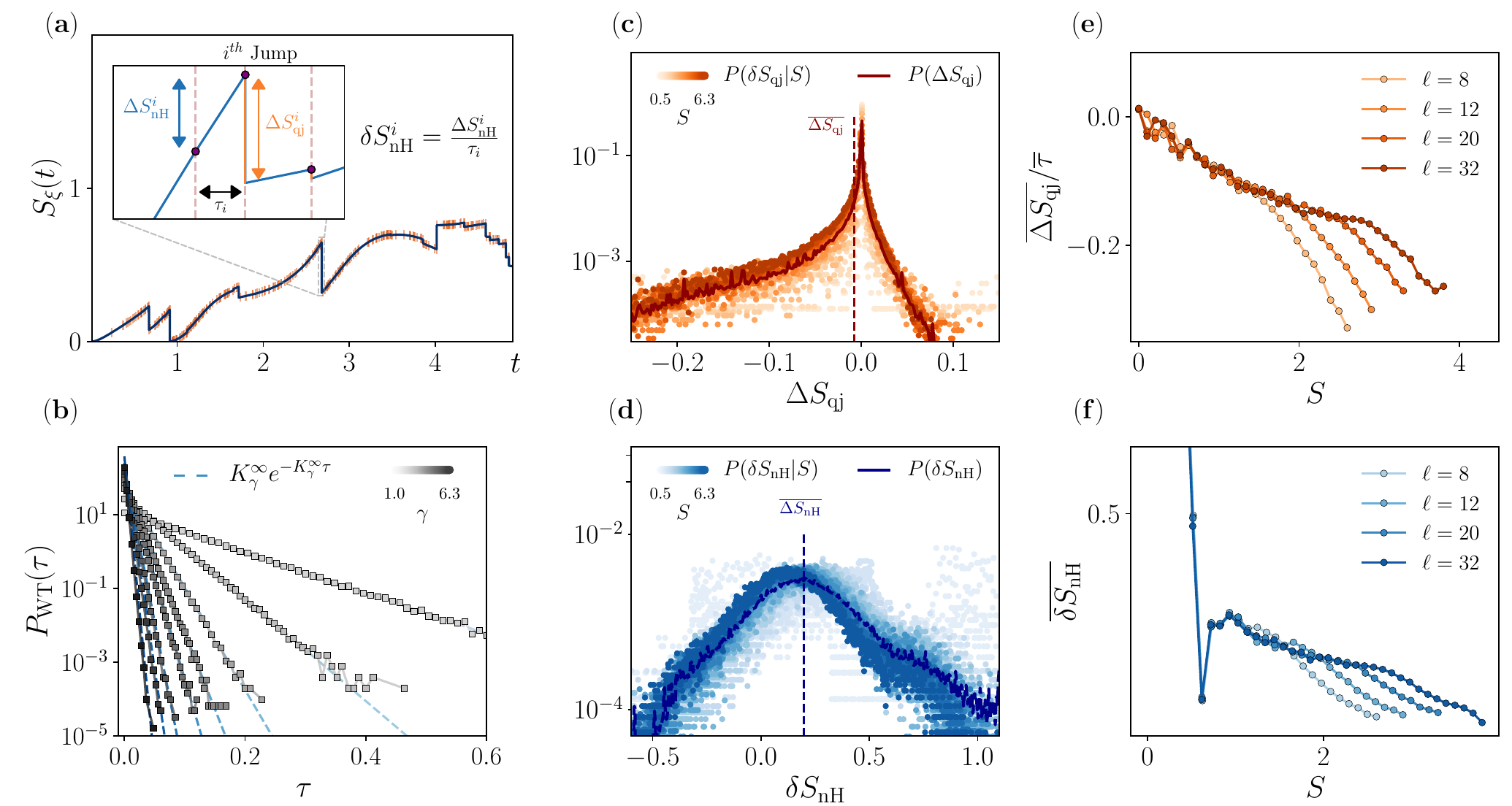}
    \caption{\label{fig:summary}  
    \textbf{Statistics of Entanglement Gain and Loss - }
Panel (a): Evolution of the entanglement entropy along a quantum trajectory, for the monitored Ising chain. The dots symbolize the quantum jumps, while the evolution in between the jumps is due to the non-Hermitian Hamiltonian. From this quantum trajectory we extract three key quantities (see inset): (i) the waiting times $\tau_i$ between QJs, (ii) the entanglement entropy change after a QJ $\Delta S_{\rm qj}$ and (iii) in between QJs $\Delta S_{\rm nh}$. We then construct the corresponding histograms shown in panels (b-c-d).
In panel (b) is plotted the Waiting Time Distribution of QJs, $P_{WT}(\tau)$ where $P_{WT}(\tau) d\tau$ is the probability of having an elapsed time between one jump and the next one in the interval $[\tau,\tau+d\tau]$. The poissonian rate $K_\gamma^\infty$ is the steady state value of the backaction (more details in Appendix \ref{sec:waiting time}). In panel (c) the statistics of entanglement change due to QJs, $P(\Delta S_{\rm qj}\vert S)$ conditioned to the entanglement content $S$ and in panel (d) the corresponding quantity for the non-Hermitian evolution.
Panel (e-f): Average change to the entanglement entropy  due to non-Hermitian evolution (panel f) and QJs (panel e). These averages are obtained from the conditional distributions, $P(\Delta S_{\rm qj}\vert S),\ P(\Delta S_{\rm nH}\vert S)$ for system sizes of $L=128$, thus are a function of the entanglement entropy $S$ and we represent in these plots the dependence from subsystem size $\ell$.}
\end{figure*}

\section{Statistics of Entanglement Gain and Loss}
\label{sec:entanglement_loss_gain}

In this Section we introduce the main 
tool that we will use throughout this work to understand the intertwined role of QJs and non-Hermitian evolution in monitored quantum system: the statistics of entanglement gain and loss. To simplify the presentation we illustrate our ideas in the context of the monitored Ising chain (defined in Sec.~\ref{sec:models}), leaving to Sec.~\ref{sec:applications} a detailed application of this tool to MIPT in different models.

Let us consider the stochastic dynamics of the entanglement entropy along a quantum trajectory, plotted in Fig.~\ref{fig:summary} (a) : the typical pattern is given by an evolution driven by the non-Hermitian Hamiltonian interrupted abruptly by a discontinuous change in entanglement entropy due to a quantum jump. From this quantum trajectory we extract three key informations. First, the waiting-time distribution of QJs, $P_{WT}(\tau)$, which represents the probability density of having a waiting time $\tau$ between QJs. As we show in 
Fig.~\ref{fig:summary}(b) the WTD displays a Poisson behavior, with a rate $K_\gamma^\infty \propto \gamma L$, that gives rise to an average waiting time $\overline{\tau}\sim 1/\gamma L$ (see also Appendix~\ref{sec:waiting time}).
Second, we extract the changes to the entanglement entropy at each jump event and in between the jumps (due to the non-Hermitian evolution). As we see from Fig.~\ref{fig:summary} (a)
the entanglement entropy can either increase or decrease after a QJ and, similarly, can either grow or diminish during the time between jumps, where non-Hermitian evolution occurs.  Given this pattern, we now ask : what is the statistics of the entanglement entropy change after a quantum jump and  \emph{in-between} quantum jumps, i.e., after an evolution step with the non-Hermitian Hamiltonian ? 
 
To answer this question, we sample along many quantum trajectories the probability density of observing a change $\Delta S_{\rm qj}$ to the entanglement entropy due to QJs and collect the resulting histogram $P(\Delta S_{\rm qj}\vert S)$. Similarly, we sample the probability density of observing a change $\delta S_{\rm nH}$ to the entanglement entropy due to the non-Hermitian evolution, that we denote  $P(\delta S_{\rm nh}\vert S)$. Crucially, these are \emph{conditional probabilities} given a certain value of entanglement entropy $S$ before the event. This dependence is a key feature of our approach: the basic idea is to understand whether QJs impact in different ways many-body states which are highly or weakly entangled. 
In practice these histograms are obtained by binning stochastic events (jumps or non-Hermitian evolution) according to the \emph{entanglement content} of the state they act on.
To be specific we say that a state has an entanglement content $S$ at a time $t$ if for $\tau \in [t-\epsilon_t , t+\epsilon_t ] $, the entanglement of a state in that time range is such that $ | S(|\psi(\tau)\rangle) -S  |  < \epsilon_S$  (for given $\epsilon_\tau,~\epsilon_S$).  

\begin{figure*}[!t] 
\centering
\includegraphics[width=0.88\textwidth]{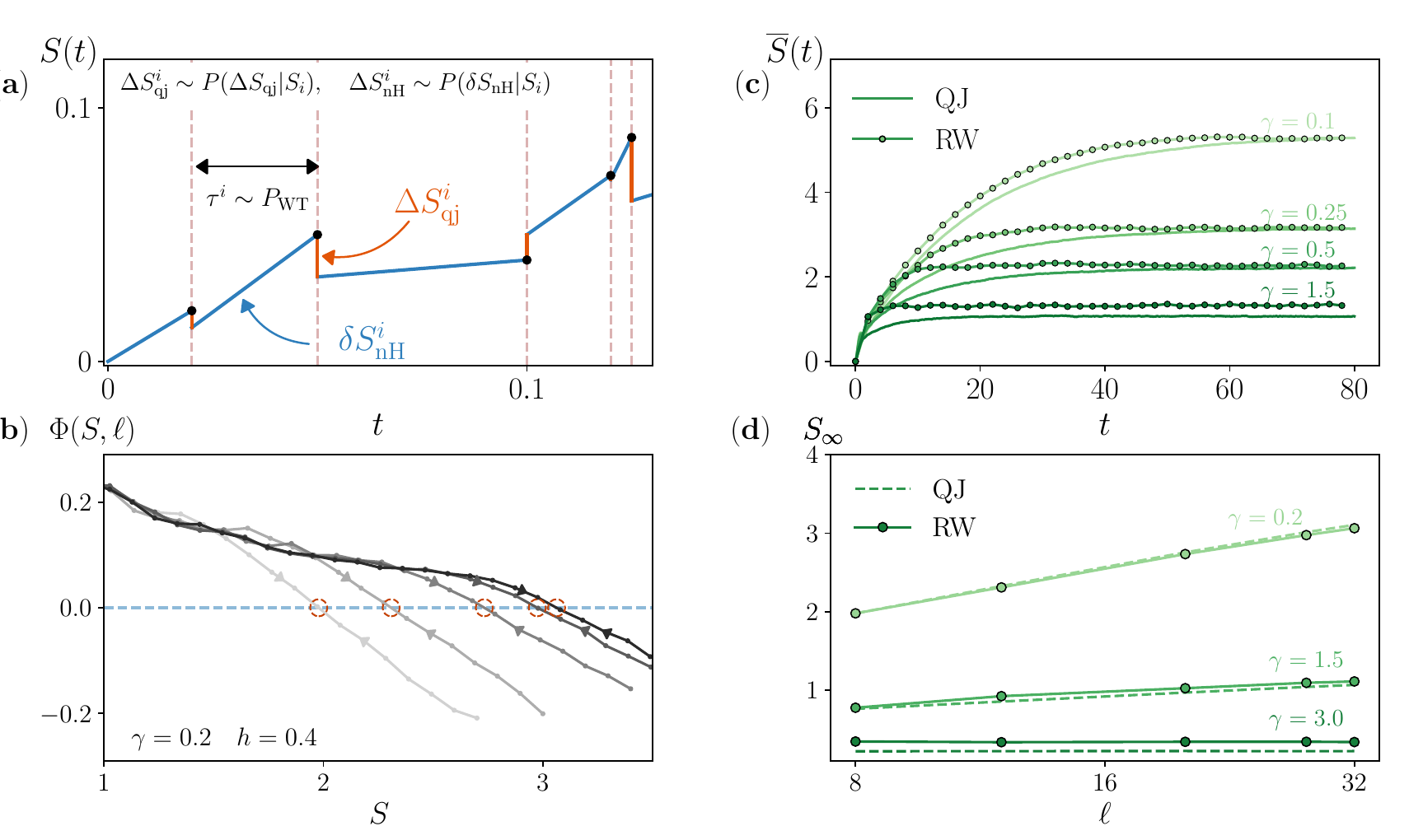}
\caption{\label{fig:classical_model} \textbf{Stochastic model for entanglement dynamics} ---  Panel (a): Cartoon depicting the entanglement dynamics as a classical random walk (RW) with stochastic drift and resetting drawn from the histograms in Fig.~\ref{fig:summary}. 
Panel (b) total rate of entanglement entropy growth obtained, for the monitored Ising chain, from the average entanglement gain and loss, according to Eq.~(\ref{eqn:S_dyn_class}). We see the rate $\Phi(S,\ell)$ vanishes at a value $S_{\infty}(\ell)$ which identifies the steady-state entanglement (circled points). 
Panel (c): Comparison between the dynamics of the average entanglement entropy obtained from the exact quantum jump dynamics (QJ) or the phenomenological classical random walk (RW). Panel (d): comparison between the steady-state $S_{\infty}$ obtained from the stochastic model with the long-time limit of the entanglement entropy obtained from the full QJ dynamics, showing a perfect agreement.
}
\end{figure*}
 
We plot these histograms in Fig.~\ref{fig:summary} (c-d) for the monitored Ising chain. The first remarkable observation is that the QJ distribution is strongly peaked around $\Delta S_{\rm qj}=0$, with broad tails, suggesting that the vast majority of quantum jumps do not substantially change the entanglement entropy. In contrast, rare quantum jumps are responsible for more significant changes, and this can be also observed at the level of the trajectory in Fig.~\ref{fig:summary} (a). The asymmetry of the distribution is also interesting to note, since it means that a single QJ is more likely to reduce the entanglement entropy, even though jumps that increase it are also possible. 
Finally, as we see in Fig.~\ref{fig:summary} (c-d), the tails of the histogram for the QJs broaden up and acquire a non-trivial $S$ dependence at least for $\Delta S_{\rm qj}<0$, indicating that highly entangled states are indeed more fragile and prone to be affected by rare QJs. For what concerns the entanglement changes during the non-Hermitian evolution in between quantum jumps, we see that the distribution appears centered around a slightly positive value and with a slight asymmetry in the tails, suggesting that the non-Hermitian evolution is primarily responsible for the growth of the entanglement.

In the remaining of this Section, we will show how in fact the statistics of entanglement changes, defined above, controls the dynamics of the average entanglement entropy, and how this can be understood in terms of a simple phenomenological model.

\subsection{Average Entanglement Gain and Loss}
To condense the rich information contained in the full statistics of entanglement changes we now focus on the first moment of those distributions, corresponding to the average change to the entanglement entropy due to quantum jumps $\overline{\Delta S}_{\rm qj}(S,\ell) $ and to the non-Hermitian evolution $\overline{\delta S}_{\rm nH}(S,\ell) $. The first moments are defined as \begin{align}\label{eqn:average_deltaSqj}
\overline{\Delta S}_{\rm qj}(S,\ell)\equiv
\int_{\Delta S_{\rm qj}}\,\Delta S_{\rm qj}\,
P(\Delta S_{\rm qj}\vert S)    
\end{align}
and similarly for $\overline{\delta S}_{\rm nH}(S,\ell) $.  We emphasize that these averages~\footnote{With a slight abuse of notation, we indicate the average over the gain/loss statistics with an overbar, as for the average over trajectories.} depend on the value of the entanglement entropy $S$ and the size of the partition $\ell$, since they are evaluated over the conditional distributions $P(\Delta S_{\rm qj}\vert S)$  and $P(\delta S_{\rm nH}\vert S)$ which itself depends on $\ell$.
In Fig.~\ref{fig:summary} (e-f), we plot these first moments for the monitored Ising chain (at a particular point of the phase diagram) for different subsystem sizes $\ell$.
In panel (f), we see that the change due to non-Hermitian evolution is on average positive, $\overline{\delta S}_{\rm nH}>0$, i.e., it induces an \emph{entanglement entropy gain}. This gain is substantial for weakly entangled states, then decreases slowly with the entanglement content. In panel (e), we plot the change due to QJs, normalized with respect to the average waiting time $\overline{\tau}$. As shown in Appendix~\ref{app:scalingL}, this ratio remains well defined upon increasing the system size $L$, despite the waiting time vanishing as $\overline{\tau}\sim 1/L$. This is consistent with the idea that a single quantum jump can affect the entanglement entropy of a quantity $O(1/L)$.
From Fig.~\ref{fig:summary} (e), we see that $\overline{\delta S}_{\rm qj}/\overline{\tau}$ is negative and decreases with $S$, i.e., QJs on average induce an \emph{entanglement entropy loss}. They do so the more the state onto which they act is entangled. Interestingly, both averages develop a subsystem-size $\ell$ dependence above a certain threshold entanglement entropy value. 
A natural question that we address next is how these average entanglement gains and losses are connected to the entanglement dynamics under QJs and its phase transition?

\subsection{Classical Stochastic Model for Entanglement Dynamics}\label{sec:classical}

We now present a classical stochastic model that builds upon the entanglement gain and loss statistics discussed in the previous section and gives a physical picture of the entanglement dynamics under QJs. In particular, we model the entanglement evolution as a random walk with random drift and partial resetting~\cite{evans2011diffusion,Evans_2020,turkeshi2022entanglementtransitionsfrom}, which is sketched in Fig.~\ref{fig:classical_model} (a) and that we will now describe. We use the previous analysis showing that the waiting time $\tau$  between jumps follows a Poisson law, which means that the random variable can be expressed as $\tau = - \log (r) /  K_\gamma^\infty $ were $r\in [0,1]$ is a random number drew for each jump. Then, during this time $\tau$, we model the change of entanglement due to the non-Hermitian evolution by $\Delta S_{\rm nH} = \tau ~\delta S_{\rm nH}$,  where $\delta S_{\rm nH}$ is picked with a probability $P(\delta S_{\rm nH}\vert S)$ from the previously computed conditional distribution of the effective non-Hermitian slopes. After the time $\tau$, the instantaneous QJ happens, and we model the change it has on the entanglement entropy by $\Delta S_{\rm qj}$, which is drawn with a probability $P(\Delta S_{\rm qj}\vert S)$ from the corresponding probability distribution. In both cases, the entropy $S$ corresponds to the entanglement before the event (i.e., before the jump or before the non-Hermitian evolution). The probability distributions we use here are taken from the exact sampling of quantum trajectories, binned according to the entanglement entropy content as discussed in the previous section.

The above stochastic process is described by the following classical master equation~(\ref{eqn:master_eq}) for the probability of having an entanglement entropy $S$ at time $t$, that we note  $\mathcal{P}_t(S)$:
\begin{equation}
    \begin{split}
        &\mathcal{P}_{t+dt}(S)=rdt \int_{\Delta S_{\rm qj}} P(\Delta S_{\rm qj}\vert S-\Delta S_{\rm qj}) \mathcal{P}_t(S-\Delta S_{\rm qj})+\\
&+(1-rdt)\int_{\delta S_{\rm nH}} P(\delta S_{\rm nH}\vert S-\delta S_{\rm nH}dt) \mathcal{P}_t(S-\delta S_{\rm nH}dt)
    \end{split}
    \label{eqn:master_eq}
\end{equation}
where the first term describes the jump, which adds a random contribution $\Delta S_{\rm qj}$ with probability $rdt\times P(\Delta S_{\rm qj}\vert S-\Delta S_{\rm qj})$, with $r$ the resetting rate, while the second one describes the non-Hermitian dynamics which increase the entanglement of a random slope $\delta S_{\rm nH}$ with probability $(1-rdt)\times P(\delta S_{\rm nH}\vert S-\delta S_{\rm nH}dt)$.

To understand the role of the first moments of the gain/loss distribution on the entanglement dynamics, it is useful to derive from the master equation~(\ref{eqn:master_eq}) a dynamical equation for the average entanglement entropy,  $S=\int dS~ S\,\mathcal{P}_t(S)$, which reads
\begin{align}\label{eqn:S_dyn_class}
\frac{dS}{dt}=\overline{\delta S_{\rm nH}}(S,\ell)+\overline{\Delta S_{\rm qj}}(S,\ell)/\overline{\tau}\equiv \Phi(S,\ell)
\end{align}
This is a straightforward rate equation for the dynamics of the average entanglement entropy $S$, which appears to be controlled by a balance between average entanglement gain $\overline{\delta S_{\rm nH}}(S,\ell)$ and loss  $\overline{\Delta S_{\rm qj}}(S,\ell)$. The fact that these gain/loss rates depend on the value of the entanglement entropy itself, as previously discussed, is crucial here and results in a non-trivial flow of the entanglement entropy with time, encapsulated in the function $\Phi(S,\ell)$ defined in Eq.~(\ref{eqn:S_dyn_class}). We emphasize that the key assumption to construct the random walk model for the entanglement entropy in Eq.~\eqref{eqn:master_eq}, from which Eq.~\eqref{eqn:S_dyn_class} directly follows, is that the rates to move depend on the value of the entanglement entropy itself.
While this assumption might seem very fine tuned, and it is likely not satisfied in the most general case, we note that it bares conceptual similarity with certain random circuit models where analogous phenomenological dynamics for entanglement have been discussed~\cite{jonay2018coarsegrained,zhou2019emergent,feng2022measurementinducedphase}. It is an interesting open question, that we leave for future work, to understand whether in certain limiting cases one could derive such an effective stochastic dynamics microscopically. In the present case this assumption must be verified a posteriori with numerical simulations, as we are going to discuss now.

In Fig.~\ref{fig:classical_model} (b), we plot $\Phi(S,\ell)$ for the Ising chain (at a particular points of the phase diagram). We see that $\Phi(S,\ell)$ 
vanishes at a value $S=S_{\infty}(\ell)$, which represents a fixed point of the entanglement stochastic dynamics~(\ref{eqn:S_dyn_class}). This fixed point is attractive, i.e., depending on the initial condition, it is approached either from the low-entangled branch where $\Phi(S,\ell)>0$ or from the high-entangled one where $\Phi(S,\ell)<0$.
According to our simple model the steady-state entanglement entropy in a monitored system satisfies the equation
\begin{align}\label{eq:steadystate}
   \Phi(S_{\infty},\ell)=\overline{\delta S_{\rm nH}}(S_{\infty},\ell)+\overline{\Delta S_{\rm qj}}(S_{\infty},\ell)/\overline{\tau}=0.
\end{align}
This relation, which is one of the main result of this work, has a clear and transparent interpretation: the steady-state entanglement entropy in a monitored system is reached when the entanglement gain due to non-Hermitian evolution and the loss due to QJs perfectly balance each other.

We can now benchmark this stochastic model for the entanglement entropy dynamics. First in Fig.~\ref{fig:classical_model} (c) we show that it can reproduce very accurately the dynamics of the average entanglement entropy in the monitored Ising chain, thus capturing the entanglement transition 
occurring in this model, as we discuss in more detail in Sec.~\ref{sec:applications}. In addition we can extract the steady-state entanglement $S_{\infty}$ and its subsystem size $\ell$ dependence from Eq.~(\ref{eq:steadystate})  and compare it with the full QJ results. As shown in Fig.~\ref{fig:classical_model}(d), the agreement is really good both at weak monitoring, where it reproduces the logarithmic scaling of the entanglement entropy, and in the area law. 
 \begin{figure*}[!t] 
\centering 
	\includegraphics[width=0.99\textwidth]{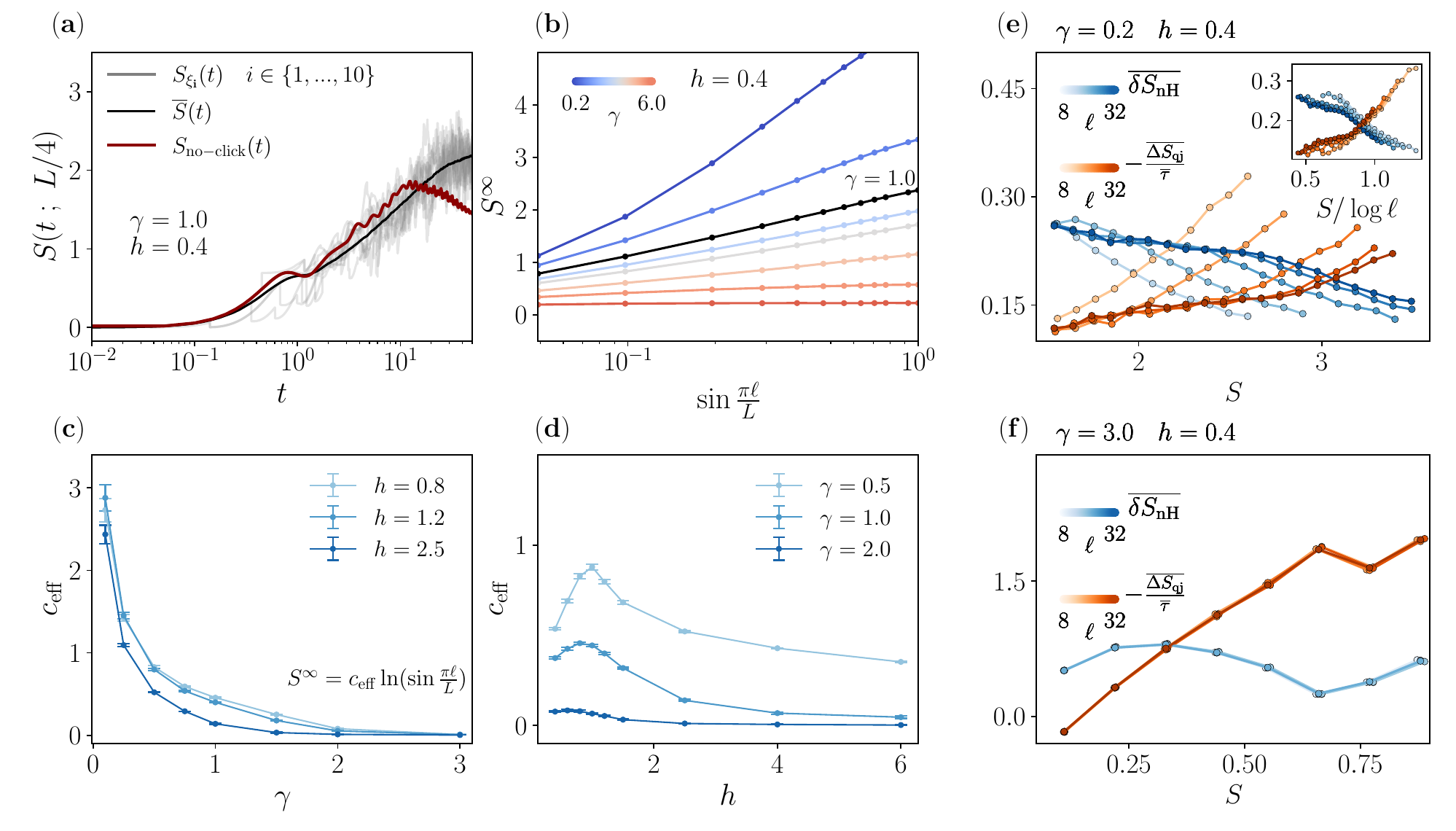}	\\
\caption{\label{fig:recap_ising}
\textbf{Entanglement dynamics under QJ monitoring for the Ising chain} --- Panel (a) Comparison between average, trajectory, and no-click evolution for  measurement rate $\gamma=1.0$ and $h=0.4$. (b) Average steady-state entanglement entropy ($\gamma=\left\{ 0.2,~0.5,~1.0,~1.5,~2.0,~ 3.0 ,~4.0 ,6.0]\right\}$). We see for weak monitoring a logarithmic scaling that evolves into an area law upon increasing $\gamma$ ; Phase diagram as obtained from the effective central charge $c_{\rm eff}$ as a function of $\gamma$ (c) and $h$ (d). Panels (e-f) Steady-State Entanglement Balance: Average entanglement gain $\overline{\delta S_{\rm nH}}$ and loss $-\overline{\delta S_{\rm qj}}/\overline{\tau}$,
as a function of the average entanglement entropy $S$ and different subsystem sizes $\ell$ (error bars are smaller than the size of the dots).In the weak monitoring phase, Panel (e), both contributions scale logarithmically with $\sin \pi \ell/L$ (see the inset), suggesting the QJs weakly renormalize the non-Hermitian dynamics. Both terms are independent of subsystem size at strong monitoring (panel f), leading to an area law.
}
\end{figure*} 
We emphasize that the classical stochastic model leading to Eq.~(\ref{eqn:S_dyn_class}) is built upon the entanglement gain and loss statistics obtained from the full QJ dynamics. Nevertheless, it still represents a considerable simplification with respect to the full quantum stochastic dynamics. The agreement between the classical model and the full QJ dynamics suggests that only few key features of the many-body state are relevant to describe the stochastic dynamics of the entanglement entropy, irrespective of the microscopic details, pointing towards a sort of universal behavior of quantum jumps and non-Hermitian evolution on entanglement. More importantly, as we will discuss in the next Section, it provides a new tool to analyze and decode the scaling of entanglement entropy in monitored systems.

\section{Applications}\label{sec:applications}

In this Section we showcase the application of our new metric, the entanglement statistics of gain/loss, to the different models of monitored fermions introduced in Sec.~\ref{sec:models} and their entanglement dynamics.  
Unless specified otherwise, throughout this Section we consider an initial product state of the form $\vert\Psi(0)\rangle=\vert0,1,0,1,\dots,0,1\rangle$ and open boundary conditions. We fix the hopping along the chains as units of energy and inverse time, $J=1$. Throughout this work we consider a system of size $L=128$ for free fermions and $L=14$ for the interacting case. We solve the QJ dynamics for non-interacting models using free-fermion techniques (see Appendix~\ref{sec:methods}) and high-order Monte Carlo wavefunction method, while for interacting fermions we use exact time-propagation with a second order Trotter expansion.

\subsection{Monitored Ising Chain}

We first present our results for the entanglement entropy dynamics of the Ising chain under QJs, which complete those  obtained in Ref.~\cite{turkeshi2022entanglementtransitionsfrom,paviglianiti2023multipartite}. In Fig.~\ref{fig:recap_ising}(a) we present the dynamics of the entanglement entropy for a certain number of stochastic trajectories and compare to the average value and the value obtained in the no-click limit.
A first observation is that the no-click limit follows
quite closely the average entanglement entropy and the fluctuations appear to remain modest in size.
 
\begin{figure*}[!t] 
\centering
\includegraphics[width=0.99\textwidth]{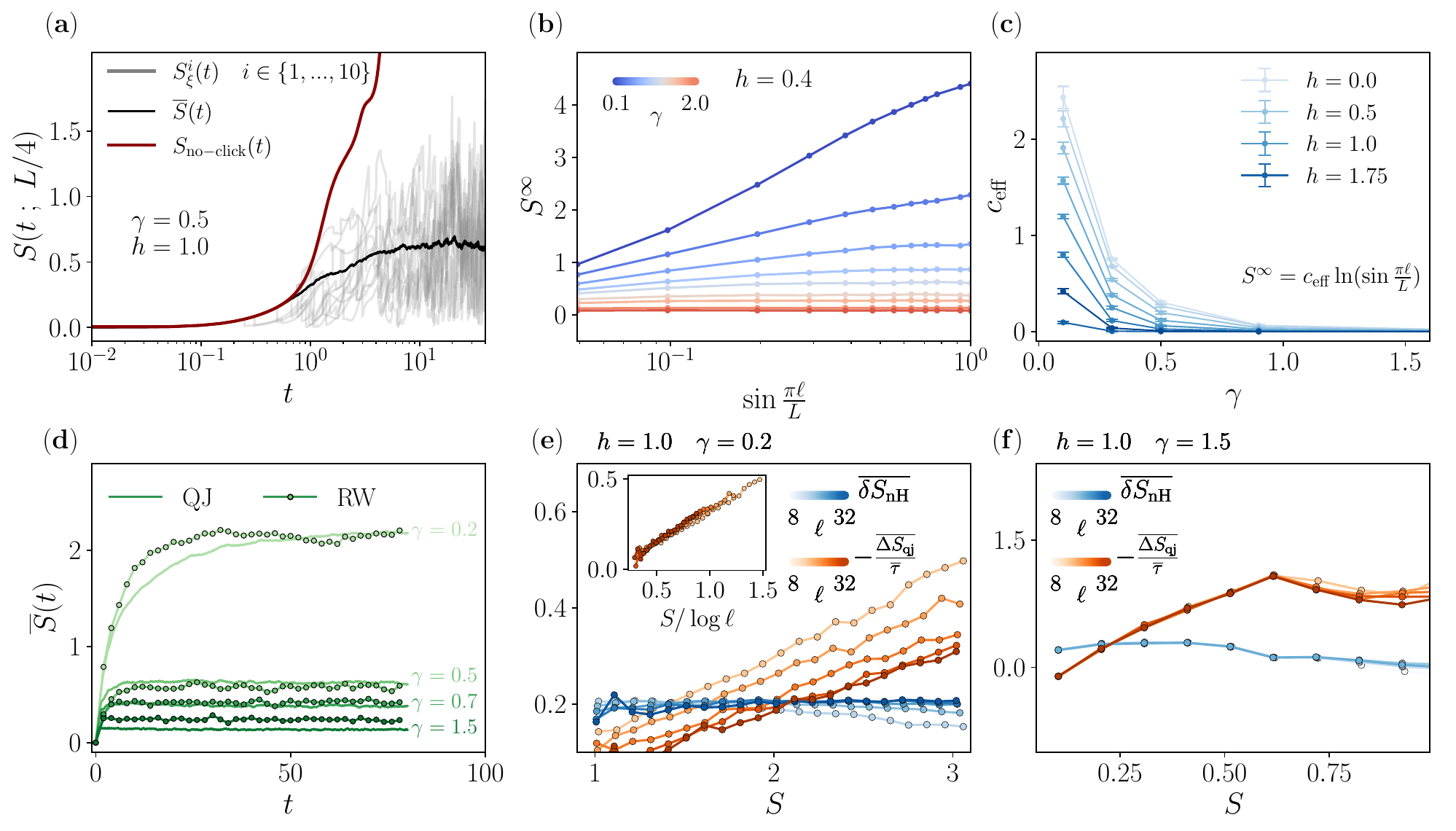}  
\caption{\label{fig:recap_SSH}
\textbf{Entanglement dynamics under QJ monitoring for the SSH chain}  --- Panel (a): Comparison between average, trajectory and no-Click evolution the measurement rate $\gamma=0.5$. Panel (b): Average steady-state entanglement for the monitored SSH model ($\gamma= \left\{0.1,~0.2,~0.3,~0.4,~0.5,~0.7,~0.9,~1.5,~2.0 \right\}$). We see for weak monitoring a logarithmic scaling that evolves into an area law upon increasing $\gamma$ ;
  Panel (c): Phase diagram as obtained from the effective central charge $c_{\rm eff}$ as a function of $\gamma$ . Panel (d): Comparison between the dynamics of the average entanglement entropy obtained from the exact quantum jump
dynamics (QJ) or the phenomenological classical random walk (RW). Panels
 (e-f):  Steady-State Entanglement Balance : Average entanglement gain $\overline{\delta S_{\rm nH}}$ and loss $-\overline{\delta S_{\rm qj}}/\overline{\tau}$, as a function of the average entanglement entropy $S$ and different subsystem sizes $\ell$ (error bars are smaller than the size of the dots). In the weak monitoring phase, Panel (e), the contribution due to QJs scale logarithmically with $\sin \pi\ell/L$ (see Inset), while the term arising from the non-Hermitian evolution is practically $\ell$-independent. In the strong monitoring phase, corresponding to the area law, both contributions do not depend on $\ell$.
  }\end{figure*}

In Fig.~\ref{fig:recap_ising} (b) we present the scaling of the steady-state entanglement entropy as a function of the subsystem size $\ell$, for different values of the parameters. We see an entanglement transition into an area law as the monitoring rate is increased. The weak-monitoring phase has an entanglement entropy compatible with a logarithmic scaling. Following previous works we extract an effective central charge $c_{\rm eff}$ by fitting the steady state entanglement entropy as  $S^{\infty}_{\ell}=c_{\rm eff}\log \sin \pi\ell/ L  $. In Fig.~\ref{fig:recap_ising} (c,d) we plot $c_{\rm eff}$ as a function of $\gamma$ and $h$. 
 
We see that increasing either $\gamma$ or $h$ drives a transition into a phase with $c_{\rm eff}=0$, corresponding to an area-law scaling of the entanglement entropy, in qualitative analogy to the no-click limit (see Appendix~\ref{app:noclick}). Still certain aspects of the phase diagram appear to be different in the QJ case, the most striking one being visible at large field $h>1$ and small $\gamma$, where the stochastic problem shows a sub-volume logarithmic scaling of the entanglement entropy in a region where according to the no-click evolution the system should be in the area law, as already noted in  Ref.~\cite{paviglianiti2023multipartite}.

We can now use the insight from the stochastic classical model, in particular the steady-state condition in Eq.~\ref{eq:steadystate}, to breakdown the entanglement entropy content of the steady-state. In particular in  Fig.~\ref{fig:recap_ising}(e-f) we plot separately  the entanglement gain $\overline{\delta S_{\rm nH}}$ and loss $-\overline{\Delta S_{\rm qj}}/\overline{\tau}$ as a function of the entanglement entropy content $S$ across the phase diagram and for different subsystem sizes $\ell$. According to our classical picture the steady-state is obtained when gain and loss are balanced, corresponding at a value of entanglement entropy where the two curves match. The scaling with subsystem size of the two contributions tells us about the mechanism driving the entanglement production. In the weak monitoring phase, for example, (panel e) both contributions scale logarithmically (see inset) suggesting that the entanglement gain due to the renormalized non-Hermitian dynamics, shows a logarithmic scaling with $\ell$, as in the no-click limit at this value of parameters. Based on these results, we conclude that in the weak monitoring phase, the non-Hermitian dynamics is only weakly renormalized by QJs. In the area law on the other side (panel f) both gain and loss contributions become essentially independent from subsystem size, crossing in a low-entanglement region as expected from an area law.
A similar analysis can be done in the region of weak monitoring rate and large field (see Appendix~\ref{app:MoreIsing}) and reveals that again for moderate field, $h\sim 1$ the non-Hermitian dynamics seems weakly renormalised and drives the logarithmic growth of entanglement entropy, while for large $h$ the non-Hermitian evolution is effectively area law while the contribution due to QJs still has a non-trivial scaling with $\ell$, which is responsible for the observed logarithmic scaling of the entanglement entropy at large values of the field $h$.

\begin{figure*}[!t]   
    \centering
   \includegraphics[width=0.99\textwidth]{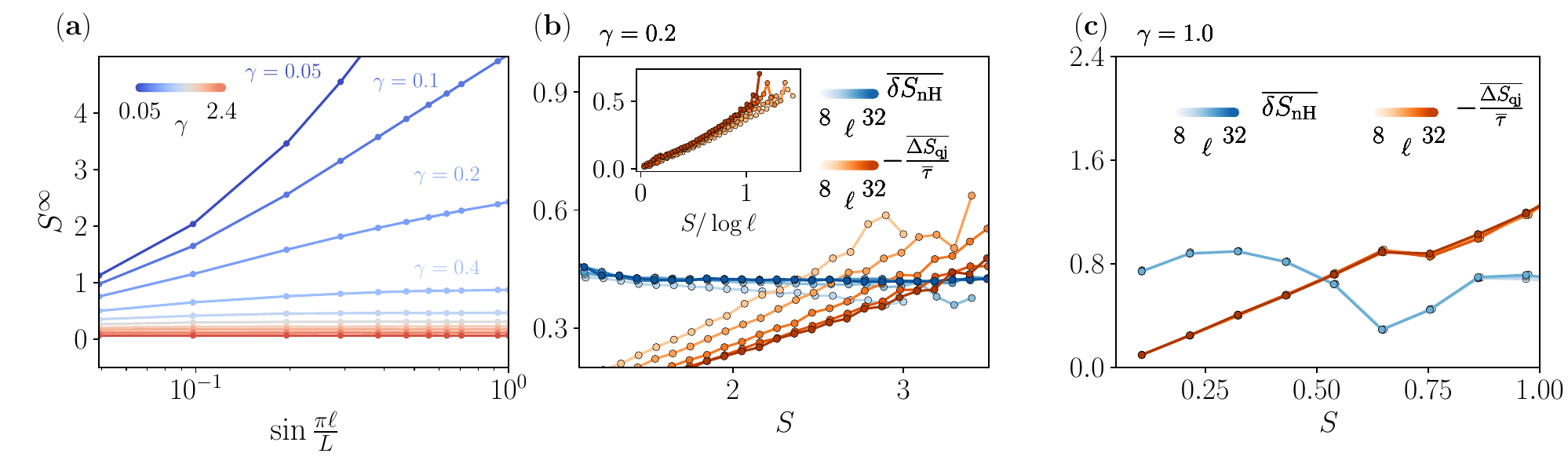}	
			\caption{
  \label{fig:recap_freeU1}  \textbf{Entanglement dynamics under QJ monitoring for monitored free fermions with $U(1)$ symmetry}  --- Panel (a) Average steady-state entanglement ($\gamma= \left\{0.05,~0.1,~0.2,~0.4,~0.6,~0.8,~1.0,~1.2,,~1.6,~2.4 \right\}$). We see for weak monitoring a logarithmic scaling that evolves into an area law upon increasing $\gamma$ ; 
   Panels
 (b-c):  Steady-State Entanglement Balance : Average entanglement gain $\overline{\delta S_{\rm nH}}$ and loss $-\overline{\Delta S_{\rm qj}}/\overline{\tau}$, as a function of the average entanglement entropy $S$ and different subsystem sizes $\ell$ (error bars are smaller than the size of the dots). In the weak monitoring phase, Panel (b), the contribution due to QJs scale logarithmically with $\ell$(see Inset), while the term arising from the non-Hermitian evolution is practically flat. In the strong monitoring phase, corresponding to the area law, both contributions do not depend on $\ell$. The crossing point corresponds again to the zero of $\Phi(S,\ell)$ i.e., to the steady state entanglement $S_{\infty}$. 
  }
\end{figure*}
\subsection{Monitored SSH Model}

We now consider the entanglement dynamics for the monitored SSH model. In Fig.~\ref{fig:recap_SSH} (a), we show the evolution of the entanglement entropy for 
a sample of quantum trajectories for weak monitoring $\gamma=0.5$, corresponding to the volume-law phase of the no-click limit (see Appendix~\ref{app:noclick}). 
Already at the level of single quantum trajectories it is evident that QJs have, in this case, a strong impact with respect to the no-click evolution. This is true in particular at weak monitoring where the no-click dynamics displays a linear growth corresponding to the volume law phase. Here, the stochastic dynamics induced by the quantum jumps effectively suppress the entanglement growth and bring the system to a stationary state which is less entangled than in the no-click limit. 

The scaling of the steady-state average entanglement entropy for different subsystem sizes and increasing values $\gamma$ is shown in Fig.~\ref{fig:recap_SSH} (b). The entanglement entropy growth in the weak monitoring regime is compatible with a logarithmic law, as for the Ising model, from which we can extract an effective central charge $c_{\rm eff}$. As we show in panel (c) this quantity vanishes as a function of $\gamma$, indicating a transition into an area-law phase.  
Although affected by the inevitable finite-size effects, our results show that the volume law phase of the non-Hermitian SSH is not stable when including quantum jumps.

We now show how looking at the statistics of entanglement gain/loss allows to obtain further insights on the impact of QJs on the monitored SSH. We first show (Fig.~\ref{fig:recap_SSH} panel d) that the classical random walk model with resetting captures perfectly the entanglement dynamics also for the SSH case. Then in panels (e-f) we look at the steady state entanglement gain and loss, respectively for weak and strong monitoring. In the former case, we see clearly that the jump contribution depends strongly on subsystem size; the bigger the subsystem, the less QJs decrease the entanglement on average.
Interestingly, the subsystem size dependence of this contribution looks logarithmic, as shown in the inset of panel (e).
On the other hand, the contribution of the non-Hermitian evolution depends weakly on both $\ell$ and $S$, at least for the values of entanglement entropy, which are allowed to be explored by the jumps, i.e., around the crossing point which corresponds to the steady-state entanglement $S_{\infty}$. 
For large monitoring (Fig~\ref{fig:recap_SSH} (f)), on the other hand, we see that both contributions are almost independent on $\ell$, and their crossing occurs in a regime of small entanglement entropy, leading to an area-law scaling. 

From this result, we conclude that for the monitored SSH, the logarithmic growth of the entanglement entropy at weak monitoring arises mainly due to QJs. The non-Hermitian dynamics on the other hand, which would lead to volume law in the no-click limit, is strongly renormalized by the effect of QJs and barely depends on subsystem size. 
In Appendix~\ref{sec:high_ent}, we discuss the dynamics from a highly entangled state to show how, in that case, one could probe the subsystem size dependence in the non-Hermitian contribution, but only at large values of entropy, far above what the system can explore under the effects of QJs in the dynamics starting from lowly entangled initial states.

\begin{figure*}[!t]   
    \centering
   \includegraphics[width=0.99\textwidth]{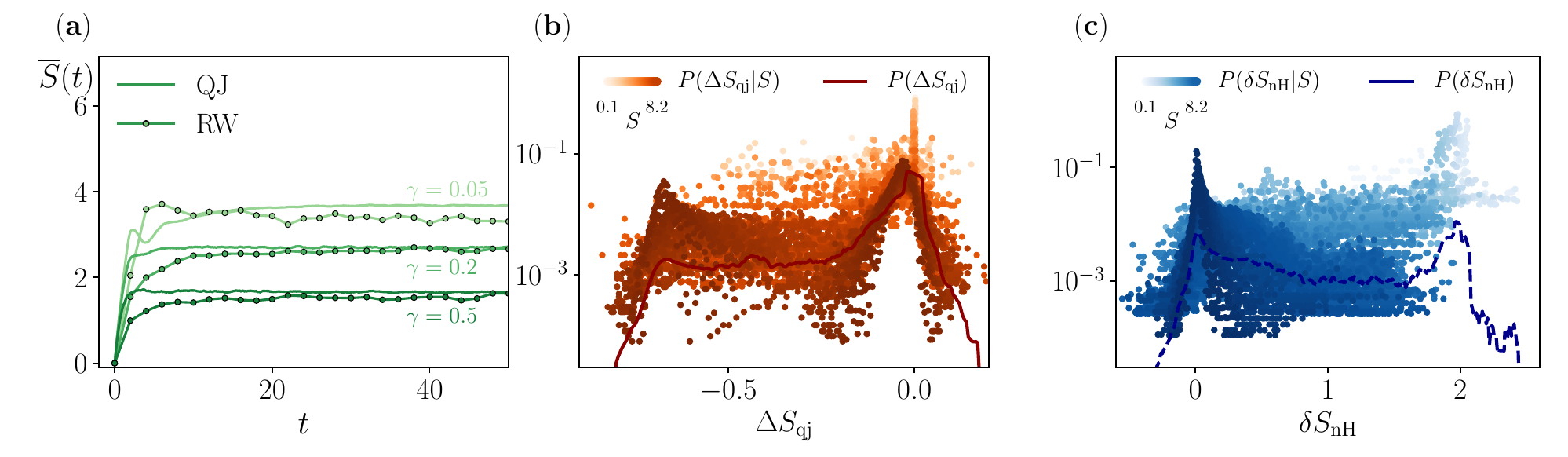}	
			\caption{
  \label{fig:recap_fermionsU1_interacting}  \textbf{Entanglement dynamics under QJ monitoring for interacting monitored fermions with $U(1)$ symmetry} --- Panel (a): Comparison between the dynamics of the average entanglement entropy obtained from the exact quantum jump dynamics (QJ) and the average of the phenomenological classical random walk (RW), 200 classical trajectories are used to perform the average. The steady state entanglement $S_{\infty}$ corresponds again to the zero of $\Phi(S,\ell)$ computed from the histograms of (b-c).
  Panel (b-c): The histogram of the entanglement entropy change after a QJ $\Delta S_{\rm qj}$ is plotted in (b) and the one in between QJs $\Delta S_{\rm nh}$ ie. due to the non-Hermitian evolution is shown in (c).
  }
\end{figure*}
\subsection{Fermions with $U(1)$ Symmetry and Monitoring of Local Density}


We now present an application of the entanglement gain/loss statistics to the case of monitored free fermions with $U(1)$ symmetry related to charge conservation, a problem that has attracted large interest in the literature. Numerical investigations with different monitoring protocols reported a transition from a critical phase with logarithmic scaling of entanglement entropy to an area-law phase, for both quantum-jumps and quantum state diffusion (QSD) type of density-monitoring~\cite{alberton2021entanglementtransitionin,vanregemortel2021entanglement}. The robustness of the weak-monitoring sub-volume phase, and hence of the transition, was questioned in the case of projective measurements, where numerical evidence supported an area-law steady-state entanglement~\cite{coppola2022growthofentanglement}. This result has been theoretically understood within a replica field theory calculation~\cite{poboiko2023theoryoffree} which identified connections with the physics of weak-localisation corrections.
Given this landscape it is therefore interesting to discuss what types of insights our analysis based on entanglement loss and gain can provide. To this extent we first show in Fig.~\ref{fig:recap_freeU1} (a) our data for the steady-state entanglement entropy versus the subsystem size, to confirm the presence of a sharp crossover from a logarithmic scaling of the 
entropy to an area law, consistent with the literature.

We then use our metric of the statistics of entanglement gain and loss to investigate the origin of this log-scaling for weak monitoring. 
Specifically we plot in panels (b-c) the contribution to the steady-state entanglement entropy due to QJs and to the renormalised non-Hermitian evolution. Quite interestingly we note a behavior very similar to the SSH case, namely the sub-volume (logarithmic, see inset) scaling of the entanglement entropy is essentially due to QJs, while the non-Hermitian Hamiltonian contribution, renormalised by the jumps, is essentially independent on $\ell$. We see therefore another case in which the no-click limit, that for this problem would describe a purely Hermitian evolution, is strongly perturbed by QJs. For large monitoring on the other hand both contributions are independent on $\ell$, compatibly with the area law scaling in panel (a). 

\subsection{Role of Interactions}

Finally we conclude this Section by discussing the generality of our findings beyond the realm of non-interacting monitored fermions. To this extent we consider the same model as in the previous section (free fermions with density monitoring) but include a nearest-neighbor interaction $V\neq0$ (See Sec.~\ref{sec:models}). In this case in the weak-monitoring phase the entanglement dynamics appears compatible with a volume law. Our interest here is to show whether the entanglement gain/loss picture and the classical stochastic model continue to work also in the case of interacting monitoring systems. 
Fig.~\ref{fig:recap_fermionsU1_interacting} shows that this is indeed the case.  In particular we plot the histograms of entanglement gain and loss for the interacting model and show that the general features, in particular the asymmetry of the distributions and the strong peak at zero persist also in the interacting case. An intriguing finding is that, as compared to the free fermionic case, here the role of the entanglement content is somewhat reversed: the probability to have a manifest entanglement change is bigger for weakly entangled states than highly entangled ones. This seems compatible with the expectation that the volume law is stable to the inclusion of measurements, indeed, since entanglement production is particularly important at low entanglement, the loss due to jumps is of lesser significance. On the other hand, having a reduced entanglement loss at large entanglement content should enable the support of a volume law phase.

\section{Discussion and Implications for Free Fermions MIPT}

Our findings for the average entanglement gain and loss have highlighted the concept of renormalized non-Hermitian dynamics. Due to quantum jumps, which effectively reshuffle the initial state onto which the non-Hermitian Hamiltonian act, the effective non-Hermitian dynamics is renormalized with respect to the bare no-click evolution. 
While for the Ising chain the logarithmic law phase is only weakly renormalized, quantum jumps can strongly impact the dynamics of weakly monitored $U(1)$ conserving models.
In particular, in the non-interacting case, quantum jumps completely wash away the no-click volume-law phase.  (Nevertheless, the inclusion of non-Gaussian interactions enhance the entanglement growth from the no-click limit, stabilizing a volume-law phase at low measurement rate. )
Our findings, therefore, highlight a fundamental difference between the weak monitoring phase of the Ising and the non-interacting $U(1)$ models.
While either cases the weak monitoring phase features a logarithmic scaling of the entanglement entropy, the origin behind this scaling stems from inequivalent mechanisms.
In the $U(1)$ symmetric models, the logarithmic scaling comes essentially from the quantum jumps action, while for the Ising chain it is the result of the combined effect of jumps and non-Hermitian evolution.
The same renormalization effect we have shown to be at play in the monitored Ising chain at intermediate and large field values and can, therefore, naturally explain the departure from the no-click limit.

We conclude with a discussion of the implications of our results for the MIPT of free fermions under quantum jumps. The irrelevance of QJs at weak monitoring in the Ising chain points towards an important role of the non-Hermitian Hamiltonian in this regime. This suggests that the logarithmic phase of entanglement entropy under QJs could be possibly stable in the large system size, as in the no-click case~\cite{turkeshi2023entanglementandcorrelation}. This is consistent with the replica field theory prediction for noisy Majorana fermions under QSD~\cite{fava2023nonlinearsigmamodels}, although we emphasize the differences in measurement protocol and unitary dynamics between the two cases. In particular based on our numerical results we cannot establish whether also in the Ising chain the entanglement scaling is $\ln^2(\ell)$. On the other hand our results for the SSH case points toward a strong effect of QJs at weak monitoring. As we have shown the logarithmic phase found numerically is mainly due to QJs. Their contribution is likely to saturate at large system sizes through a mechanism similar to the weak-localization correction~\cite{poboiko2023theoryoffree}.

\section{Conclusions}\label{sec:conclusion}

In this work we have studied the role of Quantum Jumps on the entanglement entropy dynamics of monitored quantum many-body systems. A main result is the introduction of a new tool which looks at the statistics of entanglement entropy gain and loss after or between QJs. We have shown that quite generically the resulting histograms display, particularly for the quantum jump case,  a very broad distribution and a typical value pinned close to zero, suggesting that most QJs are not significantly affecting the entanglement entropy controlled by rare jump events.
Using the full statistics of entanglement gain and loss we have built a stochastic random walk model with partial resetting, which can reproduce the full QJ dynamics of entanglement entropy. This phenomenological model offers a new light for interpreting the QJ numerics. Indeed, it suggests a natural steady-state condition for the entanglement entropy, given by the balance between gain due to the non-Hermitian evolution and losses due to quantum jumps. Remarkably, this condition accurately reproduced the scaling of entanglement entropy with subsystem size obtained by the full QJ dynamics. Furthermore, it clarified the origin of the different entanglement scaling and the mutual role of jumps and non-Hermitian evolution.  

We have used this framework to decode the mechanism controlling the entanglement dynamics and associated MIPT for three models of monitored free fermions. The outcome of this analysis reveals a compelling difference at weak monitoring between the models with $U(1)$ symmetry, where the renormalised non-Hermitian Hamiltonian does not contribute to the scaling of the entanglement entropy, and the Ising chain where this renormalisation is not present and the no-click limit remains stable to the inclusion of QJs.
Finally we have extended our entanglement gain/loss picture to the case of interacting monitored systems. Remarkably we have shown that the classical random-walk model with resetting remains valid and captures quantitatively the entanglement entropy dynamics. The statistics of entanglement gain and loss for interacting systems suggests that highly entangled states are more robust to quantum jumps - the probability to observe a rare jump which changes substantially the entanglement entropy is smaller for high entanglement content. This is in stark contrast with the case of gaussian states and can therefore suggests an interpretation for the stability of the volume law phase at weak-monitoring observed in numerical simulations on interacting monitored systems~\cite{fuji2020measurementinducedquantum}.

We envision various follow-up to this work. First, the presence of imperfect detector and dissipative interaction with the environment would require adapting the statistics of entanglement gain and loss to entanglement measures valid in open quantum systems, such as the logarithmic negativity or the quantum fisher information. 
Another venue of interest for the study of the statistics of entanglement gain and loss are higher-dimensional and long-range interacting systems, where non-conformal subextensive phases~\cite{chahine2023entanglement,muller2022measurementinduced} has been identified. 
Introducing feedback and control would drastically change the phenomenology: these elements will include new sources and sink terms in the stochastic model. We leave these generalization for future work.

\begin{acknowledgments}
We thank A. Biella, M. Buchhold, J. Dalibard, M. Dalmonte, R. Fazio, A. Paviglianiti, L. Piroli, A. Romito, P. Sierant, and A. Silva for discussions and collaborations on related topics. 
We acknowledge computational resources on the Coll\'ege de France IPH cluster. X.T. acknowledge DFG under Germany's Excellence Strategy – Cluster of Excellence Matter and Light for Quantum Computing (ML4Q) EXC 2004/1 – 390534769, and DFG Collaborative Research Center (CRC) 183 Project No. 277101999 - project B01. 
\textit{Note added. During the completion of this work, we become aware of a related manuscript studying the stability of the non-Hermitian Hamiltonian for quantum state diffusion~\cite{leung2023theory}. }

\end{acknowledgments}

\appendix
\newpage
\begin{figure}[!h]   
    \centering
   \includegraphics[width=\columnwidth]{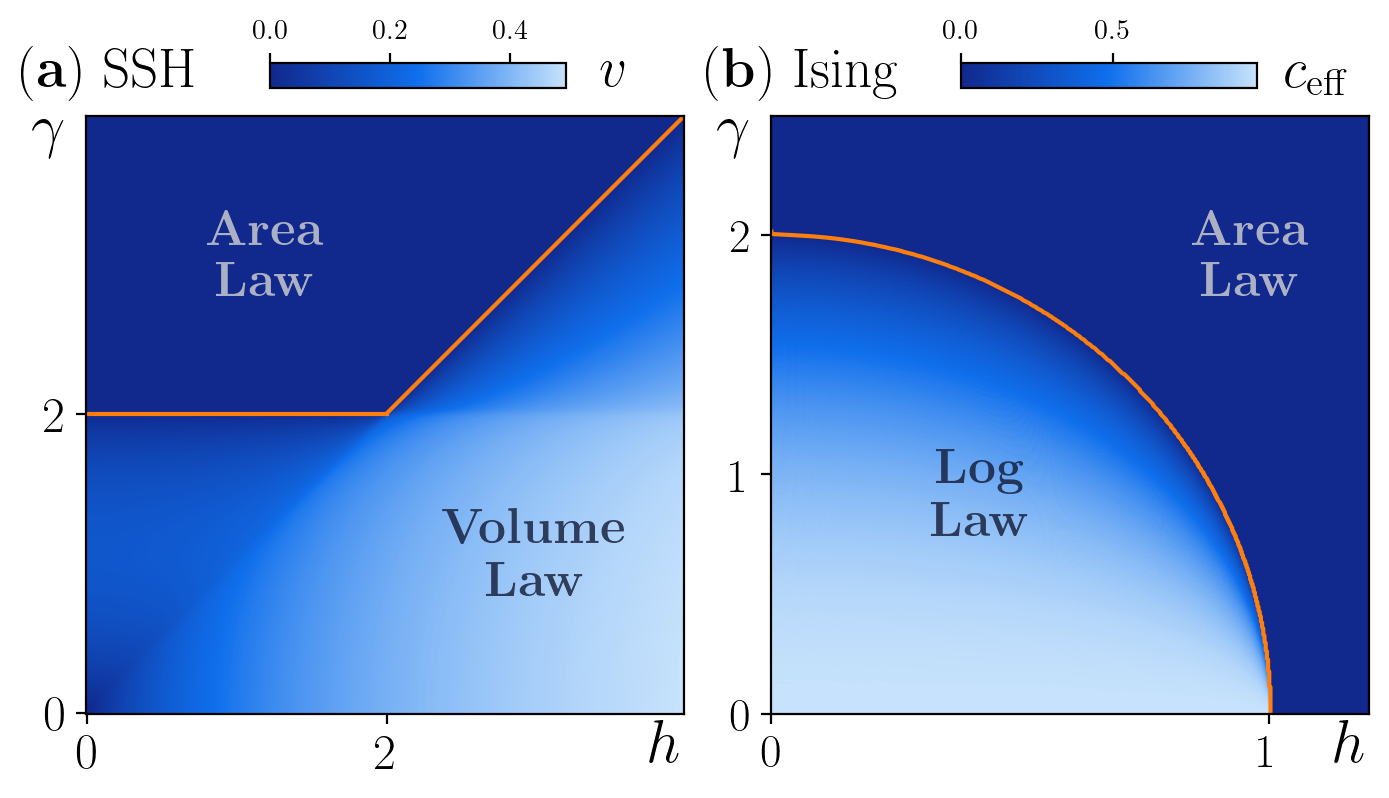}	
			\caption{
  \label{fig:recap} Entanglement phase diagram of the monitored SSH and Ising models in the no-click limit, corresponding to a deterministic non-Hermitian evolution with $H_{\rm eff}$.  Left panel: non-Hermitian SSH and volume-to area-law transition~\cite{legal2023volumetoarea}; Right panel: non-Hermitian Ising chain, featuring a transition from a critical phase with logarithmic scaling of the entanglement to an area law Zeno phase~\cite{turkeshi2023entanglementandcorrelation}. Note that our convention for the jump operators in Eq.~(\ref{eqn:jumps_Ising}) leads to a rescaling of $\gamma$ by a factor $1/2$, with respect to Ref.~\onlinecite{turkeshi2023entanglementandcorrelation}.}
\end{figure}

\section{Summary of Entanglement Dynamics in the No-Click Limit}\label{app:noclick}

Here we briefly review the results obtained for the dynamics of the entanglement entropy under purely non-Hermitian evolution driven by $H_{\rm eff}$, both for the SSH and for the Ising model, corresponding to the no-click limit of the QJ dynamics. In both cases the time evolution for the entanglement entropy ${S_{\rm nH}(t)}$ can be computed exactly in the thermodynamic limit using free-fermion techniques~\cite{turkeshi2023entanglementandcorrelation,legal2023volumetoarea}. The results have revealed a rich phase diagram as a function of the monitoring strength $\gamma$, which for a non-Hermitian problem is given by the back-action term, and the field $h$ entering the non-Hermitian Hamiltonian. We plot in Fig.~\ref{fig:recap} the phase diagram of the two models for completeness.

In the non-Hermitian SSH model in Eq.~(\ref{eqn:ssh_nh}) the weak monitoring phase for ${\gamma<\gamma^{\rm SSH}_c(h)}$ is characterized by an entanglement entropy growing linearly in time, ${S_{\rm nH}(t)\sim t}$ and saturating to a stationary value that scales linearly with the subsystem size $\ell$, i.e., ${S_{\rm nH}(t\to\infty)\simeq v \ell}$, characteristic of a volume-law phase. This is associated to the purely real spectrum protected by PT symmetry~\cite{legal2023volumetoarea}. The close-form expression of $v$ allows the complete characterization of the entanglement phases.
As the measurement strength $\gamma$ increases at fixed $h$ the prefactor $v$ of the volume law scaling decreases (cf. Fig.~\ref{fig:recap}(a)) until a critical value is reached when $v=0$ and the system enters the area-law scaling for the entanglement entropy. This entanglement transition was shown to be directly related to the spectral transition occurring in $H_{\rm eff}$~\cite{legal2023volumetoarea}. At weak monitoring the non-Hermitian quasiparticle spectrum is purely real due to the PT symmetry of $H_{\rm eff}$ and as $\gamma$ increases first PT breaks and some quasiparticle mode acquires a finite-lifetime, which induces a sharp decrease of $v$. However it is only at ${\gamma^{\rm SSH}_c(h)}$, when all the quasiparticle modes acquire a finite lifetime, the system enters the area-law scaling for the entanglement entropy.

In the non-Hermitian Ising chain in a complex transverse field, Eq.~(\ref{eqn:Ising_nh}), the entanglement entropy was found for weak monitoring $\gamma<\gamma^{\rm Ising}_c(h)$ to depend logarithmically on both time and subsystem size. The corresponding effective central charge $c_{\rm eff}$, obtained from the ansatz $S_{\rm nH}(\ell)= c_{\rm eff} \ln(\ell)$, could be obtained in closed form~\cite{turkeshi2023entanglementandcorrelation} and was shown to decrease as a function of both $\gamma$ and $h$ up to the critical line $\gamma^{\rm Ising}_c(h)$, above which the system was found to undergo an entanglement transition into an area-law phase, cf. Fig.~\ref{fig:recap}(b).  As for the SSH case, the entanglement transition for the Ising chains is directly related to a transition in the spectrum of non-Hermitian quasiparticles. The latter separates a critical gapless phase with vanishing imaginary part of the spectrum at a given point in the Brillouin zone, from a gapped phase~\cite{biella2021manybodyquantumzeno}.

\section{Statistics of Jumps Waiting Times}
\label{sec:waiting time}
In this Appendix we discuss how  QJs are distributed in time, i.e. we compute their Waiting-Time Distribution (WTD). This quantity has been studied in the early days of quantum jumps, motivated by resonance fluorescence spectra~\cite{Cohen-Tannoudji_1986,carmichael1989photoelectron}, and have found a multitude of applications from solid-state quantum information~\cite{delteil2014observation} to quantum transport~\cite{brandes2008waiting,albert2012electron,landi2021waiting,landi2023patterns,coppola2023conditional,landi2023current} to laser cooling~\cite{bardou1994subrecoil} and is also a key quantity in this work.  

In general, computing the waiting time distribution is a non-trivial task~\cite{landi2023current}. There is, however, a special case where this can be done straightforwardly. Whenever the imaginary part of the non-Hermitian Hamiltonian commutes with the real part (and that the initial state is an eigenstate of the non-Hermitian part), then one can conclude that the norm decay is exponential and therefore so the cumulative waiting time-distribution, $F=1-\exp\left(- K \tau\right)$ where $K=\sum_i \langle L^{\dagger}_i L_i\rangle$. In this case, the time for the next jump can be directly obtained as a function of a random number $r$, as $\tau=-\frac{1}{K}\mbox{ln} r$. Such an instance occurs for a system with local density monitoring and conservation of global particle number~\cite{alberton2021entanglementtransitionin,coppola2022growthofentanglement}.

Here this quantity has to be extracted numerically, but emerges naturally from the numerical implementation of stochastic QJ dynamics beyond the first-order Monte Carlo schemes~\cite{daley2014quantum}. We start recalling that the WTD is obtained from the decay of the norm $\mathcal{N}(t)\equiv ||e^{-i H_\mathrm{eff}t}|\Psi\rangle||$, which reads in general $\mathcal{N}(t) = \exp(- K(t) t ) $ where
$K(t)$ represents the back-action associated to the non-Hermitian Hamiltonian $H_\mathrm{eff}$, i.e. $K(t)=\sum_i \langle L^{\dagger}_i L_i\rangle$. At short times the rate of decay of the norm can display fluctuations (depending on the model and the initial state) which can result in biasing the WTD. However at sufficiently long-times one expect a Poisson law
\begin{equation}
P_{WT}(\tau;t\rightarrow\infty)\sim e^{-K_\gamma^\infty \tau}    
\end{equation}
at least for systems where the monitoring process is local on each sites (where $K_\gamma^\infty$ denotes the steady state value of $K(t)$ which eventually depends on the monitoring rate $\gamma$).

\begin{figure}[t!]
    \includegraphics[width=0.95\columnwidth]{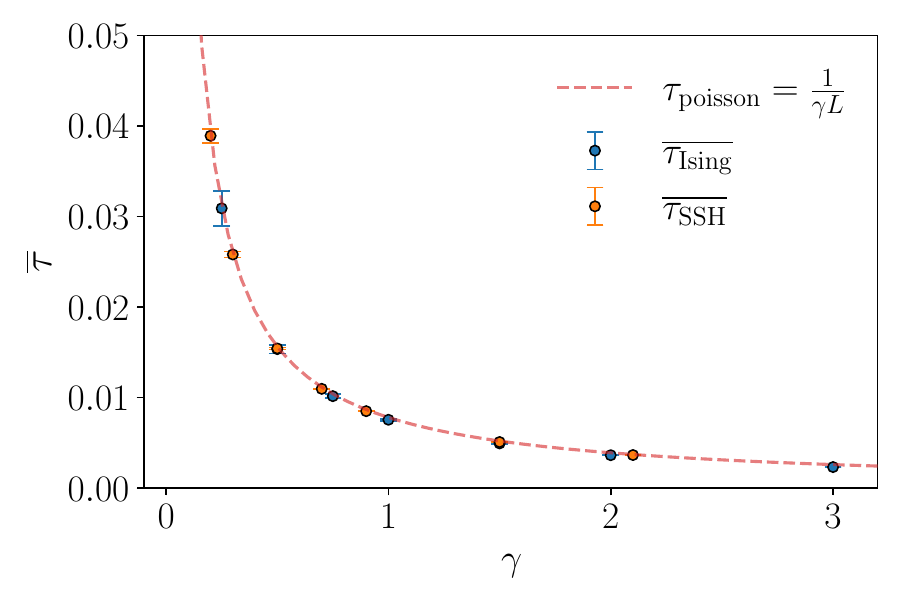}
    \caption{\label{fig:WTD_Ising} Average waiting time of quantum jumps in the Ising and SSH models, showing perfect agreement. In both cases $\overline{\tau}\sim 1/\gamma L$.}
\end{figure}

Our numerical analysis on the monitored Ising and SSH models confirm the expectation that QJs are Poisson distributed. The SSH model displays more pronounced tails in the short-time WTD, caused by fluctuations due to the initial state. Indeed for the SSH model the observable $K(t)$ driving the decay of the norm is $K_\mathrm{SSH}(t) = 2\gamma \sum_i [n_{A,i}(t) + 1 - n_{B,i}(t)]$, where $n_{A/B,i}(t) = \langle \Psi(t)|c_{A/B,i}^\dagger c_{A/B,i}|\Psi(t)\rangle$. At short times, the observable $K_\mathrm{SSH}(t)$ displays significant temporal fluctuations due to the chosen initial condition, and the  WTD shows longer tails. On the other hand in the Ising model  jumps try to refill empty sites and keep the total number of particles effectively constant. Thus, $K(t)$ is centered around the average number of particles with small fluctuations.  A Poisson distribution of QJ is typical for systems in which the backaction term is a constant of motion. Interestingly, both models feature the same average waiting time of QJs, as shown in Fig.~\ref{fig:WTD_Ising} (b), and given by $\tau\sim 1/\gamma L$.

\section{System size scaling of entanglement loss}\label{app:scalingL}

\begin{figure}[b]
\includegraphics[width=0.99\columnwidth]{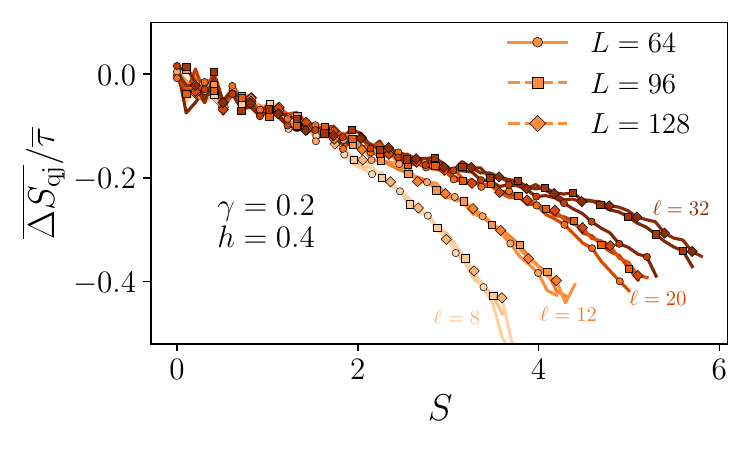} 	
	\caption{\label{fig:loss_scaling}  Loss distribution $ \overline{\Delta S_\mathrm{qj}}/ \overline{\tau} $ obtained in the Ising model for different system sizes $L=64,~96,~128$. The distribution is invariant regarding the system size; we observe a slight disagreement at large subsystem sizes $\ell$ in the small system $L=64$, which is due to the finite size effect. }
\end{figure}

In the main text we introduced a classical stochastic model for the entanglement entropy dynamics, which builds upon the entanglement gain and loss statistics. In particular, the role of QJs is to induce an entanglement loss that, in the stationary state, is balanced by the gain provided by the non-Hermitian evolution, as described in Eq.~(\ref{eqn:S_dyn_class}) of the main text. This equation divides the entanglement loss by the average waiting time $\overline{\tau}$, which considers the instantaneous nature of QJs.  As we have seen in Appendix.~\ref{sec:waiting time}, the average waiting time scales as $1/L$, which raises the question of the stability of the steady-state condition in the large system size limit. In this Appendix, we provide evidence supporting the statement that the average entanglement loss due to the jump scales as $1/L$, so the ratio with the averaging waiting time remains finite when $L\rightarrow \infty$. In Fig.~\ref{fig:loss_scaling}, we plot, for the monitored Ising chain at a representative value of the parameters, the average entanglement loss divided by the waiting time, $\overline{\Delta S_{\rm qj}}(S,\ell)/\overline{\tau}$ as a function of $S,\ell$ and different system sizes. We see that the data corresponding to different system sizes collapses onto each other for each subsystem sizes $\ell$, which demonstrate this invariance of the ratio $ \overline{\Delta S_\mathrm{qj}}/ \overline{\tau} $.

\section{Quantum jumps from highly entangled initial states}
\label{sec:high_ent}

\begin{figure}[h]
\includegraphics[width=0.95\columnwidth]{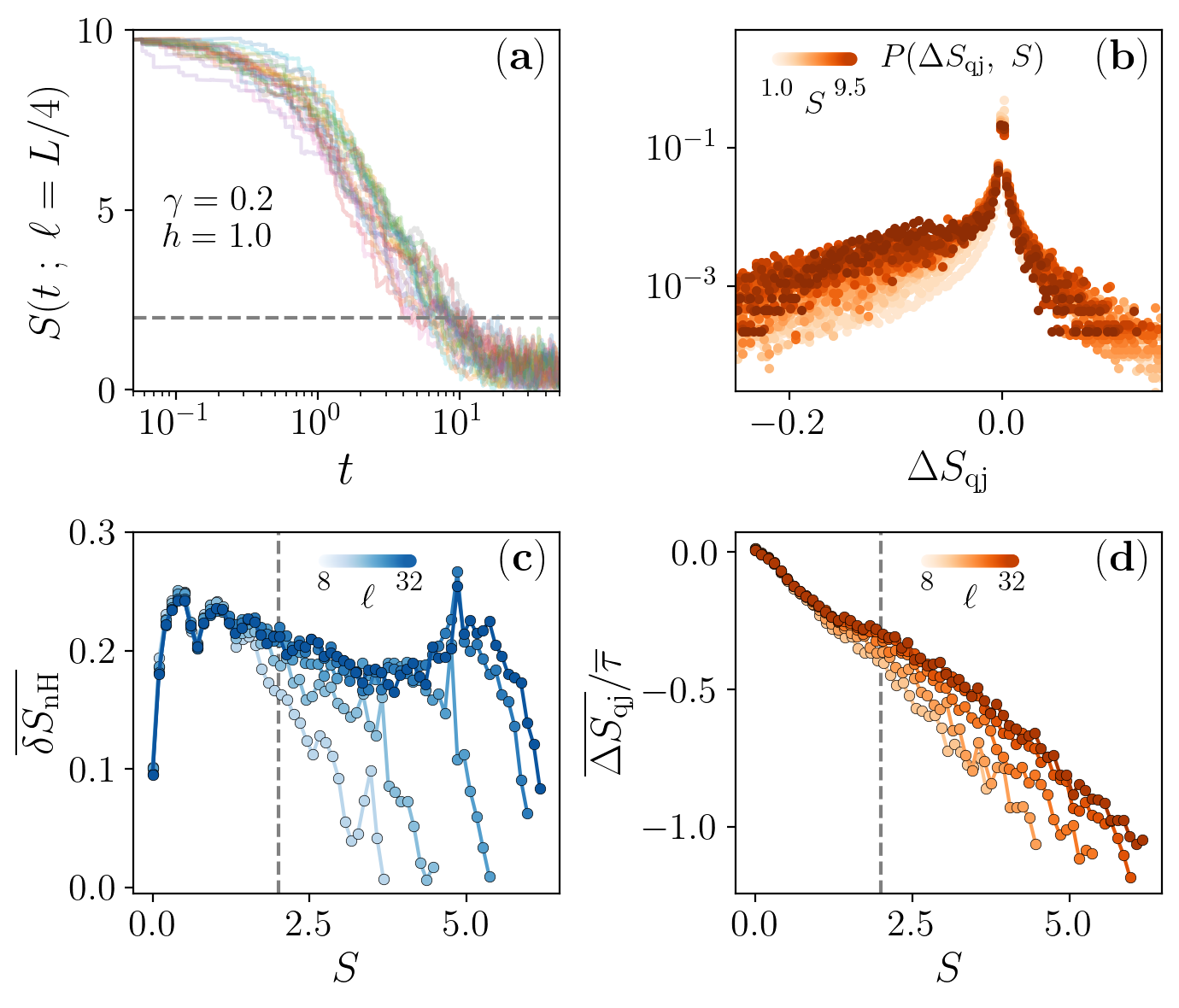} 
	\caption{\label{fig:ssh_high_ent} Analysis of trajectories obtained from a highly entangled initial state in the SSH model. Panel (a) : Entanglement  Dynamics in the monitored SSH chain under QJ dynamics with an initial state highly entangled (namely a steady state of the no-click limit). In the inset we show the conditioned version of the distribution  $\Delta S_{\rm qj}$ with respect to the entanglement content when jumps happen. Panel (b) :  First moment of the conditioned distribution $\delta S_{\rm nH}$ in function of the entanglement content $S$. This quantity is obtained from these trajectories starting from the highly entangled initial state and which allow to probe the large entanglement behaviour of the distribution}
\end{figure}

In this Appendix we provide evidence to support the conjecture that highly entangled gaussian states are more fragile to measurements than weakly entangled ones. To this extent, we consider the monitored SSH chain starting from an initial condition corresponding to the long-time limit of the associated non-Hermitian Hamiltonian, which is known to support volume-law entangled states for small values of $\gamma$~\cite{legal2023volumetoarea}.  In Fig.~\ref{fig:ssh_high_ent} (a) we show a sample of trajectories for the entanglement entropy, all converging towards a steady-state value with low entanglement (and equal to the steady-state reached from a product state initial condition). In other words, the system under monitoring cannot sustain volume-law entanglement. In panel (b) we plot the conditional distribution of entanglement loss $P(\Delta S_{\rm qj}, S)$, which now displays a broadening of its tails indicating that the role of jumps become more relevant. In particular, we see that the probability of a large entanglement loss due to jumps at atypical (i.e. high) entanglement content  increases, which explains why the initial state entanglement cannot be preserved. In panels (c) we present the same analysis given in the main text on the first moment of the gain conditional distribution $\overline{\delta S_{\rm nH}}$, but for this particular initial condition which allows us to probe entanglement content of higher values. We note that in the region of the steady state (at relatively low entanglement  $S\leq 2$ delimited by the dashed line), the average entanglement gain from non-Hermitian dynamics is  essentially independent from subsystem size $\ell$, a behavior similar to what observed in Fig.~\ref{fig:entanglement_breakdown_ssh} for a different initial condition. However, as the system dynamics explores larger values of entanglement entropy ($S\ge 2$) we see important subsystem size effects in $\overline{\delta S_{\rm nH}}$. This behaviour is usually hidden when starting from a low entangled initial condition because the jumps are confining the dynamics in this lowly entangled space.

\section{Monitored Ising Chain at Large Field}\label{app:MoreIsing}

\begin{figure}[b]
    \includegraphics[width=0.95\columnwidth]{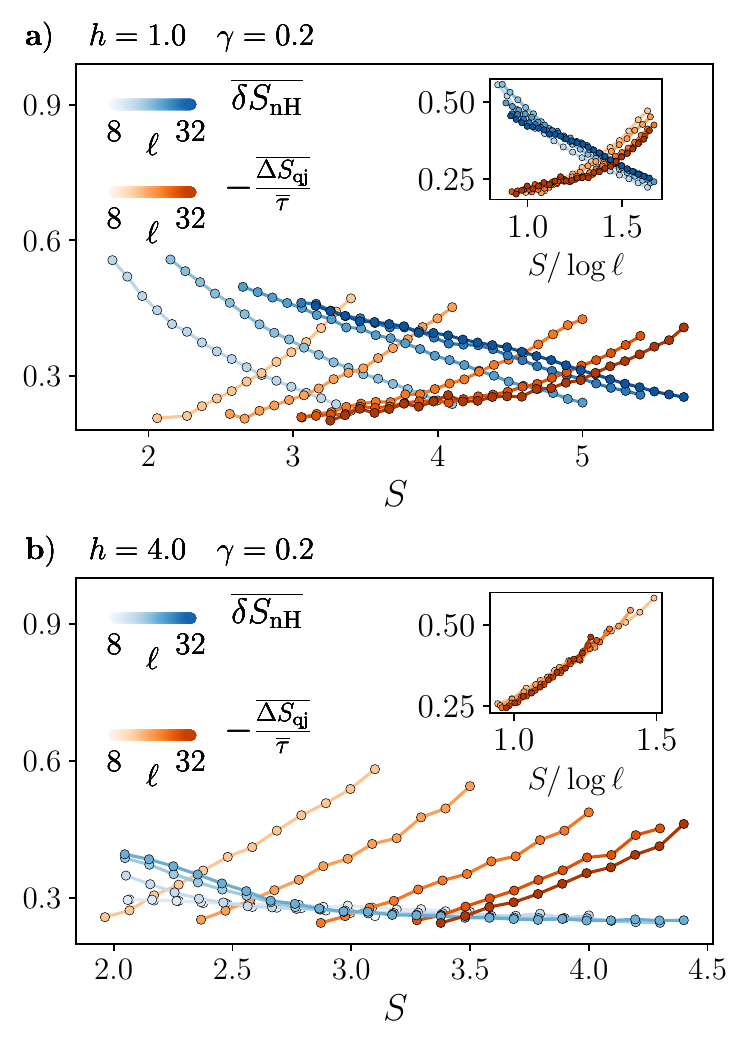}
    \caption{\label{fig:Ising_strange_part} Steady-State Entanglement Balance for the Monitored Ising chain with large transverse field $h > 1$: Average entanglement gain $\overline{\delta S_\mathrm{nH}}$ and loss $\overline{\delta S_\mathrm{qj}}/\overline{\tau}$, as a function of the average entanglement entropy S and different subsystem sizes $\ell$. Panel (a): the gain contribution due to the non-Hermitian dynamics acquires a non-trivial $\ell$ dependence scaling logarithmically (see inset). On the other hand, the QJ loss term scales also logarithmically which leads to a sub-volume steady-state entanglement absent in the no-click limit. For even larger $h$ (Panel (b), $h = 4$), the non-Hermitian gain term becomes $\ell$ independent, as in the no-click limit. At the same time, the QJ retains a non-trivial scaling (inset), leading to the observed logarithmic scaling.  }
\end{figure}

In this Appendix we discuss a particular feature of the monitored Ising chain phase diagram. Indeed, when considering the region of weak monitoring and large transverse field, deviations from the no-click limit were reported \cite{paviglianiti2023multipartite, turkeshi2021measurementinducedentanglement}. In particular, we fix here $\gamma = 0.2$ and scan the transverse field for increasing values of $h > 1$, where the no-click evolution predicts an area-law for the entanglement entropy but the QJ dynamics still showed a logarithmic scaling. In Fig.~\ref{fig:Ising_strange_part} (a-b), we repeat our analysis of average entanglement gain and loss for different subsystem sizes $\ell$. For $h = 1.0$, we again observe strong $\ell$ dependence in the QJ contribution and the non-Hermitian one. Our steady-state condition Eq.~(\ref{eq:steadystate}) predicts a logarithmic scaling for the entanglement entropy in  agreement with the QJ simulation. Moreover like in the weak transverse field limit by breaking down the entanglement content into gain and loss, we notice that this log-phase is due to a combined effect of logarithmic scale of both : the quantum jumps loss and the renormalized non-Hermitian gain. Upon increasing the transverse field to $h = 4$, we see that the entanglement gain due to the non- Hermitian dynamics becomes essentially $\ell$ independent, in agreement with the no-click limit. On the other hand, the loss due to QJs scales logarithmically with $\ell$ (see inset), which results in a steady-state entanglement showing a log-scaling. This logarithmic phase, as for the weak monitoring of the SSH model, is solely due to the QJs. 

This analysis therefore highlights two mechanisms at play behind the deviations from the no-click limit observed in the monitored Ising chain phase diagram for $h > 1$. At moderate fields there is a non-trivial renormalization of the non-Hermitian dynamics due to QJs, that leads to a log-scaling of its average gain contribution, as in the weak-monitoring regime of the model. At larger fields instead this renormalisation is washed away, the non-Hermitian dynamics behaves as in the no-click limit and the logarithmic scaling of the entropy arises only from the jumps. While our numerical results does not allow to conclude about the large system size limit, the above analysis suggests that, in absence of a non-trivial non-Hermitian dynamics, the large field log-phase might eventually saturate at large system sizes into an area law, similarly to the weak monitoring phase of the SSH model.

\section{Free Fermion Techniques with Quantum Jumps}\label{sec:methods}

This Section provides details on solving the stochastic Schrödinger dynamics using free-fermion techniques. Since the Hamiltonian we consider for both models is quadratic, and the monitoring process preserves Gaussianity, the state $\vert\psi \rangle$ is entirely determined by the 2-point correlation matrix $C$ thanks to Wick's theorem. We can either store the state in this correlation matrix or use another representation based on pure states.

\subsection{Correlation matrix}
In general, to solve the Schrödinger equation in a free fermions problem, we use Wick's theorem to rewrite the equation in terms of the correlation matrix
\begin{equation}
    C(t) = \begin{pmatrix}
        G(t) & F(t) \\
        F(t)^\dagger & 1- G(t)^T
    \end{pmatrix}
\end{equation}
where $ C_{m,n}(t) = \langle \Psi(t) | \mathbf{c}_m^\dagger  \mathbf{c}_n | \Psi(t)  \rangle$ is a matrix of size $2L \times 2L $ and $\mathbf{c}^T = ( c_1~c_2~ ... ~c_L~ c_1^\dagger~ c_2^\dagger ~...~ c_L^\dagger ) $. We get for the deterministic (non-Hermitian) part when measuring the local density, the equation~\cite{turkeshi2022entanglementtransitionsfrom} 
\begin{equation}
\label{eq:deterministic_evol_corr}
    \frac{dC(t)}{dt} = 2i[\mathbb{H},C(t)] + \gamma C \Lambda C - \gamma \Lambda_+  C \Lambda_+ + \gamma \Lambda_-  C \Lambda_-
\end{equation}
where $\Lambda_+ = \mathbb{1}_{L} \otimes (\sigma_z + 1)/2$, $\Lambda_- = \mathbb{1}_{2L} - \Lambda_+$  and $\Lambda = \mathbb{1}_{L} \otimes \sigma_z$, $\sigma_z=\mbox{diag}(1,-1)$ is the Pauli matrix and  $\mathbb{H}$ is defined such that $H = \mathbf{c}^\dagger\mathbb{H}\mathbf{c}$.

The von Neumann entanglement of a subsystem of size $\ell$ can be extracted through the Majorana fermions correlation matrix \cite{turkeshi2022entanglementtransitionsfrom,Vidal2003}.   Effectively we get it by diagonalising the following matrix 
\begin{equation}
    \mathcal{A}(t) = \begin{pmatrix}
        \mathcal{G}_\ell^I + \mathcal{F}_\ell^I & \mathcal{G}_\ell^R - \mathcal{F}_\ell^R - \mathbb{1} \\  \mathbb{1}- \mathcal{G}_\ell^R - \mathcal{F}_\ell^R & \mathcal{G}_\ell^I + \mathcal{F}_\ell^I
    \end{pmatrix}
\end{equation}
with $2G_{\ell \times \ell } = \mathcal{G}_\ell^R + i\mathcal{G}_\ell^I $ and $2F_{\ell \times \ell } = \mathcal{F}_\ell^R + i\mathcal{F}_\ell^I $, which has an imaginary spectrum. The entanglement is given by the formula 
\begin{equation}
    \label{eq:entanglement_corr}
    S_A(t) = -\sum_{j=1}^{\ell} \nu_j(t) \ln\left[\nu_j(t)\right] +(1-\nu_j(t))\ln\left[1-\nu_j(t)\right]
\end{equation}
with $\nu_j(t) = (1-\lambda_j(t))/2$ where $\lambda_j(t)$ are the $\ell$ eigenvalues of $\mathcal{A}(t)$ with a positive imaginary part. 

Jumps such that $L_j \propto n_j$ (i.e., we measure the local density) can be implemented at the level of the correlation matrix. For a jump on site $l$ we have
\begin{align}
\label{eq:jump_corr_density}
    \nonumber G_{m,n}^J &= 
\left\{\begin{array}{l}
         1, \qquad \qquad  \mathrm{if} \quad  n = m = l  \\
         0, \qquad \qquad 
             \mathrm{if} \quad m = l,  n \neq l  \quad 
             \mathrm{or} \quad
             m \neq l,  n = l
         \\
    G_{m,n} - \frac{G_{m,l}G_{l,n} + F_{m,l}(F^\dagger)_{n,l}}{G_{l,l}}, \quad \mathrm{otherwise}
    \end{array} \right.
     \\ dF_{m,n} &= \frac{G_{n,l}F_{l,m} - G_{m,l}F_{l,n}}{G_{l,l}}.
\end{align}
To avoid numerical instabilities, it is advisable to explicitly write the zeros in the matrix update.

In this work we implement the sampling of the waiting-time distribution, therefore the goal is to find the time at which the jump happens. Thus we need to be able to evaluate the decay of the norm during the non-Hermitian evolution and find the time for the next jump. To do so, we solve numerically the equation 
\begin{equation}
\frac{d}{dt}\langle \tilde{\Psi}(t) | \tilde{\Psi}(t) \rangle = - \gamma \langle \tilde{\Psi}(t) | \tilde{\Psi}(t) \rangle \sum_j G_{j,j}(t)
\end{equation}
where $G_{j,j}(t)$ is obtain by solving the correlation matrix equation of motion in parallel and $| \tilde{\Psi}(t) \rangle$ is the unormalized wavefunction. Then we can find by bisection for each jump the time at which it happens~\cite{daley2014quantum}. 

A simplification arises when the model presents a $U(1)$ symmetry because the conservation of the number of particles reduces by a factor of 2 the size of the correlation matrix to consider. Indeed, we have to consider only the matrix 
$ G_{m,n}(t) = \langle \psi(t) | \mathbf{c}_m^\dagger  \mathbf{c}_n | \psi(t)  \rangle$  with $\mathbf{c}^T = ( c_1~c_2~ ... ~c_L ) $. Here we give the equation that can be used for the SSH model and for convenience we will use a basis respecting the site $A$ and $B$ of the model, thus we consider the basis where $c_{2k-1} = c_{A,k}$ and $c_{2k} = c_{B,k}$ for $1 \leq k \leq L/2 $. 
 In that case we get for the deterministic part (when measuring $n_{A,i}$ and $1-n_{B,i}$), the equation~\cite{turkeshi2022entanglementtransitionsfrom} 
\begin{equation}
\label{eq:deterministic_evol_corr_ssh}
    \frac{dG(t)}{dt} = 2i[\mathbb{H},G(t)] + \gamma G \Lambda G - \gamma \Lambda_A  C \Lambda_A + \gamma \Lambda_B  G \Lambda_B
\end{equation}
where $\Lambda_A = \mathbb{1}_{L/2} \otimes (\sigma_z + 1)/2$, $\Lambda_B = \mathbb{1}_{L} - \Lambda_A$  and $\Lambda = \mathbb{1}_{L/2} \otimes \sigma_z$ and $\mathbb{H}$ defined such that $H = \mathbf{c}^\dagger\mathbb{H}\mathbf{c}$.
Then the other methods to do the jumps apply directly.

\subsection{Wavefunction representation}

When the system conserves the number of particles, as in the monitored SSH case, not only the correlation matrix approach described above simplifies but one can also use a different approach based on representing
the wavefunction as 
\begin{equation}
    |\psi(t) \rangle = P(t) c_{i_1}^\dagger ... c_{i_N}^\dagger |0 \rangle = \prod_{k=1}^N \sum_{m=1}^L U_{m,k}(t) c_m^\dagger |0 \rangle
\end{equation}
where we consider $N$ particles. 
Then the $U(t)$ matrix can be updated through $\tilde{U}(t+dt) = e^{-i\mathbb{H}_\mathrm{eff} dt }U(t) $ \cite{cao2019entanglementina}. Nonetheless, when $H_\mathrm{eff}$ is non Hermitian, the state should be normalised. This is done by performing a QR decomposition on $\tilde{U}(t+dt)  $, which normalize the state without modifying it. The correlation matrix $G(t)$ is then obtained through 
\begin{equation}
    \label{eq:QR_dec}
    G(t) = U(t)^* U(t)^T \qquad \mathrm{with} \qquad \tilde{U}(t) = U(t) R(t).
\end{equation}
We can implement the jumps as previously at the level of this correlation matrix like in Eq.~\ref{eq:jump_corr_density}. In the case of the SSH model we are also considering jumps of the form $L_j \propto 1-n_j$ in that case the update of $G(t)$ is such that 
\begin{equation}
     \nonumber G_{m,n}^J = 
\left\{\begin{array}{l}
         0, \qquad \qquad  \mathrm{if} \quad  n = m = l  \\
         0, \qquad \qquad 
             \mathrm{if} \quad m = l,  n \neq l  \quad 
             \mathrm{or} \quad
             m \neq l,  n = l
         \\
    G_{m,n} + \frac{G_{m,l}G_{l,n} }{1-G_{l,l}}, \quad \mathrm{otherwise}.
    \end{array} \right.
\end{equation}

The jumps preserve the gaussianity and $U(1)$ symmetry, thus the new correlation matrix is of the form $G(t) = U(t)^*U(t)^T$ thus is semi-definite positive hermitian, and the $k^{th}$ vector column of $U(t)$ called $U_k$ satisfy 
\begin{equation}
    G(t)U_k^* = U_k^* \qquad 1 \leq k \leq N.
\end{equation}
Then the SVD decomposition $G(t) = U_{\mathrm{SVD}} D U_{\mathrm{SVD}}^\dagger$ gives directly these eigenvectors, and the restriction of the first $N$ columns with eigenvalue 1 (i.e. the $N$ particles) gives
\begin{equation}
    U(t) = (U_{\mathrm{SVD}})_{L \times N}.
\end{equation}
Since the columns of $U_{\mathrm{SVD}}$ are orthogonal the state is well normalized and correctly defined. Like in the previous subsection we need the decay of the norm to evaluate the time at which the jump is happening. In that case, the norm can be directly obtained from the QR decomposition, indeed 
\begin{equation}
    \label{eq:normU1}
    \langle \tilde{\Psi}(t) | \tilde{\Psi}(t) \rangle = \prod_{j=1}^N R(t)_{j,j}R(t)_{j,j}^*,
\end{equation}
where $R(t)$ is the matrix from Eq.~\ref{eq:QR_dec}. 

This kind of representation of the wavefunction can be extended to the general case without particle number conservation. In that case we have that ~\cite{bravyi2005lagrangian}

\begin{equation}
    | \psi(t) \rangle = \mathcal{N} \mathrm{exp} \left( -\frac{1}{2}\sum_{i,j} \left[(U(t)^\dagger)^{-1} V(t)^\dagger\right]_{i,j} c_i^\dagger c_j^\dagger \right) | 0 \rangle
\end{equation}
where $\mathcal{N}$ enforces the normalization, and the evolution is given by imposing that $\gamma_k(t) |\psi(t) \rangle = 0 $ with $\gamma_k(t) = \sum_j V_{j,k}^*(t) c_j^\dagger + U_{j,k}^*(t) c_j$. And we now evolve the 2 matrices $U(t)$ and $V(t)$ through 
\begin{equation} 
 \begin{pmatrix}
   U(t)  \\ V(t)  
 \end{pmatrix} =  e^{-2i \mathbb{H}_\mathrm{eff} t  } \begin{pmatrix}
   U(0)  \\ V(0)  
 \end{pmatrix} .
\end{equation}

Since the evolution is non Hermitian the state is here again not normalized, but this can be enforced through the factor $\mathcal{N}$. As significantly, to guarantee that $\gamma_k(t)$ is a well defined fermionic operator we have to impose 
\begin{equation}
\label{eq:fermionic_cond}
    \begin{cases}
        U(t)^\dagger U(t) + V(t)^\dagger V(t)  = \mathbb{1} \\ 
        V^T(t)U(t) + U^T(t)V(t) = 0 
    \end{cases}.
\end{equation}
This can be done by performing a QR decomposition on  
\begin{equation}
    \mathbb{U}(t) = \begin{pmatrix}
        U(t) & V(t)^* \\ 
        V(t) & U(t)^*
    \end{pmatrix}
\end{equation} as proved in Ref.~\cite{fava2023nonlinearsigmamodels}.To implement the jump within this framework, we can compute the correlation matrix $C(t)$ which is given by 
\begin{equation}
\label{eq:corr_mat_UV}
     C(t) = \begin{pmatrix}
        U(t)U(t)^\dagger & U(t)V^\dagger(t) \\ 
    V(t)U^\dagger(t) & V(t)V^\dagger(t)
    \end{pmatrix}
\end{equation}

As in the $U(1)$ case, we then need to retrieve the state matrix, which is done using a SVD decomposition of the correlation matrix.

\bibliography{jumps_bib}

\begin{thebibliography}{98}%
\makeatletter
\providecommand \@ifxundefined [1]{%
 \@ifx{#1\undefined}
}%
\providecommand \@ifnum [1]{%
 \ifnum #1\expandafter \@firstoftwo
 \else \expandafter \@secondoftwo
 \fi
}%
\providecommand \@ifx [1]{%
 \ifx #1\expandafter \@firstoftwo
 \else \expandafter \@secondoftwo
 \fi
}%
\providecommand \natexlab [1]{#1}%
\providecommand \enquote  [1]{``#1''}%
\providecommand \bibnamefont  [1]{#1}%
\providecommand \bibfnamefont [1]{#1}%
\providecommand \citenamefont [1]{#1}%
\providecommand \href@noop [0]{\@secondoftwo}%
\providecommand \href [0]{\begingroup \@sanitize@url \@href}%
\providecommand \@href[1]{\@@startlink{#1}\@@href}%
\providecommand \@@href[1]{\endgroup#1\@@endlink}%
\providecommand \@sanitize@url [0]{\catcode `\\12\catcode `\$12\catcode
  `\&12\catcode `\#12\catcode `\^12\catcode `\_12\catcode `\%12\relax}%
\providecommand \@@startlink[1]{}%
\providecommand \@@endlink[0]{}%
\providecommand \url  [0]{\begingroup\@sanitize@url \@url }%
\providecommand \@url [1]{\endgroup\@href {#1}{\urlprefix }}%
\providecommand \urlprefix  [0]{URL }%
\providecommand \Eprint [0]{\href }%
\providecommand \doibase [0]{https://doi.org/}%
\providecommand \selectlanguage [0]{\@gobble}%
\providecommand \bibinfo  [0]{\@secondoftwo}%
\providecommand \bibfield  [0]{\@secondoftwo}%
\providecommand \translation [1]{[#1]}%
\providecommand \BibitemOpen [0]{}%
\providecommand \bibitemStop [0]{}%
\providecommand \bibitemNoStop [0]{.\EOS\space}%
\providecommand \EOS [0]{\spacefactor3000\relax}%
\providecommand \BibitemShut  [1]{\csname bibitem#1\endcsname}%
\let\auto@bib@innerbib\@empty
\bibitem [{\citenamefont {Calabrese}\ and\ \citenamefont
  {Cardy}(2005)}]{Calabrese_2005}%
  \BibitemOpen
  \bibfield  {author} {\bibinfo {author} {\bibfnamefont {P.}~\bibnamefont
  {Calabrese}}\ and\ \bibinfo {author} {\bibfnamefont {J.}~\bibnamefont
  {Cardy}},\ }\href {https://doi.org/10.1088/1742-5468/2005/04/P04010}
  {\bibfield  {journal} {\bibinfo  {journal} {Journal of Statistical Mechanics:
  Theory and Experiment}\ }\textbf {\bibinfo {volume} {2005}},\ \bibinfo
  {pages} {P04010} (\bibinfo {year} {2005})}\BibitemShut {NoStop}%
\bibitem [{\citenamefont {Liu}\ and\ \citenamefont
  {Suh}(2014)}]{liu2014entanglement}%
  \BibitemOpen
  \bibfield  {author} {\bibinfo {author} {\bibfnamefont {H.}~\bibnamefont
  {Liu}}\ and\ \bibinfo {author} {\bibfnamefont {S.~J.}\ \bibnamefont {Suh}},\
  }\href {https://doi.org/10.1103/PhysRevLett.112.011601} {\bibfield  {journal}
  {\bibinfo  {journal} {Phys. Rev. Lett.}\ }\textbf {\bibinfo {volume} {112}},\
  \bibinfo {pages} {011601} (\bibinfo {year} {2014})}\BibitemShut {NoStop}%
\bibitem [{\citenamefont {Nahum}\ \emph {et~al.}(2017)\citenamefont {Nahum},
  \citenamefont {Ruhman}, \citenamefont {Vijay},\ and\ \citenamefont
  {Haah}}]{nahum2017quantum}%
  \BibitemOpen
  \bibfield  {author} {\bibinfo {author} {\bibfnamefont {A.}~\bibnamefont
  {Nahum}}, \bibinfo {author} {\bibfnamefont {J.}~\bibnamefont {Ruhman}},
  \bibinfo {author} {\bibfnamefont {S.}~\bibnamefont {Vijay}},\ and\ \bibinfo
  {author} {\bibfnamefont {J.}~\bibnamefont {Haah}},\ }\href
  {https://doi.org/10.1103/PhysRevX.7.031016} {\bibfield  {journal} {\bibinfo
  {journal} {Phys. Rev. X}\ }\textbf {\bibinfo {volume} {7}},\ \bibinfo {pages}
  {031016} (\bibinfo {year} {2017})}\BibitemShut {NoStop}%
\bibitem [{\citenamefont {Chiara}\ \emph {et~al.}(2006)\citenamefont {Chiara},
  \citenamefont {Montangero}, \citenamefont {Calabrese},\ and\ \citenamefont
  {Fazio}}]{DeChiara_2006}%
  \BibitemOpen
  \bibfield  {author} {\bibinfo {author} {\bibfnamefont {G.~D.}\ \bibnamefont
  {Chiara}}, \bibinfo {author} {\bibfnamefont {S.}~\bibnamefont {Montangero}},
  \bibinfo {author} {\bibfnamefont {P.}~\bibnamefont {Calabrese}},\ and\
  \bibinfo {author} {\bibfnamefont {R.}~\bibnamefont {Fazio}},\ }\href
  {https://doi.org/10.1088/1742-5468/2006/03/P03001} {\bibfield  {journal}
  {\bibinfo  {journal} {Journal of Statistical Mechanics: Theory and
  Experiment}\ }\textbf {\bibinfo {volume} {2006}},\ \bibinfo {pages} {P03001}
  (\bibinfo {year} {2006})}\BibitemShut {NoStop}%
\bibitem [{\citenamefont {Kim}\ and\ \citenamefont
  {Huse}(2013)}]{kim2013ballistic}%
  \BibitemOpen
  \bibfield  {author} {\bibinfo {author} {\bibfnamefont {H.}~\bibnamefont
  {Kim}}\ and\ \bibinfo {author} {\bibfnamefont {D.~A.}\ \bibnamefont {Huse}},\
  }\href {https://doi.org/10.1103/PhysRevLett.111.127205} {\bibfield  {journal}
  {\bibinfo  {journal} {Phys. Rev. Lett.}\ }\textbf {\bibinfo {volume} {111}},\
  \bibinfo {pages} {127205} (\bibinfo {year} {2013})}\BibitemShut {NoStop}%
\bibitem [{\citenamefont {Bardarson}\ \emph {et~al.}(2012)\citenamefont
  {Bardarson}, \citenamefont {Pollmann},\ and\ \citenamefont
  {Moore}}]{bardarson2012unbounded}%
  \BibitemOpen
  \bibfield  {author} {\bibinfo {author} {\bibfnamefont {J.~H.}\ \bibnamefont
  {Bardarson}}, \bibinfo {author} {\bibfnamefont {F.}~\bibnamefont
  {Pollmann}},\ and\ \bibinfo {author} {\bibfnamefont {J.~E.}\ \bibnamefont
  {Moore}},\ }\href {https://doi.org/10.1103/PhysRevLett.109.017202} {\bibfield
   {journal} {\bibinfo  {journal} {Phys. Rev. Lett.}\ }\textbf {\bibinfo
  {volume} {109}},\ \bibinfo {pages} {017202} (\bibinfo {year}
  {2012})}\BibitemShut {NoStop}%
\bibitem [{\citenamefont {Fisher}\ \emph {et~al.}(2023)\citenamefont {Fisher},
  \citenamefont {Khemani}, \citenamefont {Nahum},\ and\ \citenamefont
  {Vijay}}]{fisher2023randomquantumcircuits}%
  \BibitemOpen
  \bibfield  {author} {\bibinfo {author} {\bibfnamefont {M.~P.}\ \bibnamefont
  {Fisher}}, \bibinfo {author} {\bibfnamefont {V.}~\bibnamefont {Khemani}},
  \bibinfo {author} {\bibfnamefont {A.}~\bibnamefont {Nahum}},\ and\ \bibinfo
  {author} {\bibfnamefont {S.}~\bibnamefont {Vijay}},\ }\href
  {https://doi.org/10.1146/annurev-conmatphys-031720-030658} {\bibfield
  {journal} {\bibinfo  {journal} {Annu. Rev. Condens. Matter Phys.}\ }\textbf
  {\bibinfo {volume} {14}},\ \bibinfo {pages} {335} (\bibinfo {year}
  {2023})}\BibitemShut {NoStop}%
\bibitem [{\citenamefont {Potter}\ and\ \citenamefont
  {Vasseur}(2022)}]{potter2022quantumsciencesandtechnology}%
  \BibitemOpen
  \bibfield  {author} {\bibinfo {author} {\bibfnamefont {A.~C.}\ \bibnamefont
  {Potter}}\ and\ \bibinfo {author} {\bibfnamefont {R.}~\bibnamefont
  {Vasseur}},\ }\href@noop {} {\emph {\bibinfo {title} {Quantum Sciences and
  Technology}}}\ (\bibinfo  {publisher} {Springer, Cham},\ \bibinfo {year}
  {2022})\ p.\ \bibinfo {pages} {211}\BibitemShut {NoStop}%
\bibitem [{\citenamefont {Lunt}\ \emph {et~al.}(2022)\citenamefont {Lunt},
  \citenamefont {Richter},\ and\ \citenamefont
  {Pal}}]{lunt2022quantumsciencesandtechnology}%
  \BibitemOpen
  \bibfield  {author} {\bibinfo {author} {\bibfnamefont {O.}~\bibnamefont
  {Lunt}}, \bibinfo {author} {\bibfnamefont {J.}~\bibnamefont {Richter}},\ and\
  \bibinfo {author} {\bibfnamefont {A.}~\bibnamefont {Pal}},\ }\href@noop {}
  {\emph {\bibinfo {title} {Quantum Sciences and Technology}}}\ (\bibinfo
  {publisher} {Springer, Cham},\ \bibinfo {year} {2022})\ p.\ \bibinfo {pages}
  {251}\BibitemShut {NoStop}%
\bibitem [{\citenamefont {Gullans}\ and\ \citenamefont
  {Huse}(2020)}]{gullans2020dynamicalpurificationphase}%
  \BibitemOpen
  \bibfield  {author} {\bibinfo {author} {\bibfnamefont {M.~J.}\ \bibnamefont
  {Gullans}}\ and\ \bibinfo {author} {\bibfnamefont {D.~A.}\ \bibnamefont
  {Huse}},\ }\href {https://doi.org/10.1103/PhysRevX.10.041020} {\bibfield
  {journal} {\bibinfo  {journal} {Phys. Rev. X}\ }\textbf {\bibinfo {volume}
  {10}},\ \bibinfo {pages} {041020} (\bibinfo {year} {2020})}\BibitemShut
  {NoStop}%
\bibitem [{\citenamefont {Noel}\ \emph {et~al.}(2022)\citenamefont {Noel},
  \citenamefont {Niroula}, \citenamefont {Zhu}, \citenamefont {Risinger},
  \citenamefont {Egan}, \citenamefont {Biswas}, \citenamefont {Cetina},
  \citenamefont {Gorshkov}, \citenamefont {Gullans}, \citenamefont {Huse},\
  and\ \citenamefont {Monroe}}]{noel2021measurementinducedquantum}%
  \BibitemOpen
  \bibfield  {author} {\bibinfo {author} {\bibfnamefont {C.}~\bibnamefont
  {Noel}}, \bibinfo {author} {\bibfnamefont {P.}~\bibnamefont {Niroula}},
  \bibinfo {author} {\bibfnamefont {D.}~\bibnamefont {Zhu}}, \bibinfo {author}
  {\bibfnamefont {A.}~\bibnamefont {Risinger}}, \bibinfo {author}
  {\bibfnamefont {L.}~\bibnamefont {Egan}}, \bibinfo {author} {\bibfnamefont
  {D.}~\bibnamefont {Biswas}}, \bibinfo {author} {\bibfnamefont
  {M.}~\bibnamefont {Cetina}}, \bibinfo {author} {\bibfnamefont {A.~V.}\
  \bibnamefont {Gorshkov}}, \bibinfo {author} {\bibfnamefont {M.~J.}\
  \bibnamefont {Gullans}}, \bibinfo {author} {\bibfnamefont {D.~A.}\
  \bibnamefont {Huse}},\ and\ \bibinfo {author} {\bibfnamefont
  {C.}~\bibnamefont {Monroe}},\ }\href
  {https://www.nature.com/articles/s41567-022-01619-7} {\bibfield  {journal}
  {\bibinfo  {journal} {Nature Phys.}\ }\textbf {\bibinfo {volume} {18}},\
  \bibinfo {pages} {760} (\bibinfo {year} {2022})}\BibitemShut {NoStop}%
\bibitem [{\citenamefont {Koh}\ \emph {et~al.}(2023)\citenamefont {Koh},
  \citenamefont {Sun}, \citenamefont {Motta},\ and\ \citenamefont
  {Minnich}}]{koh2022experimentalrealizationof}%
  \BibitemOpen
  \bibfield  {author} {\bibinfo {author} {\bibfnamefont {J.~M.}\ \bibnamefont
  {Koh}}, \bibinfo {author} {\bibfnamefont {S.-N.}\ \bibnamefont {Sun}},
  \bibinfo {author} {\bibfnamefont {M.}~\bibnamefont {Motta}},\ and\ \bibinfo
  {author} {\bibfnamefont {A.~J.}\ \bibnamefont {Minnich}},\ }\href
  {https://doi.org/10.1038/s41567-023-02076-6} {\bibfield  {journal} {\bibinfo
  {journal} {Nature Phys.}\ }\textbf {\bibinfo {volume} {19}},\ \bibinfo
  {pages} {1314} (\bibinfo {year} {2023})}\BibitemShut {NoStop}%
\bibitem [{\citenamefont {{Google AI and
  Collaborators}}(2023)}]{hoke2023quantuminformationphases}%
  \BibitemOpen
  \bibfield  {author} {\bibinfo {author} {\bibnamefont {{Google AI and
  Collaborators}}},\ }\href {https://doi.org/10.1038/s41586-023-06505-7}
  {\bibfield  {journal} {\bibinfo  {journal} {Nature}\ }\textbf {\bibinfo
  {volume} {622}},\ \bibinfo {pages} {481–486} (\bibinfo {year}
  {2023})}\BibitemShut {NoStop}%
\bibitem [{\citenamefont {Li}\ \emph {et~al.}(2018)\citenamefont {Li},
  \citenamefont {Chen},\ and\ \citenamefont
  {Fisher}}]{li2018quantumzenoeffect}%
  \BibitemOpen
  \bibfield  {author} {\bibinfo {author} {\bibfnamefont {Y.}~\bibnamefont
  {Li}}, \bibinfo {author} {\bibfnamefont {X.}~\bibnamefont {Chen}},\ and\
  \bibinfo {author} {\bibfnamefont {M.~P.~A.}\ \bibnamefont {Fisher}},\ }\href
  {https://doi.org/10.1103/PhysRevB.98.205136} {\bibfield  {journal} {\bibinfo
  {journal} {Phys. Rev. B}\ }\textbf {\bibinfo {volume} {98}},\ \bibinfo
  {pages} {205136} (\bibinfo {year} {2018})}\BibitemShut {NoStop}%
\bibitem [{\citenamefont {Li}\ \emph {et~al.}(2019)\citenamefont {Li},
  \citenamefont {Chen},\ and\ \citenamefont
  {Fisher}}]{li2019measurementdrivenentanglement}%
  \BibitemOpen
  \bibfield  {author} {\bibinfo {author} {\bibfnamefont {Y.}~\bibnamefont
  {Li}}, \bibinfo {author} {\bibfnamefont {X.}~\bibnamefont {Chen}},\ and\
  \bibinfo {author} {\bibfnamefont {M.~P.~A.}\ \bibnamefont {Fisher}},\ }\href
  {https://doi.org/10.1103/PhysRevB.100.134306} {\bibfield  {journal} {\bibinfo
   {journal} {Phys. Rev. B}\ }\textbf {\bibinfo {volume} {100}},\ \bibinfo
  {pages} {134306} (\bibinfo {year} {2019})}\BibitemShut {NoStop}%
\bibitem [{\citenamefont {Skinner}\ \emph {et~al.}(2019)\citenamefont
  {Skinner}, \citenamefont {Ruhman},\ and\ \citenamefont
  {Nahum}}]{skinner2019measurementinducedphase}%
  \BibitemOpen
  \bibfield  {author} {\bibinfo {author} {\bibfnamefont {B.}~\bibnamefont
  {Skinner}}, \bibinfo {author} {\bibfnamefont {J.}~\bibnamefont {Ruhman}},\
  and\ \bibinfo {author} {\bibfnamefont {A.}~\bibnamefont {Nahum}},\ }\href
  {https://doi.org/10.1103/PhysRevX.9.031009} {\bibfield  {journal} {\bibinfo
  {journal} {Phys. Rev. X}\ }\textbf {\bibinfo {volume} {9}},\ \bibinfo {pages}
  {031009} (\bibinfo {year} {2019})}\BibitemShut {NoStop}%
\bibitem [{\citenamefont {Szyniszewski}\ \emph {et~al.}(2019)\citenamefont
  {Szyniszewski}, \citenamefont {Romito},\ and\ \citenamefont
  {Schomerus}}]{szyniszewski2019entanglementtransitionfrom}%
  \BibitemOpen
  \bibfield  {author} {\bibinfo {author} {\bibfnamefont {M.}~\bibnamefont
  {Szyniszewski}}, \bibinfo {author} {\bibfnamefont {A.}~\bibnamefont
  {Romito}},\ and\ \bibinfo {author} {\bibfnamefont {H.}~\bibnamefont
  {Schomerus}},\ }\href {https://doi.org/10.1103/PhysRevB.100.064204}
  {\bibfield  {journal} {\bibinfo  {journal} {Phys. Rev. B}\ }\textbf {\bibinfo
  {volume} {100}},\ \bibinfo {pages} {064204} (\bibinfo {year}
  {2019})}\BibitemShut {NoStop}%
\bibitem [{\citenamefont {Jian}\ \emph {et~al.}(2020)\citenamefont {Jian},
  \citenamefont {You}, \citenamefont {Vasseur},\ and\ \citenamefont
  {Ludwig}}]{jian2020measurementinducedcriticality}%
  \BibitemOpen
  \bibfield  {author} {\bibinfo {author} {\bibfnamefont {C.-M.}\ \bibnamefont
  {Jian}}, \bibinfo {author} {\bibfnamefont {Y.-Z.}\ \bibnamefont {You}},
  \bibinfo {author} {\bibfnamefont {R.}~\bibnamefont {Vasseur}},\ and\ \bibinfo
  {author} {\bibfnamefont {A.~W.~W.}\ \bibnamefont {Ludwig}},\ }\href
  {https://doi.org/10.1103/PhysRevB.101.104302} {\bibfield  {journal} {\bibinfo
   {journal} {Phys. Rev. B}\ }\textbf {\bibinfo {volume} {101}},\ \bibinfo
  {pages} {104302} (\bibinfo {year} {2020})}\BibitemShut {NoStop}%
\bibitem [{\citenamefont {Choi}\ \emph {et~al.}(2020)\citenamefont {Choi},
  \citenamefont {Bao}, \citenamefont {Qi},\ and\ \citenamefont
  {Altman}}]{choi2020quantumerrorcorrection}%
  \BibitemOpen
  \bibfield  {author} {\bibinfo {author} {\bibfnamefont {S.}~\bibnamefont
  {Choi}}, \bibinfo {author} {\bibfnamefont {Y.}~\bibnamefont {Bao}}, \bibinfo
  {author} {\bibfnamefont {X.-L.}\ \bibnamefont {Qi}},\ and\ \bibinfo {author}
  {\bibfnamefont {E.}~\bibnamefont {Altman}},\ }\href
  {https://doi.org/10.1103/PhysRevLett.125.030505} {\bibfield  {journal}
  {\bibinfo  {journal} {Phys. Rev. Lett.}\ }\textbf {\bibinfo {volume} {125}},\
  \bibinfo {pages} {030505} (\bibinfo {year} {2020})}\BibitemShut {NoStop}%
\bibitem [{\citenamefont {Zabalo}\ \emph {et~al.}(2022)\citenamefont {Zabalo},
  \citenamefont {Gullans}, \citenamefont {Wilson}, \citenamefont {Vasseur},
  \citenamefont {Ludwig}, \citenamefont {Gopalakrishnan}, \citenamefont
  {Huse},\ and\ \citenamefont {Pixley}}]{zabalo2022operatorscalingdimensions}%
  \BibitemOpen
  \bibfield  {author} {\bibinfo {author} {\bibfnamefont {A.}~\bibnamefont
  {Zabalo}}, \bibinfo {author} {\bibfnamefont {M.~J.}\ \bibnamefont {Gullans}},
  \bibinfo {author} {\bibfnamefont {J.~H.}\ \bibnamefont {Wilson}}, \bibinfo
  {author} {\bibfnamefont {R.}~\bibnamefont {Vasseur}}, \bibinfo {author}
  {\bibfnamefont {A.~W.~W.}\ \bibnamefont {Ludwig}}, \bibinfo {author}
  {\bibfnamefont {S.}~\bibnamefont {Gopalakrishnan}}, \bibinfo {author}
  {\bibfnamefont {D.~A.}\ \bibnamefont {Huse}},\ and\ \bibinfo {author}
  {\bibfnamefont {J.~H.}\ \bibnamefont {Pixley}},\ }\href
  {https://doi.org/10.1103/PhysRevLett.128.050602} {\bibfield  {journal}
  {\bibinfo  {journal} {Phys. Rev. Lett.}\ }\textbf {\bibinfo {volume} {128}},\
  \bibinfo {pages} {050602} (\bibinfo {year} {2022})}\BibitemShut {NoStop}%
\bibitem [{\citenamefont {Sierant}\ and\ \citenamefont
  {Turkeshi}(2022)}]{sierant2022universalbehaviorbeyond}%
  \BibitemOpen
  \bibfield  {author} {\bibinfo {author} {\bibfnamefont {P.}~\bibnamefont
  {Sierant}}\ and\ \bibinfo {author} {\bibfnamefont {X.}~\bibnamefont
  {Turkeshi}},\ }\href {https://doi.org/10.1103/PhysRevLett.128.130605}
  {\bibfield  {journal} {\bibinfo  {journal} {Phys. Rev. Lett.}\ }\textbf
  {\bibinfo {volume} {128}},\ \bibinfo {pages} {130605} (\bibinfo {year}
  {2022})}\BibitemShut {NoStop}%
\bibitem [{\citenamefont {Sierant}\ \emph {et~al.}(2022)\citenamefont
  {Sierant}, \citenamefont {Schir\`o}, \citenamefont {Lewenstein},\ and\
  \citenamefont {Turkeshi}}]{sierant2022measurementinducedphase}%
  \BibitemOpen
  \bibfield  {author} {\bibinfo {author} {\bibfnamefont {P.}~\bibnamefont
  {Sierant}}, \bibinfo {author} {\bibfnamefont {M.}~\bibnamefont {Schir\`o}},
  \bibinfo {author} {\bibfnamefont {M.}~\bibnamefont {Lewenstein}},\ and\
  \bibinfo {author} {\bibfnamefont {X.}~\bibnamefont {Turkeshi}},\ }\href
  {https://doi.org/10.1103/PhysRevB.106.214316} {\bibfield  {journal} {\bibinfo
   {journal} {Phys. Rev. B}\ }\textbf {\bibinfo {volume} {106}},\ \bibinfo
  {pages} {214316} (\bibinfo {year} {2022})}\BibitemShut {NoStop}%
\bibitem [{\citenamefont {Klocke}\ and\ \citenamefont
  {Buchhold}(2023)}]{klocke2023majorana}%
  \BibitemOpen
  \bibfield  {author} {\bibinfo {author} {\bibfnamefont {K.}~\bibnamefont
  {Klocke}}\ and\ \bibinfo {author} {\bibfnamefont {M.}~\bibnamefont
  {Buchhold}},\ }\href {https://doi.org/10.1103/PhysRevX.13.041028} {\bibfield
  {journal} {\bibinfo  {journal} {Phys. Rev. X}\ }\textbf {\bibinfo {volume}
  {13}},\ \bibinfo {pages} {041028} (\bibinfo {year} {2023})}\BibitemShut
  {NoStop}%
\bibitem [{\citenamefont {Fuji}\ and\ \citenamefont
  {Ashida}(2020)}]{fuji2020measurementinducedquantum}%
  \BibitemOpen
  \bibfield  {author} {\bibinfo {author} {\bibfnamefont {Y.}~\bibnamefont
  {Fuji}}\ and\ \bibinfo {author} {\bibfnamefont {Y.}~\bibnamefont {Ashida}},\
  }\href {https://doi.org/10.1103/PhysRevB.102.054302} {\bibfield  {journal}
  {\bibinfo  {journal} {Phys. Rev. B}\ }\textbf {\bibinfo {volume} {102}},\
  \bibinfo {pages} {054302} (\bibinfo {year} {2020})}\BibitemShut {NoStop}%
\bibitem [{\citenamefont {Lunt}\ and\ \citenamefont
  {Pal}(2020)}]{lunt2020measurementinducedentanglement}%
  \BibitemOpen
  \bibfield  {author} {\bibinfo {author} {\bibfnamefont {O.}~\bibnamefont
  {Lunt}}\ and\ \bibinfo {author} {\bibfnamefont {A.}~\bibnamefont {Pal}},\
  }\href {https://doi.org/10.1103/PhysRevResearch.2.043072} {\bibfield
  {journal} {\bibinfo  {journal} {Phys. Rev. Res.}\ }\textbf {\bibinfo {volume}
  {2}},\ \bibinfo {pages} {043072} (\bibinfo {year} {2020})}\BibitemShut
  {NoStop}%
\bibitem [{\citenamefont {Doggen}\ \emph {et~al.}(2022)\citenamefont {Doggen},
  \citenamefont {Gefen}, \citenamefont {Gornyi}, \citenamefont {Mirlin},\ and\
  \citenamefont {Polyakov}}]{dogger2022generalizedquantummeasurements}%
  \BibitemOpen
  \bibfield  {author} {\bibinfo {author} {\bibfnamefont {E.~V.~H.}\
  \bibnamefont {Doggen}}, \bibinfo {author} {\bibfnamefont {Y.}~\bibnamefont
  {Gefen}}, \bibinfo {author} {\bibfnamefont {I.~V.}\ \bibnamefont {Gornyi}},
  \bibinfo {author} {\bibfnamefont {A.~D.}\ \bibnamefont {Mirlin}},\ and\
  \bibinfo {author} {\bibfnamefont {D.~G.}\ \bibnamefont {Polyakov}},\ }\href
  {https://doi.org/10.1103/PhysRevResearch.4.023146} {\bibfield  {journal}
  {\bibinfo  {journal} {Phys. Rev. Res.}\ }\textbf {\bibinfo {volume} {4}},\
  \bibinfo {pages} {023146} (\bibinfo {year} {2022})}\BibitemShut {NoStop}%
\bibitem [{\citenamefont {Xing}\ \emph {et~al.}(2023)\citenamefont {Xing},
  \citenamefont {Turkeshi}, \citenamefont {Schiró}, \citenamefont {Fazio},\
  and\ \citenamefont {Poletti}}]{xing2023interactions}%
  \BibitemOpen
  \bibfield  {author} {\bibinfo {author} {\bibfnamefont {B.}~\bibnamefont
  {Xing}}, \bibinfo {author} {\bibfnamefont {X.}~\bibnamefont {Turkeshi}},
  \bibinfo {author} {\bibfnamefont {M.}~\bibnamefont {Schiró}}, \bibinfo
  {author} {\bibfnamefont {R.}~\bibnamefont {Fazio}},\ and\ \bibinfo {author}
  {\bibfnamefont {D.}~\bibnamefont {Poletti}},\ }\href@noop {} {} (\bibinfo
  {year} {2023}),\ \Eprint {https://arxiv.org/abs/2308.09133} {arXiv:2308.09133
  [quant-ph]} \BibitemShut {NoStop}%
\bibitem [{\citenamefont {Altland}\ \emph {et~al.}(2022)\citenamefont
  {Altland}, \citenamefont {Buchhold}, \citenamefont {Diehl},\ and\
  \citenamefont {Micklitz}}]{altland2022dynamics}%
  \BibitemOpen
  \bibfield  {author} {\bibinfo {author} {\bibfnamefont {A.}~\bibnamefont
  {Altland}}, \bibinfo {author} {\bibfnamefont {M.}~\bibnamefont {Buchhold}},
  \bibinfo {author} {\bibfnamefont {S.}~\bibnamefont {Diehl}},\ and\ \bibinfo
  {author} {\bibfnamefont {T.}~\bibnamefont {Micklitz}},\ }\href
  {https://doi.org/10.1103/PhysRevResearch.4.L022066} {\bibfield  {journal}
  {\bibinfo  {journal} {Phys. Rev. Res.}\ }\textbf {\bibinfo {volume} {4}},\
  \bibinfo {pages} {L022066} (\bibinfo {year} {2022})}\BibitemShut {NoStop}%
\bibitem [{\citenamefont {Cao}\ \emph {et~al.}(2019)\citenamefont {Cao},
  \citenamefont {Tilloy},\ and\ \citenamefont {{De
  Luca}}}]{cao2019entanglementina}%
  \BibitemOpen
  \bibfield  {author} {\bibinfo {author} {\bibfnamefont {X.}~\bibnamefont
  {Cao}}, \bibinfo {author} {\bibfnamefont {A.}~\bibnamefont {Tilloy}},\ and\
  \bibinfo {author} {\bibfnamefont {A.}~\bibnamefont {{De Luca}}},\ }\href
  {https://doi.org/10.21468/SciPostPhys.7.2.024} {\bibfield  {journal}
  {\bibinfo  {journal} {SciPost Phys.}\ }\textbf {\bibinfo {volume} {7}},\
  \bibinfo {pages} {024} (\bibinfo {year} {2019})}\BibitemShut {NoStop}%
\bibitem [{\citenamefont {Fidkowski}\ \emph {et~al.}(2021)\citenamefont
  {Fidkowski}, \citenamefont {Haah},\ and\ \citenamefont
  {Hastings}}]{fidkowski2021howdynamicalquantum}%
  \BibitemOpen
  \bibfield  {author} {\bibinfo {author} {\bibfnamefont {L.}~\bibnamefont
  {Fidkowski}}, \bibinfo {author} {\bibfnamefont {J.}~\bibnamefont {Haah}},\
  and\ \bibinfo {author} {\bibfnamefont {M.~B.}\ \bibnamefont {Hastings}},\
  }\href {https://doi.org/10.22331/q-2021-01-17-382} {\bibfield  {journal}
  {\bibinfo  {journal} {{Quantum}}\ }\textbf {\bibinfo {volume} {5}},\ \bibinfo
  {pages} {382} (\bibinfo {year} {2021})}\BibitemShut {NoStop}%
\bibitem [{\citenamefont {Coppola}\ \emph {et~al.}(2022)\citenamefont
  {Coppola}, \citenamefont {Tirrito}, \citenamefont {Karevski},\ and\
  \citenamefont {Collura}}]{coppola2022growthofentanglement}%
  \BibitemOpen
  \bibfield  {author} {\bibinfo {author} {\bibfnamefont {M.}~\bibnamefont
  {Coppola}}, \bibinfo {author} {\bibfnamefont {E.}~\bibnamefont {Tirrito}},
  \bibinfo {author} {\bibfnamefont {D.}~\bibnamefont {Karevski}},\ and\
  \bibinfo {author} {\bibfnamefont {M.}~\bibnamefont {Collura}},\ }\href
  {https://doi.org/10.1103/PhysRevB.105.094303} {\bibfield  {journal} {\bibinfo
   {journal} {Phys. Rev. B}\ }\textbf {\bibinfo {volume} {105}},\ \bibinfo
  {pages} {094303} (\bibinfo {year} {2022})}\BibitemShut {NoStop}%
\bibitem [{\citenamefont {Lóio}\ \emph {et~al.}(2023)\citenamefont {Lóio},
  \citenamefont {De~Luca}, \citenamefont {De~Nardis},\ and\ \citenamefont
  {Turkeshi}}]{loio2023purificationtimescalesin}%
  \BibitemOpen
  \bibfield  {author} {\bibinfo {author} {\bibfnamefont {H.}~\bibnamefont
  {Lóio}}, \bibinfo {author} {\bibfnamefont {A.}~\bibnamefont {De~Luca}},
  \bibinfo {author} {\bibfnamefont {J.}~\bibnamefont {De~Nardis}},\ and\
  \bibinfo {author} {\bibfnamefont {X.}~\bibnamefont {Turkeshi}},\ }\href
  {http://dx.doi.org/10.1103/PhysRevB.108.L020306} {\bibfield  {journal}
  {\bibinfo  {journal} {Phys. Rev. B}\ }\textbf {\bibinfo {volume} {108}}
  (\bibinfo {year} {2023})}\BibitemShut {NoStop}%
\bibitem [{\citenamefont {Poboiko}\ \emph {et~al.}(2023)\citenamefont
  {Poboiko}, \citenamefont {P\"opperl}, \citenamefont {Gornyi},\ and\
  \citenamefont {Mirlin}}]{poboiko2023theoryoffree}%
  \BibitemOpen
  \bibfield  {author} {\bibinfo {author} {\bibfnamefont {I.}~\bibnamefont
  {Poboiko}}, \bibinfo {author} {\bibfnamefont {P.}~\bibnamefont {P\"opperl}},
  \bibinfo {author} {\bibfnamefont {I.~V.}\ \bibnamefont {Gornyi}},\ and\
  \bibinfo {author} {\bibfnamefont {A.~D.}\ \bibnamefont {Mirlin}},\ }\href
  {https://doi.org/10.1103/PhysRevX.13.041046} {\bibfield  {journal} {\bibinfo
  {journal} {Phys. Rev. X}\ }\textbf {\bibinfo {volume} {13}},\ \bibinfo
  {pages} {041046} (\bibinfo {year} {2023})}\BibitemShut {NoStop}%
\bibitem [{\citenamefont {Jian}\ \emph {et~al.}(2023)\citenamefont {Jian},
  \citenamefont {Shapourian}, \citenamefont {Bauer},\ and\ \citenamefont
  {Ludwig}}]{jian2023measurementinducedentanglement}%
  \BibitemOpen
  \bibfield  {author} {\bibinfo {author} {\bibfnamefont {C.-M.}\ \bibnamefont
  {Jian}}, \bibinfo {author} {\bibfnamefont {H.}~\bibnamefont {Shapourian}},
  \bibinfo {author} {\bibfnamefont {B.}~\bibnamefont {Bauer}},\ and\ \bibinfo
  {author} {\bibfnamefont {A.~W.~W.}\ \bibnamefont {Ludwig}},\ }\href@noop {}
  {} (\bibinfo {year} {2023}),\ \Eprint {https://arxiv.org/abs/2302.09094}
  {arXiv:2302.09094} \BibitemShut {NoStop}%
\bibitem [{\citenamefont {Fava}\ \emph {et~al.}(2023)\citenamefont {Fava},
  \citenamefont {Piroli}, \citenamefont {Swann}, \citenamefont {Bernard},\ and\
  \citenamefont {Nahum}}]{fava2023nonlinearsigmamodels}%
  \BibitemOpen
  \bibfield  {author} {\bibinfo {author} {\bibfnamefont {M.}~\bibnamefont
  {Fava}}, \bibinfo {author} {\bibfnamefont {L.}~\bibnamefont {Piroli}},
  \bibinfo {author} {\bibfnamefont {T.}~\bibnamefont {Swann}}, \bibinfo
  {author} {\bibfnamefont {D.}~\bibnamefont {Bernard}},\ and\ \bibinfo {author}
  {\bibfnamefont {A.}~\bibnamefont {Nahum}},\ }\href
  {https://doi.org/10.1103/PhysRevX.13.041045} {\bibfield  {journal} {\bibinfo
  {journal} {Phys. Rev. X}\ }\textbf {\bibinfo {volume} {13}},\ \bibinfo
  {pages} {041045} (\bibinfo {year} {2023})}\BibitemShut {NoStop}%
\bibitem [{\citenamefont {Carisch}\ \emph {et~al.}(2023)\citenamefont
  {Carisch}, \citenamefont {Romito},\ and\ \citenamefont
  {Zilberberg}}]{carisch2023quantifying}%
  \BibitemOpen
  \bibfield  {author} {\bibinfo {author} {\bibfnamefont {C.}~\bibnamefont
  {Carisch}}, \bibinfo {author} {\bibfnamefont {A.}~\bibnamefont {Romito}},\
  and\ \bibinfo {author} {\bibfnamefont {O.}~\bibnamefont {Zilberberg}},\
  }\href@noop {} {} (\bibinfo {year} {2023}),\ \Eprint
  {https://arxiv.org/abs/2304.02965} {arXiv:2304.02965 [quant-ph]} \BibitemShut
  {NoStop}%
\bibitem [{\citenamefont {Jin}\ and\ \citenamefont
  {Martin}(2023)}]{jin2023measurementinduced}%
  \BibitemOpen
  \bibfield  {author} {\bibinfo {author} {\bibfnamefont {T.}~\bibnamefont
  {Jin}}\ and\ \bibinfo {author} {\bibfnamefont {D.~G.}\ \bibnamefont
  {Martin}},\ }\href@noop {} {} (\bibinfo {year} {2023}),\ \Eprint
  {https://arxiv.org/abs/2309.15034} {arXiv:2309.15034 [quant-ph]} \BibitemShut
  {NoStop}%
\bibitem [{\citenamefont {Alberton}\ \emph {et~al.}(2021)\citenamefont
  {Alberton}, \citenamefont {Buchhold},\ and\ \citenamefont
  {Diehl}}]{alberton2021entanglementtransitionin}%
  \BibitemOpen
  \bibfield  {author} {\bibinfo {author} {\bibfnamefont {O.}~\bibnamefont
  {Alberton}}, \bibinfo {author} {\bibfnamefont {M.}~\bibnamefont {Buchhold}},\
  and\ \bibinfo {author} {\bibfnamefont {S.}~\bibnamefont {Diehl}},\ }\href
  {https://doi.org/10.1103/PhysRevLett.126.170602} {\bibfield  {journal}
  {\bibinfo  {journal} {Phys. Rev. Lett.}\ }\textbf {\bibinfo {volume} {126}},\
  \bibinfo {pages} {170602} (\bibinfo {year} {2021})}\BibitemShut {NoStop}%
\bibitem [{\citenamefont {Van~Regemortel}\ \emph {et~al.}(2021)\citenamefont
  {Van~Regemortel}, \citenamefont {Cian}, \citenamefont {Seif}, \citenamefont
  {Dehghani},\ and\ \citenamefont {Hafezi}}]{vanregemortel2021entanglement}%
  \BibitemOpen
  \bibfield  {author} {\bibinfo {author} {\bibfnamefont {M.}~\bibnamefont
  {Van~Regemortel}}, \bibinfo {author} {\bibfnamefont {Z.-P.}\ \bibnamefont
  {Cian}}, \bibinfo {author} {\bibfnamefont {A.}~\bibnamefont {Seif}}, \bibinfo
  {author} {\bibfnamefont {H.}~\bibnamefont {Dehghani}},\ and\ \bibinfo
  {author} {\bibfnamefont {M.}~\bibnamefont {Hafezi}},\ }\href
  {http://dx.doi.org/10.1103/PhysRevLett.126.123604} {\bibfield  {journal}
  {\bibinfo  {journal} {Phys. Rev. B}\ }\textbf {\bibinfo {volume} {126}}
  (\bibinfo {year} {2021})}\BibitemShut {NoStop}%
\bibitem [{\citenamefont {Turkeshi}\ \emph {et~al.}(2021)\citenamefont
  {Turkeshi}, \citenamefont {Biella}, \citenamefont {Fazio}, \citenamefont
  {Dalmonte},\ and\ \citenamefont
  {Schir\'o}}]{turkeshi2021measurementinducedentanglement}%
  \BibitemOpen
  \bibfield  {author} {\bibinfo {author} {\bibfnamefont {X.}~\bibnamefont
  {Turkeshi}}, \bibinfo {author} {\bibfnamefont {A.}~\bibnamefont {Biella}},
  \bibinfo {author} {\bibfnamefont {R.}~\bibnamefont {Fazio}}, \bibinfo
  {author} {\bibfnamefont {M.}~\bibnamefont {Dalmonte}},\ and\ \bibinfo
  {author} {\bibfnamefont {M.}~\bibnamefont {Schir\'o}},\ }\href
  {https://doi.org/10.1103/PhysRevB.103.224210} {\bibfield  {journal} {\bibinfo
   {journal} {Phys. Rev. B}\ }\textbf {\bibinfo {volume} {103}},\ \bibinfo
  {pages} {224210} (\bibinfo {year} {2021})}\BibitemShut {NoStop}%
\bibitem [{\citenamefont {Botzung}\ \emph {et~al.}(2021)\citenamefont
  {Botzung}, \citenamefont {Diehl},\ and\ \citenamefont
  {M\"uller}}]{botzung2021engineereddissipationinduced}%
  \BibitemOpen
  \bibfield  {author} {\bibinfo {author} {\bibfnamefont {T.}~\bibnamefont
  {Botzung}}, \bibinfo {author} {\bibfnamefont {S.}~\bibnamefont {Diehl}},\
  and\ \bibinfo {author} {\bibfnamefont {M.}~\bibnamefont {M\"uller}},\ }\href
  {https://doi.org/10.1103/PhysRevB.104.184422} {\bibfield  {journal} {\bibinfo
   {journal} {Phys. Rev. B}\ }\textbf {\bibinfo {volume} {104}},\ \bibinfo
  {pages} {184422} (\bibinfo {year} {2021})}\BibitemShut {NoStop}%
\bibitem [{\citenamefont {Bao}\ \emph {et~al.}(2021)\citenamefont {Bao},
  \citenamefont {Choi},\ and\ \citenamefont
  {Altman}}]{bao2021symmetryenrichedphases}%
  \BibitemOpen
  \bibfield  {author} {\bibinfo {author} {\bibfnamefont {Y.}~\bibnamefont
  {Bao}}, \bibinfo {author} {\bibfnamefont {S.}~\bibnamefont {Choi}},\ and\
  \bibinfo {author} {\bibfnamefont {E.}~\bibnamefont {Altman}},\ }\href
  {https://doi.org/10.1016/j.aop.2021.168618} {\bibfield  {journal} {\bibinfo
  {journal} {Ann. Phys.}\ }\textbf {\bibinfo {volume} {435}},\ \bibinfo {pages}
  {168618} (\bibinfo {year} {2021})}\BibitemShut {NoStop}%
\bibitem [{\citenamefont {Turkeshi}\ \emph {et~al.}(2022)\citenamefont
  {Turkeshi}, \citenamefont {Dalmonte}, \citenamefont {Fazio},\ and\
  \citenamefont {Schir{\`o}}}]{turkeshi2022entanglementtransitionsfrom}%
  \BibitemOpen
  \bibfield  {author} {\bibinfo {author} {\bibfnamefont {X.}~\bibnamefont
  {Turkeshi}}, \bibinfo {author} {\bibfnamefont {M.}~\bibnamefont {Dalmonte}},
  \bibinfo {author} {\bibfnamefont {R.}~\bibnamefont {Fazio}},\ and\ \bibinfo
  {author} {\bibfnamefont {M.}~\bibnamefont {Schir{\`o}}},\ }\href
  {https://doi.org/10.1103/PhysRevB.105.L241114} {\bibfield  {journal}
  {\bibinfo  {journal} {Phys. Rev. B}\ }\textbf {\bibinfo {volume} {105}},\
  \bibinfo {pages} {L241114} (\bibinfo {year} {2022})}\BibitemShut {NoStop}%
\bibitem [{\citenamefont {Piccitto}\ \emph {et~al.}(2022)\citenamefont
  {Piccitto}, \citenamefont {Russomanno},\ and\ \citenamefont
  {Rossini}}]{piccitto2022entanglementtransitionsin}%
  \BibitemOpen
  \bibfield  {author} {\bibinfo {author} {\bibfnamefont {G.}~\bibnamefont
  {Piccitto}}, \bibinfo {author} {\bibfnamefont {A.}~\bibnamefont
  {Russomanno}},\ and\ \bibinfo {author} {\bibfnamefont {D.}~\bibnamefont
  {Rossini}},\ }\href {https://doi.org/10.1103/PhysRevB.105.064305} {\bibfield
  {journal} {\bibinfo  {journal} {Phys. Rev. B}\ }\textbf {\bibinfo {volume}
  {105}},\ \bibinfo {pages} {064305} (\bibinfo {year} {2022})}\BibitemShut
  {NoStop}%
\bibitem [{\citenamefont {Kells}\ \emph {et~al.}(2023)\citenamefont {Kells},
  \citenamefont {Meidan},\ and\ \citenamefont
  {Romito}}]{kells2021topologicaltransitionswith}%
  \BibitemOpen
  \bibfield  {author} {\bibinfo {author} {\bibfnamefont {G.}~\bibnamefont
  {Kells}}, \bibinfo {author} {\bibfnamefont {D.}~\bibnamefont {Meidan}},\ and\
  \bibinfo {author} {\bibfnamefont {A.}~\bibnamefont {Romito}},\ }\href
  {https://doi.org/10.21468/SciPostPhys.14.3.031} {\bibfield  {journal}
  {\bibinfo  {journal} {SciPost Phys.}\ }\textbf {\bibinfo {volume} {14}},\
  \bibinfo {pages} {031} (\bibinfo {year} {2023})}\BibitemShut {NoStop}%
\bibitem [{\citenamefont {Paviglianiti}\ and\ \citenamefont
  {Silva}(2023)}]{paviglianiti2023multipartite}%
  \BibitemOpen
  \bibfield  {author} {\bibinfo {author} {\bibfnamefont {A.}~\bibnamefont
  {Paviglianiti}}\ and\ \bibinfo {author} {\bibfnamefont {A.}~\bibnamefont
  {Silva}},\ }\href {https://doi.org/10.1103/PhysRevB.108.184302} {\bibfield
  {journal} {\bibinfo  {journal} {Phys. Rev. B}\ }\textbf {\bibinfo {volume}
  {108}},\ \bibinfo {pages} {184302} (\bibinfo {year} {2023})}\BibitemShut
  {NoStop}%
\bibitem [{\citenamefont {M\"uller}\ \emph {et~al.}(2022)\citenamefont
  {M\"uller}, \citenamefont {Diehl},\ and\ \citenamefont
  {Buchhold}}]{muller2022measurementinduced}%
  \BibitemOpen
  \bibfield  {author} {\bibinfo {author} {\bibfnamefont {T.}~\bibnamefont
  {M\"uller}}, \bibinfo {author} {\bibfnamefont {S.}~\bibnamefont {Diehl}},\
  and\ \bibinfo {author} {\bibfnamefont {M.}~\bibnamefont {Buchhold}},\ }\href
  {https://doi.org/10.1103/PhysRevLett.128.010605} {\bibfield  {journal}
  {\bibinfo  {journal} {Phys. Rev. Lett.}\ }\textbf {\bibinfo {volume} {128}},\
  \bibinfo {pages} {010605} (\bibinfo {year} {2022})}\BibitemShut {NoStop}%
\bibitem [{\citenamefont {Buchhold}\ \emph {et~al.}(2022)\citenamefont
  {Buchhold}, \citenamefont {Müller},\ and\ \citenamefont
  {Diehl}}]{buchhold2022revealing}%
  \BibitemOpen
  \bibfield  {author} {\bibinfo {author} {\bibfnamefont {M.}~\bibnamefont
  {Buchhold}}, \bibinfo {author} {\bibfnamefont {T.}~\bibnamefont {Müller}},\
  and\ \bibinfo {author} {\bibfnamefont {S.}~\bibnamefont {Diehl}},\
  }\href@noop {} {} (\bibinfo {year} {2022}),\ \Eprint
  {https://arxiv.org/abs/2208.10506} {arXiv:2208.10506 [cond-mat.dis-nn]}
  \BibitemShut {NoStop}%
\bibitem [{\citenamefont {Dalibard}\ \emph {et~al.}(1992)\citenamefont
  {Dalibard}, \citenamefont {Castin},\ and\ \citenamefont
  {M\o{}lmer}}]{dalibard1992wavefunction}%
  \BibitemOpen
  \bibfield  {author} {\bibinfo {author} {\bibfnamefont {J.}~\bibnamefont
  {Dalibard}}, \bibinfo {author} {\bibfnamefont {Y.}~\bibnamefont {Castin}},\
  and\ \bibinfo {author} {\bibfnamefont {K.}~\bibnamefont {M\o{}lmer}},\ }\href
  {https://doi.org/10.1103/PhysRevLett.68.580} {\bibfield  {journal} {\bibinfo
  {journal} {Phys. Rev. Lett.}\ }\textbf {\bibinfo {volume} {68}},\ \bibinfo
  {pages} {580} (\bibinfo {year} {1992})}\BibitemShut {NoStop}%
\bibitem [{\citenamefont {Plenio}\ and\ \citenamefont
  {Knight}(1998)}]{plenio1998quantum}%
  \BibitemOpen
  \bibfield  {author} {\bibinfo {author} {\bibfnamefont {M.~B.}\ \bibnamefont
  {Plenio}}\ and\ \bibinfo {author} {\bibfnamefont {P.~L.}\ \bibnamefont
  {Knight}},\ }\href {https://doi.org/10.1103/RevModPhys.70.101} {\bibfield
  {journal} {\bibinfo  {journal} {Rev. Mod. Phys.}\ }\textbf {\bibinfo {volume}
  {70}},\ \bibinfo {pages} {101} (\bibinfo {year} {1998})}\BibitemShut
  {NoStop}%
\bibitem [{\citenamefont {Wiseman}\ and\ \citenamefont
  {Milburn}(2009)}]{wiseman2009quantummeasurementand}%
  \BibitemOpen
  \bibfield  {author} {\bibinfo {author} {\bibfnamefont {H.~M.}\ \bibnamefont
  {Wiseman}}\ and\ \bibinfo {author} {\bibfnamefont {G.~J.}\ \bibnamefont
  {Milburn}},\ }\href@noop {} {\emph {\bibinfo {title} {Quantum Measurement and
  Control}}}\ (\bibinfo  {publisher} {Cambridge University Press},\ \bibinfo
  {year} {Cambridge, England, 2009})\BibitemShut {NoStop}%
\bibitem [{\citenamefont {Biella}\ and\ \citenamefont
  {Schir{\'{o}}}(2021)}]{biella2021manybodyquantumzeno}%
  \BibitemOpen
  \bibfield  {author} {\bibinfo {author} {\bibfnamefont {A.}~\bibnamefont
  {Biella}}\ and\ \bibinfo {author} {\bibfnamefont {M.}~\bibnamefont
  {Schir{\'{o}}}},\ }\href {https://doi.org/10.22331/q-2021-08-19-528}
  {\bibfield  {journal} {\bibinfo  {journal} {{Quantum}}\ }\textbf {\bibinfo
  {volume} {5}},\ \bibinfo {pages} {528} (\bibinfo {year} {2021})}\BibitemShut
  {NoStop}%
\bibitem [{\citenamefont {Gopalakrishnan}\ and\ \citenamefont
  {Gullans}(2021)}]{gopalakrishnan2021entanglementandpurification}%
  \BibitemOpen
  \bibfield  {author} {\bibinfo {author} {\bibfnamefont {S.}~\bibnamefont
  {Gopalakrishnan}}\ and\ \bibinfo {author} {\bibfnamefont {M.~J.}\
  \bibnamefont {Gullans}},\ }\href
  {https://doi.org/10.1103/PhysRevLett.126.170503} {\bibfield  {journal}
  {\bibinfo  {journal} {Phys. Rev. Lett.}\ }\textbf {\bibinfo {volume} {126}},\
  \bibinfo {pages} {170503} (\bibinfo {year} {2021})}\BibitemShut {NoStop}%
\bibitem [{\citenamefont {Jian}\ \emph {et~al.}(2021)\citenamefont {Jian},
  \citenamefont {Yang}, \citenamefont {Bi},\ and\ \citenamefont
  {Chen}}]{jian2021yangleeedge}%
  \BibitemOpen
  \bibfield  {author} {\bibinfo {author} {\bibfnamefont {S.-K.}\ \bibnamefont
  {Jian}}, \bibinfo {author} {\bibfnamefont {Z.-C.}\ \bibnamefont {Yang}},
  \bibinfo {author} {\bibfnamefont {Z.}~\bibnamefont {Bi}},\ and\ \bibinfo
  {author} {\bibfnamefont {X.}~\bibnamefont {Chen}},\ }\href
  {https://doi.org/10.1103/PhysRevB.104.L161107} {\bibfield  {journal}
  {\bibinfo  {journal} {Phys. Rev. B}\ }\textbf {\bibinfo {volume} {104}},\
  \bibinfo {pages} {L161107} (\bibinfo {year} {2021})}\BibitemShut {NoStop}%
\bibitem [{\citenamefont {Turkeshi}\ and\ \citenamefont
  {Schir\'o}(2023)}]{turkeshi2023entanglementandcorrelation}%
  \BibitemOpen
  \bibfield  {author} {\bibinfo {author} {\bibfnamefont {X.}~\bibnamefont
  {Turkeshi}}\ and\ \bibinfo {author} {\bibfnamefont {M.}~\bibnamefont
  {Schir\'o}},\ }\href {https://doi.org/10.1103/PhysRevB.107.L020403}
  {\bibfield  {journal} {\bibinfo  {journal} {Phys. Rev. B}\ }\textbf {\bibinfo
  {volume} {107}},\ \bibinfo {pages} {L020403} (\bibinfo {year}
  {2023})}\BibitemShut {NoStop}%
\bibitem [{\citenamefont {Gal}\ \emph {et~al.}(2023)\citenamefont {Gal},
  \citenamefont {Turkeshi},\ and\ \citenamefont
  {Schirò}}]{legal2023volumetoarea}%
  \BibitemOpen
  \bibfield  {author} {\bibinfo {author} {\bibfnamefont {Y.~L.}\ \bibnamefont
  {Gal}}, \bibinfo {author} {\bibfnamefont {X.}~\bibnamefont {Turkeshi}},\ and\
  \bibinfo {author} {\bibfnamefont {M.}~\bibnamefont {Schirò}},\ }\href
  {https://doi.org/10.21468/SciPostPhys.14.5.138} {\bibfield  {journal}
  {\bibinfo  {journal} {SciPost Phys.}\ }\textbf {\bibinfo {volume} {14}},\
  \bibinfo {pages} {138} (\bibinfo {year} {2023})}\BibitemShut {NoStop}%
\bibitem [{\citenamefont {Kawabata}\ \emph {et~al.}(2023)\citenamefont
  {Kawabata}, \citenamefont {Numasawa},\ and\ \citenamefont
  {Ryu}}]{kawabata2023entanglementphasetransition}%
  \BibitemOpen
  \bibfield  {author} {\bibinfo {author} {\bibfnamefont {K.}~\bibnamefont
  {Kawabata}}, \bibinfo {author} {\bibfnamefont {T.}~\bibnamefont {Numasawa}},\
  and\ \bibinfo {author} {\bibfnamefont {S.}~\bibnamefont {Ryu}},\ }\href
  {https://doi.org/10.1103/PhysRevX.13.021007} {\bibfield  {journal} {\bibinfo
  {journal} {Phys. Rev. X}\ }\textbf {\bibinfo {volume} {13}},\ \bibinfo
  {pages} {021007} (\bibinfo {year} {2023})}\BibitemShut {NoStop}%
\bibitem [{\citenamefont {Orito}\ and\ \citenamefont
  {Imura}(2023)}]{orito2023entanglement}%
  \BibitemOpen
  \bibfield  {author} {\bibinfo {author} {\bibfnamefont {T.}~\bibnamefont
  {Orito}}\ and\ \bibinfo {author} {\bibfnamefont {K.-I.}\ \bibnamefont
  {Imura}},\ }\href {https://doi.org/10.1103/PhysRevB.108.214308} {\bibfield
  {journal} {\bibinfo  {journal} {Phys. Rev. B}\ }\textbf {\bibinfo {volume}
  {108}},\ \bibinfo {pages} {214308} (\bibinfo {year} {2023})}\BibitemShut
  {NoStop}%
\bibitem [{\citenamefont {Zerba}\ and\ \citenamefont
  {Silva}(2023)}]{zerba2023measurement}%
  \BibitemOpen
  \bibfield  {author} {\bibinfo {author} {\bibfnamefont {C.}~\bibnamefont
  {Zerba}}\ and\ \bibinfo {author} {\bibfnamefont {A.}~\bibnamefont {Silva}},\
  }\href {https://doi.org/10.21468/SciPostPhysCore.6.3.051} {\bibfield
  {journal} {\bibinfo  {journal} {SciPost Phys. Core}\ }\textbf {\bibinfo
  {volume} {6}},\ \bibinfo {pages} {051} (\bibinfo {year} {2023})}\BibitemShut
  {NoStop}%
\bibitem [{\citenamefont {Granet}\ \emph {et~al.}(2023)\citenamefont {Granet},
  \citenamefont {Zhang},\ and\ \citenamefont {Dreyer}}]{granet2023volume}%
  \BibitemOpen
  \bibfield  {author} {\bibinfo {author} {\bibfnamefont {E.}~\bibnamefont
  {Granet}}, \bibinfo {author} {\bibfnamefont {C.}~\bibnamefont {Zhang}},\ and\
  \bibinfo {author} {\bibfnamefont {H.}~\bibnamefont {Dreyer}},\ }\href
  {https://doi.org/10.1103/PhysRevLett.130.230401} {\bibfield  {journal}
  {\bibinfo  {journal} {Phys. Rev. Lett.}\ }\textbf {\bibinfo {volume} {130}},\
  \bibinfo {pages} {230401} (\bibinfo {year} {2023})}\BibitemShut {NoStop}%
\bibitem [{\citenamefont {Su}\ \emph {et~al.}(2023)\citenamefont {Su},
  \citenamefont {Clerk},\ and\ \citenamefont {Martin}}]{su2023dynamics}%
  \BibitemOpen
  \bibfield  {author} {\bibinfo {author} {\bibfnamefont {L.}~\bibnamefont
  {Su}}, \bibinfo {author} {\bibfnamefont {A.}~\bibnamefont {Clerk}},\ and\
  \bibinfo {author} {\bibfnamefont {I.}~\bibnamefont {Martin}},\ }\href@noop {}
  {} (\bibinfo {year} {2023}),\ \Eprint {https://arxiv.org/abs/2306.07428}
  {arXiv:2306.07428 [quant-ph]} \BibitemShut {NoStop}%
\bibitem [{\citenamefont {Banerjee}\ and\ \citenamefont
  {Sengupta}(2023)}]{banerjee2023entanglement}%
  \BibitemOpen
  \bibfield  {author} {\bibinfo {author} {\bibfnamefont {T.}~\bibnamefont
  {Banerjee}}\ and\ \bibinfo {author} {\bibfnamefont {K.}~\bibnamefont
  {Sengupta}},\ }\href@noop {} {} (\bibinfo {year} {2023}),\ \Eprint
  {https://arxiv.org/abs/2309.07661} {arXiv:2309.07661 [cond-mat.str-el]}
  \BibitemShut {NoStop}%
\bibitem [{\citenamefont {Lee}\ \emph {et~al.}(2023)\citenamefont {Lee},
  \citenamefont {Jin}, \citenamefont {Wang}, \citenamefont {McDonald},\ and\
  \citenamefont {Clerk}}]{lee2023entanglement}%
  \BibitemOpen
  \bibfield  {author} {\bibinfo {author} {\bibfnamefont {G.}~\bibnamefont
  {Lee}}, \bibinfo {author} {\bibfnamefont {T.}~\bibnamefont {Jin}}, \bibinfo
  {author} {\bibfnamefont {Y.-X.}\ \bibnamefont {Wang}}, \bibinfo {author}
  {\bibfnamefont {A.}~\bibnamefont {McDonald}},\ and\ \bibinfo {author}
  {\bibfnamefont {A.}~\bibnamefont {Clerk}},\ }\href@noop {} {} (\bibinfo
  {year} {2023}),\ \Eprint {https://arxiv.org/abs/2308.14614} {arXiv:2308.14614
  [quant-ph]} \BibitemShut {NoStop}%
\bibitem [{\citenamefont {Yuto~Ashida}\ and\ \citenamefont
  {Ueda}(2020)}]{ashida2020review}%
  \BibitemOpen
  \bibfield  {author} {\bibinfo {author} {\bibfnamefont {Z.~G.}\ \bibnamefont
  {Yuto~Ashida}}\ and\ \bibinfo {author} {\bibfnamefont {M.}~\bibnamefont
  {Ueda}},\ }\href {https://doi.org/10.1080/00018732.2021.1876991} {\bibfield
  {journal} {\bibinfo  {journal} {Advances in Physics}\ }\textbf {\bibinfo
  {volume} {69}},\ \bibinfo {pages} {249} (\bibinfo {year} {2020})},\ \Eprint
  {https://arxiv.org/abs/https://doi.org/10.1080/00018732.2021.1876991}
  {https://doi.org/10.1080/00018732.2021.1876991} \BibitemShut {NoStop}%
\bibitem [{\citenamefont {Lau}\ and\ \citenamefont
  {Clerk}(2018)}]{lau2018fundamental}%
  \BibitemOpen
  \bibfield  {author} {\bibinfo {author} {\bibfnamefont {H.-K.}\ \bibnamefont
  {Lau}}\ and\ \bibinfo {author} {\bibfnamefont {A.~A.}\ \bibnamefont
  {Clerk}},\ }\href {https://doi.org/10.1038/s41467-018-06477-7} {\bibfield
  {journal} {\bibinfo  {journal} {Nature Communications}\ }\textbf {\bibinfo
  {volume} {9}},\ \bibinfo {pages} {4320} (\bibinfo {year} {2018})}\BibitemShut
  {NoStop}%
\bibitem [{\citenamefont {Minganti}\ \emph {et~al.}(2019)\citenamefont
  {Minganti}, \citenamefont {Miranowicz}, \citenamefont {Chhajlany},\ and\
  \citenamefont {Nori}}]{minganti2019quantum}%
  \BibitemOpen
  \bibfield  {author} {\bibinfo {author} {\bibfnamefont {F.}~\bibnamefont
  {Minganti}}, \bibinfo {author} {\bibfnamefont {A.}~\bibnamefont
  {Miranowicz}}, \bibinfo {author} {\bibfnamefont {R.~W.}\ \bibnamefont
  {Chhajlany}},\ and\ \bibinfo {author} {\bibfnamefont {F.}~\bibnamefont
  {Nori}},\ }\href {https://doi.org/10.1103/PhysRevA.100.062131} {\bibfield
  {journal} {\bibinfo  {journal} {Phys. Rev. A}\ }\textbf {\bibinfo {volume}
  {100}},\ \bibinfo {pages} {062131} (\bibinfo {year} {2019})}\BibitemShut
  {NoStop}%
\bibitem [{\citenamefont {Naghiloo}\ \emph {et~al.}(2019)\citenamefont
  {Naghiloo}, \citenamefont {Abbasi}, \citenamefont {Joglekar},\ and\
  \citenamefont {Murch}}]{naghiloo2019quantum}%
  \BibitemOpen
  \bibfield  {author} {\bibinfo {author} {\bibfnamefont {M.}~\bibnamefont
  {Naghiloo}}, \bibinfo {author} {\bibfnamefont {M.}~\bibnamefont {Abbasi}},
  \bibinfo {author} {\bibfnamefont {Y.~N.}\ \bibnamefont {Joglekar}},\ and\
  \bibinfo {author} {\bibfnamefont {K.~W.}\ \bibnamefont {Murch}},\ }\href
  {https://doi.org/10.1038/s41567-019-0652-z} {\bibfield  {journal} {\bibinfo
  {journal} {Nature Physics}\ }\textbf {\bibinfo {volume} {15}},\ \bibinfo
  {pages} {1232} (\bibinfo {year} {2019})}\BibitemShut {NoStop}%
\bibitem [{\citenamefont {Minganti}\ \emph {et~al.}(2020)\citenamefont
  {Minganti}, \citenamefont {Miranowicz}, \citenamefont {Chhajlany},
  \citenamefont {Arkhipov},\ and\ \citenamefont {Nori}}]{minganti2020hybrid}%
  \BibitemOpen
  \bibfield  {author} {\bibinfo {author} {\bibfnamefont {F.}~\bibnamefont
  {Minganti}}, \bibinfo {author} {\bibfnamefont {A.}~\bibnamefont
  {Miranowicz}}, \bibinfo {author} {\bibfnamefont {R.~W.}\ \bibnamefont
  {Chhajlany}}, \bibinfo {author} {\bibfnamefont {I.~I.}\ \bibnamefont
  {Arkhipov}},\ and\ \bibinfo {author} {\bibfnamefont {F.}~\bibnamefont
  {Nori}},\ }\href {https://doi.org/10.1103/PhysRevA.101.062112} {\bibfield
  {journal} {\bibinfo  {journal} {Phys. Rev. A}\ }\textbf {\bibinfo {volume}
  {101}},\ \bibinfo {pages} {062112} (\bibinfo {year} {2020})}\BibitemShut
  {NoStop}%
\bibitem [{\citenamefont {Han}\ \emph {et~al.}(2023)\citenamefont {Han},
  \citenamefont {Wu}, \citenamefont {Huang}, \citenamefont {Wu}, \citenamefont
  {Zou}, \citenamefont {Yi}, \citenamefont {Zhang}, \citenamefont {Li},
  \citenamefont {Xu}, \citenamefont {Zheng}, \citenamefont {Fan}, \citenamefont
  {Wen}, \citenamefont {Yang},\ and\ \citenamefont
  {Zheng}}]{han2023exceptional}%
  \BibitemOpen
  \bibfield  {author} {\bibinfo {author} {\bibfnamefont {P.-R.}\ \bibnamefont
  {Han}}, \bibinfo {author} {\bibfnamefont {F.}~\bibnamefont {Wu}}, \bibinfo
  {author} {\bibfnamefont {X.-J.}\ \bibnamefont {Huang}}, \bibinfo {author}
  {\bibfnamefont {H.-Z.}\ \bibnamefont {Wu}}, \bibinfo {author} {\bibfnamefont
  {C.-L.}\ \bibnamefont {Zou}}, \bibinfo {author} {\bibfnamefont
  {W.}~\bibnamefont {Yi}}, \bibinfo {author} {\bibfnamefont {M.}~\bibnamefont
  {Zhang}}, \bibinfo {author} {\bibfnamefont {H.}~\bibnamefont {Li}}, \bibinfo
  {author} {\bibfnamefont {K.}~\bibnamefont {Xu}}, \bibinfo {author}
  {\bibfnamefont {D.}~\bibnamefont {Zheng}}, \bibinfo {author} {\bibfnamefont
  {H.}~\bibnamefont {Fan}}, \bibinfo {author} {\bibfnamefont {J.}~\bibnamefont
  {Wen}}, \bibinfo {author} {\bibfnamefont {Z.-B.}\ \bibnamefont {Yang}},\ and\
  \bibinfo {author} {\bibfnamefont {S.-B.}\ \bibnamefont {Zheng}},\ }\href
  {https://doi.org/10.1103/PhysRevLett.131.260201} {\bibfield  {journal}
  {\bibinfo  {journal} {Phys. Rev. Lett.}\ }\textbf {\bibinfo {volume} {131}},\
  \bibinfo {pages} {260201} (\bibinfo {year} {2023})}\BibitemShut {NoStop}%
\bibitem [{\citenamefont {Ehrhardt}\ and\ \citenamefont
  {Larson}(2023)}]{ehrhardt2023exploring}%
  \BibitemOpen
  \bibfield  {author} {\bibinfo {author} {\bibfnamefont {C.}~\bibnamefont
  {Ehrhardt}}\ and\ \bibinfo {author} {\bibfnamefont {J.}~\bibnamefont
  {Larson}},\ }\href@noop {} {\bibinfo {title} {Exploring the impact of
  fluctuation-induced criticality on non-hermitian skin effect and quantum
  sensors}} (\bibinfo {year} {2023}),\ \Eprint
  {https://arxiv.org/abs/2310.18259} {arXiv:2310.18259 [quant-ph]} \BibitemShut
  {NoStop}%
\bibitem [{\citenamefont {Ueda}(1990)}]{ueda1990}%
  \BibitemOpen
  \bibfield  {author} {\bibinfo {author} {\bibfnamefont {M.}~\bibnamefont
  {Ueda}},\ }\href {https://doi.org/10.1103/PhysRevA.41.3875} {\bibfield
  {journal} {\bibinfo  {journal} {Phys. Rev. A}\ }\textbf {\bibinfo {volume}
  {41}},\ \bibinfo {pages} {3875} (\bibinfo {year} {1990})}\BibitemShut
  {NoStop}%
\bibitem [{\citenamefont {Gardiner}\ \emph {et~al.}(1992)\citenamefont
  {Gardiner}, \citenamefont {Parkins},\ and\ \citenamefont
  {Zoller}}]{gardiner1992wave}%
  \BibitemOpen
  \bibfield  {author} {\bibinfo {author} {\bibfnamefont {C.~W.}\ \bibnamefont
  {Gardiner}}, \bibinfo {author} {\bibfnamefont {A.~S.}\ \bibnamefont
  {Parkins}},\ and\ \bibinfo {author} {\bibfnamefont {P.}~\bibnamefont
  {Zoller}},\ }\href {https://doi.org/10.1103/PhysRevA.46.4363} {\bibfield
  {journal} {\bibinfo  {journal} {Phys. Rev. A}\ }\textbf {\bibinfo {volume}
  {46}},\ \bibinfo {pages} {4363} (\bibinfo {year} {1992})}\BibitemShut
  {NoStop}%
\bibitem [{\citenamefont {Daley}(2014)}]{daley2014quantum}%
  \BibitemOpen
  \bibfield  {author} {\bibinfo {author} {\bibfnamefont {A.~J.}\ \bibnamefont
  {Daley}},\ }\href {https://doi.org/10.1080%2F00018732.2014.933502} {\bibfield
   {journal} {\bibinfo  {journal} {Adv. Phys.}\ }\textbf {\bibinfo {volume}
  {63}},\ \bibinfo {pages} {77} (\bibinfo {year} {2014})}\BibitemShut {NoStop}%
\bibitem [{\citenamefont {Breuer}\ and\ \citenamefont
  {Petruccione}(2002)}]{breuer2002thetheoryof}%
  \BibitemOpen
  \bibfield  {author} {\bibinfo {author} {\bibfnamefont {H.-P.}\ \bibnamefont
  {Breuer}}\ and\ \bibinfo {author} {\bibfnamefont {F.}~\bibnamefont
  {Petruccione}},\ }\href@noop {} {\emph {\bibinfo {title} {The Theory of Open
  Quantum Systems}}}\ (\bibinfo  {publisher} {Oxford University Press},\
  \bibinfo {year} {Oxford, England, 2002})\BibitemShut {NoStop}%
\bibitem [{\citenamefont {Radaelli}\ \emph {et~al.}(2023)\citenamefont
  {Radaelli}, \citenamefont {Landi},\ and\ \citenamefont
  {Binder}}]{radaelli2023gillespie}%
  \BibitemOpen
  \bibfield  {author} {\bibinfo {author} {\bibfnamefont {M.}~\bibnamefont
  {Radaelli}}, \bibinfo {author} {\bibfnamefont {G.~T.}\ \bibnamefont
  {Landi}},\ and\ \bibinfo {author} {\bibfnamefont {F.~C.}\ \bibnamefont
  {Binder}},\ }\href@noop {} {\bibinfo {title} {A gillespie algorithm for
  efficient simulation of quantum jump trajectories}} (\bibinfo {year}
  {2023}),\ \Eprint {https://arxiv.org/abs/2303.15405} {arXiv:2303.15405
  [quant-ph]} \BibitemShut {NoStop}%
\bibitem [{\citenamefont {Cohen-Tannoudji}\ and\ \citenamefont
  {Dalibard}(1986)}]{Cohen-Tannoudji_1986}%
  \BibitemOpen
  \bibfield  {author} {\bibinfo {author} {\bibfnamefont {C.}~\bibnamefont
  {Cohen-Tannoudji}}\ and\ \bibinfo {author} {\bibfnamefont {J.}~\bibnamefont
  {Dalibard}},\ }\href {https://doi.org/10.1209/0295-5075/1/9/004} {\bibfield
  {journal} {\bibinfo  {journal} {Europhysics Letters}\ }\textbf {\bibinfo
  {volume} {1}},\ \bibinfo {pages} {441} (\bibinfo {year} {1986})}\BibitemShut
  {NoStop}%
\bibitem [{\citenamefont {Landi}\ \emph {et~al.}(2023)\citenamefont {Landi},
  \citenamefont {Kewming}, \citenamefont {Mitchison},\ and\ \citenamefont
  {Potts}}]{landi2023current}%
  \BibitemOpen
  \bibfield  {author} {\bibinfo {author} {\bibfnamefont {G.~T.}\ \bibnamefont
  {Landi}}, \bibinfo {author} {\bibfnamefont {M.~J.}\ \bibnamefont {Kewming}},
  \bibinfo {author} {\bibfnamefont {M.~T.}\ \bibnamefont {Mitchison}},\ and\
  \bibinfo {author} {\bibfnamefont {P.~P.}\ \bibnamefont {Potts}},\ }\href@noop
  {} {} (\bibinfo {year} {2023}),\ \Eprint {https://arxiv.org/abs/2303.04270}
  {arXiv:2303.04270 [quant-ph]} \BibitemShut {NoStop}%
\bibitem [{\citenamefont
  {Carmichael}(1999)}]{carmichael1999statisticalmethodsin}%
  \BibitemOpen
  \bibfield  {author} {\bibinfo {author} {\bibfnamefont {H.}~\bibnamefont
  {Carmichael}},\ }\href@noop {} {\emph {\bibinfo {title} {Statistical Methods
  in Quantum Optics 1}}}\ (\bibinfo  {publisher} {Springer Science \& Business
  Media},\ \bibinfo {year} {Berlin, Germany, 1999})\BibitemShut {NoStop}%
\bibitem [{\citenamefont {Calabrese}\ and\ \citenamefont
  {Cardy}(2004)}]{calabrese2004entanglemententropyand}%
  \BibitemOpen
  \bibfield  {author} {\bibinfo {author} {\bibfnamefont {P.}~\bibnamefont
  {Calabrese}}\ and\ \bibinfo {author} {\bibfnamefont {J.}~\bibnamefont
  {Cardy}},\ }\href {https://doi.org/10.1088/1742-5468/2004/06/p06002}
  {\bibfield  {journal} {\bibinfo  {journal} {J. Stat. Mech.}\ }\textbf
  {\bibinfo {volume} {2004}},\ \bibinfo {pages} {P06002} (\bibinfo {year}
  {2004})}\BibitemShut {NoStop}%
\bibitem [{\citenamefont {Amico}\ \emph {et~al.}(2008)\citenamefont {Amico},
  \citenamefont {Fazio}, \citenamefont {Osterloh},\ and\ \citenamefont
  {Vedral}}]{amico2008entanglementinmanybody}%
  \BibitemOpen
  \bibfield  {author} {\bibinfo {author} {\bibfnamefont {L.}~\bibnamefont
  {Amico}}, \bibinfo {author} {\bibfnamefont {R.}~\bibnamefont {Fazio}},
  \bibinfo {author} {\bibfnamefont {A.}~\bibnamefont {Osterloh}},\ and\
  \bibinfo {author} {\bibfnamefont {V.}~\bibnamefont {Vedral}},\ }\href
  {https://doi.org/10.1103/RevModPhys.80.517} {\bibfield  {journal} {\bibinfo
  {journal} {Rev. Mod. Phys.}\ }\textbf {\bibinfo {volume} {80}},\ \bibinfo
  {pages} {517} (\bibinfo {year} {2008})}\BibitemShut {NoStop}%
\bibitem [{\citenamefont {Hickey}\ \emph {et~al.}(2013)\citenamefont {Hickey},
  \citenamefont {Genway}, \citenamefont {Lesanovsky},\ and\ \citenamefont
  {Garrahan}}]{hickey2013time}%
  \BibitemOpen
  \bibfield  {author} {\bibinfo {author} {\bibfnamefont {J.~M.}\ \bibnamefont
  {Hickey}}, \bibinfo {author} {\bibfnamefont {S.}~\bibnamefont {Genway}},
  \bibinfo {author} {\bibfnamefont {I.}~\bibnamefont {Lesanovsky}},\ and\
  \bibinfo {author} {\bibfnamefont {J.~P.}\ \bibnamefont {Garrahan}},\ }\href
  {https://doi.org/10.1103/PhysRevB.87.184303} {\bibfield  {journal} {\bibinfo
  {journal} {Phys. Rev. B}\ }\textbf {\bibinfo {volume} {87}},\ \bibinfo
  {pages} {184303} (\bibinfo {year} {2013})}\BibitemShut {NoStop}%
\bibitem [{\citenamefont {Evans}\ and\ \citenamefont
  {Majumdar}(2011)}]{evans2011diffusion}%
  \BibitemOpen
  \bibfield  {author} {\bibinfo {author} {\bibfnamefont {M.~R.}\ \bibnamefont
  {Evans}}\ and\ \bibinfo {author} {\bibfnamefont {S.~N.}\ \bibnamefont
  {Majumdar}},\ }\href {https://doi.org/10.1103/PhysRevLett.106.160601}
  {\bibfield  {journal} {\bibinfo  {journal} {Phys. Rev. Lett.}\ }\textbf
  {\bibinfo {volume} {106}},\ \bibinfo {pages} {160601} (\bibinfo {year}
  {2011})}\BibitemShut {NoStop}%
\bibitem [{\citenamefont {Evans}\ \emph {et~al.}(2020)\citenamefont {Evans},
  \citenamefont {Majumdar},\ and\ \citenamefont {Schehr}}]{Evans_2020}%
  \BibitemOpen
  \bibfield  {author} {\bibinfo {author} {\bibfnamefont {M.~R.}\ \bibnamefont
  {Evans}}, \bibinfo {author} {\bibfnamefont {S.~N.}\ \bibnamefont
  {Majumdar}},\ and\ \bibinfo {author} {\bibfnamefont {G.}~\bibnamefont
  {Schehr}},\ }\href {https://doi.org/10.1088/1751-8121/ab7cfe} {\bibfield
  {journal} {\bibinfo  {journal} {J. Phys. A: Math. Theor.}\ }\textbf {\bibinfo
  {volume} {53}},\ \bibinfo {pages} {193001} (\bibinfo {year}
  {2020})}\BibitemShut {NoStop}%
\bibitem [{\citenamefont {Jonay}\ \emph {et~al.}(2018)\citenamefont {Jonay},
  \citenamefont {Huse},\ and\ \citenamefont {Nahum}}]{jonay2018coarsegrained}%
  \BibitemOpen
  \bibfield  {author} {\bibinfo {author} {\bibfnamefont {C.}~\bibnamefont
  {Jonay}}, \bibinfo {author} {\bibfnamefont {D.~A.}\ \bibnamefont {Huse}},\
  and\ \bibinfo {author} {\bibfnamefont {A.}~\bibnamefont {Nahum}},\
  }\href@noop {} {\bibinfo {title} {Coarse-grained dynamics of operator and
  state entanglement}} (\bibinfo {year} {2018}),\ \Eprint
  {https://arxiv.org/abs/1803.00089} {arXiv:1803.00089} \BibitemShut {NoStop}%
\bibitem [{\citenamefont {Zhou}\ and\ \citenamefont
  {Nahum}(2019)}]{zhou2019emergent}%
  \BibitemOpen
  \bibfield  {author} {\bibinfo {author} {\bibfnamefont {T.}~\bibnamefont
  {Zhou}}\ and\ \bibinfo {author} {\bibfnamefont {A.}~\bibnamefont {Nahum}},\
  }\href {https://doi.org/10.1103/PhysRevB.99.174205} {\bibfield  {journal}
  {\bibinfo  {journal} {Phys. Rev. B}\ }\textbf {\bibinfo {volume} {99}},\
  \bibinfo {pages} {174205} (\bibinfo {year} {2019})}\BibitemShut {NoStop}%
\bibitem [{\citenamefont {Feng}\ \emph {et~al.}(2022)\citenamefont {Feng},
  \citenamefont {Skinner},\ and\ \citenamefont
  {Nahum}}]{feng2022measurementinducedphase}%
  \BibitemOpen
  \bibfield  {author} {\bibinfo {author} {\bibfnamefont {X.}~\bibnamefont
  {Feng}}, \bibinfo {author} {\bibfnamefont {B.}~\bibnamefont {Skinner}},\ and\
  \bibinfo {author} {\bibfnamefont {A.}~\bibnamefont {Nahum}},\ }\href@noop {}
  {} (\bibinfo {year} {2022}),\ \Eprint {https://arxiv.org/abs/2210.07264}
  {arXiv:2210.07264 [cond-mat.stat-mech]} \BibitemShut {NoStop}%
\bibitem [{\citenamefont {Chahine}\ and\ \citenamefont
  {Buchhold}(2023)}]{chahine2023entanglement}%
  \BibitemOpen
  \bibfield  {author} {\bibinfo {author} {\bibfnamefont {K.}~\bibnamefont
  {Chahine}}\ and\ \bibinfo {author} {\bibfnamefont {M.}~\bibnamefont
  {Buchhold}},\ }\href@noop {} {} (\bibinfo {year} {2023}),\ \Eprint
  {https://arxiv.org/abs/2309.12391} {arXiv:2309.12391 [cond-mat.str-el]}
  \BibitemShut {NoStop}%
\bibitem [{\citenamefont {Leung}\ \emph {et~al.}(2023)\citenamefont {Leung},
  \citenamefont {Meidan},\ and\ \citenamefont {Romito}}]{leung2023theory}%
  \BibitemOpen
  \bibfield  {author} {\bibinfo {author} {\bibfnamefont {C.~Y.}\ \bibnamefont
  {Leung}}, \bibinfo {author} {\bibfnamefont {D.}~\bibnamefont {Meidan}},\ and\
  \bibinfo {author} {\bibfnamefont {A.}~\bibnamefont {Romito}},\ }\href@noop {}
  {\bibinfo {title} {Theory of free fermions dynamics under partial
  post-selected monitoring}} (\bibinfo {year} {2023}),\ \Eprint
  {https://arxiv.org/abs/2312.14022} {arXiv:2312.14022 [quant-ph]} \BibitemShut
  {NoStop}%
\bibitem [{\citenamefont {Carmichael}\ \emph {et~al.}(1989)\citenamefont
  {Carmichael}, \citenamefont {Singh}, \citenamefont {Vyas},\ and\
  \citenamefont {Rice}}]{carmichael1989photoelectron}%
  \BibitemOpen
  \bibfield  {author} {\bibinfo {author} {\bibfnamefont {H.~J.}\ \bibnamefont
  {Carmichael}}, \bibinfo {author} {\bibfnamefont {S.}~\bibnamefont {Singh}},
  \bibinfo {author} {\bibfnamefont {R.}~\bibnamefont {Vyas}},\ and\ \bibinfo
  {author} {\bibfnamefont {P.~R.}\ \bibnamefont {Rice}},\ }\href
  {https://doi.org/10.1103/PhysRevA.39.1200} {\bibfield  {journal} {\bibinfo
  {journal} {Phys. Rev. A}\ }\textbf {\bibinfo {volume} {39}},\ \bibinfo
  {pages} {1200} (\bibinfo {year} {1989})}\BibitemShut {NoStop}%
\bibitem [{\citenamefont {Delteil}\ \emph {et~al.}(2014)\citenamefont
  {Delteil}, \citenamefont {Gao}, \citenamefont {Fallahi}, \citenamefont
  {Miguel-Sanchez},\ and\ \citenamefont {Imamo\ifmmode~\breve{g}\else
  \u{g}\fi{}lu}}]{delteil2014observation}%
  \BibitemOpen
  \bibfield  {author} {\bibinfo {author} {\bibfnamefont {A.}~\bibnamefont
  {Delteil}}, \bibinfo {author} {\bibfnamefont {W.-b.}\ \bibnamefont {Gao}},
  \bibinfo {author} {\bibfnamefont {P.}~\bibnamefont {Fallahi}}, \bibinfo
  {author} {\bibfnamefont {J.}~\bibnamefont {Miguel-Sanchez}},\ and\ \bibinfo
  {author} {\bibfnamefont {A.}~\bibnamefont {Imamo\ifmmode~\breve{g}\else
  \u{g}\fi{}lu}},\ }\href {https://doi.org/10.1103/PhysRevLett.112.116802}
  {\bibfield  {journal} {\bibinfo  {journal} {Phys. Rev. Lett.}\ }\textbf
  {\bibinfo {volume} {112}},\ \bibinfo {pages} {116802} (\bibinfo {year}
  {2014})}\BibitemShut {NoStop}%
\bibitem [{\citenamefont {Brandes}(2008)}]{brandes2008waiting}%
  \BibitemOpen
  \bibfield  {author} {\bibinfo {author} {\bibfnamefont {T.}~\bibnamefont
  {Brandes}},\ }\href {https://doi.org/10.1063/1.3037125} {\bibfield  {journal}
  {\bibinfo  {journal} {AIP Conference Proceedings}\ }\textbf {\bibinfo
  {volume} {1074}},\ \bibinfo {pages} {102} (\bibinfo {year}
  {2008})}\BibitemShut {NoStop}%
\bibitem [{\citenamefont {Albert}\ \emph {et~al.}(2012)\citenamefont {Albert},
  \citenamefont {Haack}, \citenamefont {Flindt},\ and\ \citenamefont
  {B\"uttiker}}]{albert2012electron}%
  \BibitemOpen
  \bibfield  {author} {\bibinfo {author} {\bibfnamefont {M.}~\bibnamefont
  {Albert}}, \bibinfo {author} {\bibfnamefont {G.}~\bibnamefont {Haack}},
  \bibinfo {author} {\bibfnamefont {C.}~\bibnamefont {Flindt}},\ and\ \bibinfo
  {author} {\bibfnamefont {M.}~\bibnamefont {B\"uttiker}},\ }\href
  {https://doi.org/10.1103/PhysRevLett.108.186806} {\bibfield  {journal}
  {\bibinfo  {journal} {Phys. Rev. Lett.}\ }\textbf {\bibinfo {volume} {108}},\
  \bibinfo {pages} {186806} (\bibinfo {year} {2012})}\BibitemShut {NoStop}%
\bibitem [{\citenamefont {Landi}(2021)}]{landi2021waiting}%
  \BibitemOpen
  \bibfield  {author} {\bibinfo {author} {\bibfnamefont {G.~T.}\ \bibnamefont
  {Landi}},\ }\href {https://doi.org/10.1103/PhysRevB.104.195408} {\bibfield
  {journal} {\bibinfo  {journal} {Phys. Rev. B}\ }\textbf {\bibinfo {volume}
  {104}},\ \bibinfo {pages} {195408} (\bibinfo {year} {2021})}\BibitemShut
  {NoStop}%
\bibitem [{\citenamefont {Landi}(2023)}]{landi2023patterns}%
  \BibitemOpen
  \bibfield  {author} {\bibinfo {author} {\bibfnamefont {G.~T.}\ \bibnamefont
  {Landi}},\ }\href@noop {} {} (\bibinfo {year} {2023}),\ \Eprint
  {https://arxiv.org/abs/2305.07957} {arXiv:2305.07957 [quant-ph]} \BibitemShut
  {NoStop}%
\bibitem [{\citenamefont {Coppola}\ \emph {et~al.}(2023)\citenamefont
  {Coppola}, \citenamefont {Karevski},\ and\ \citenamefont
  {Landi}}]{coppola2023conditional}%
  \BibitemOpen
  \bibfield  {author} {\bibinfo {author} {\bibfnamefont {M.}~\bibnamefont
  {Coppola}}, \bibinfo {author} {\bibfnamefont {D.}~\bibnamefont {Karevski}},\
  and\ \bibinfo {author} {\bibfnamefont {G.~T.}\ \bibnamefont {Landi}},\
  }\href@noop {} {\bibinfo {title} {Conditional no-jump dynamics of
  non-interacting quantum chains}} (\bibinfo {year} {2023}),\ \Eprint
  {https://arxiv.org/abs/2311.05515} {arXiv:2311.05515 [cond-mat.stat-mech]}
  \BibitemShut {NoStop}%
\bibitem [{\citenamefont {Bardou}\ \emph {et~al.}(1994)\citenamefont {Bardou},
  \citenamefont {Bouchaud}, \citenamefont {Emile}, \citenamefont {Aspect},\
  and\ \citenamefont {Cohen-Tannoudji}}]{bardou1994subrecoil}%
  \BibitemOpen
  \bibfield  {author} {\bibinfo {author} {\bibfnamefont {F.}~\bibnamefont
  {Bardou}}, \bibinfo {author} {\bibfnamefont {J.~P.}\ \bibnamefont
  {Bouchaud}}, \bibinfo {author} {\bibfnamefont {O.}~\bibnamefont {Emile}},
  \bibinfo {author} {\bibfnamefont {A.}~\bibnamefont {Aspect}},\ and\ \bibinfo
  {author} {\bibfnamefont {C.}~\bibnamefont {Cohen-Tannoudji}},\ }\href
  {https://doi.org/10.1103/PhysRevLett.72.203} {\bibfield  {journal} {\bibinfo
  {journal} {Phys. Rev. Lett.}\ }\textbf {\bibinfo {volume} {72}},\ \bibinfo
  {pages} {203} (\bibinfo {year} {1994})}\BibitemShut {NoStop}%
\bibitem [{\citenamefont {Vidal}\ \emph {et~al.}(2003)\citenamefont {Vidal},
  \citenamefont {Latorre}, \citenamefont {Rico},\ and\ \citenamefont
  {Kitaev}}]{Vidal2003}%
  \BibitemOpen
  \bibfield  {author} {\bibinfo {author} {\bibfnamefont {G.}~\bibnamefont
  {Vidal}}, \bibinfo {author} {\bibfnamefont {J.~I.}\ \bibnamefont {Latorre}},
  \bibinfo {author} {\bibfnamefont {E.}~\bibnamefont {Rico}},\ and\ \bibinfo
  {author} {\bibfnamefont {A.}~\bibnamefont {Kitaev}},\ }\href
  {https://doi.org/10.1103/physrevlett.90.227902} {\bibfield  {journal}
  {\bibinfo  {journal} {Phys. Rev. Letters}\ }\textbf {\bibinfo {volume} {90}}
  (\bibinfo {year} {2003})}\BibitemShut {NoStop}%
\bibitem [{\citenamefont {Bravyi}(2005)}]{bravyi2005lagrangian}%
  \BibitemOpen
  \bibfield  {author} {\bibinfo {author} {\bibfnamefont {S.}~\bibnamefont
  {Bravyi}},\ }\href@noop {} {\bibfield  {journal} {\bibinfo  {journal}
  {Quantum Info. Comput.}\ }\textbf {\bibinfo {volume} {5}},\ \bibinfo {pages}
  {216–238} (\bibinfo {year} {2005})}\BibitemShut {NoStop}%
\end{thebibliography}%
\bibliographystyle{apsrev4-2}

\end{document}